\documentclass[aps,prd,
nofootinbib,groupedaddress]{revtex4-2}

\usepackage{amsmath}
\usepackage{amsfonts}
\usepackage{graphicx}
\usepackage{hyperref}
\usepackage{caption}
\usepackage{slashed}
\usepackage[mathscr]{euscript}
\usepackage{amssymb}

\usepackage{amsmath}
\usepackage{graphicx}
\usepackage{dcolumn}
\usepackage{bm}
\usepackage{epsfig}
\usepackage{amssymb} 
\usepackage{color}
\usepackage{datetime}
\usepackage{wasysym}
\usepackage{slashed} 
\usepackage[mathscr]{euscript}
\usepackage{ulem}

\usepackage{microtype}

\newcommand{\bal}{\begin{align}}
\newcommand{\eal}{\end{align}}
\newcommand{\beq}{\begin{eqnarray}}
\newcommand{\eeq}{\end{eqnarray}}
\newcommand{\nneeq}{\nonumber \end{eqnarray}}
\newcommand{\nnn}{\nonumber}
\newcommand{\nn}{\nonumber \\}
\newcommand{\es}{& = &}
\newcommand{\eqv}{& \equiv &}
\newcommand{\rs}{\, = \,}
\newcommand{\ps}{& + &}
\newcommand{\ms}{& - &}
\newcommand{\ts}{& \times &}

\newcommand{\nt}{\nn \ts}
\newcommand{\np}{\nn \ps}
\newcommand{\nm}{\nn \ms}

\newcommand{\tl}{ & \to & }

\newcommand{\cC}{ {\cal C} }

\newcommand{\cM}{ {\cal M} }
\newcommand{\cH}{ {\cal H} }

\newcommand{\cG}{ {\cal G} }

\newcommand{\cU}{ {\cal U} }
\newcommand{\cL}{ {\cal L} }

\newcommand{\f}{\hspace{-3pt} \not\!}

\newcommand{\2}{ \, { 1 \over 2} \, }
\newcommand{\4}{ \, { 1 \over 4} \, }

\newcommand{\red}{\color[rgb]{1,0,0}}

\begin{document}
\title{Gluon mass and small-$x$ dynamics in hadrons}
\author{ Stanis{\l}aw D. G{\l}azek }
\affiliation{ Institute of Theoretical Physics \\ 
Faculty of Physics, University of Warsaw \\  
ul. Pasteura 5, 02-093 Warsaw, Poland }
\date{20260628 08:00 NY}
\begin{abstract} 
Precise derivation of the logarithmically scale-dependent Hamiltonian eigenstate picture for hadrons 
in the space of virtual quark and gluon states of the canonical front form of QCD requires addressing 
first the problem of divergences stronger than logarithmic, and especially the small-x divergences
in the dynamics of gluons. We propose to facilitate the regularization and cancellation of these 
divergences using a gluon mass parameter and an auxiliary color-octet scalar field corresponding to 
the longitudinally polarized gluons. The auxiliary field decouples from the hadronic constituent 
dynamics when the mass parameter tends to zero, as required in the gauge theory. The same method 
applies in the cancellation of the quadratic ultraviolet transverse divergences in the self-interactions. 
After explaining how the method works in computations of the scattering amplitudes, we describe 
its application to the bound-state eigenvalue problems. We focus on the results it leads to already in 
the second-order weak-coupling expansion for effective Hamiltonians of heavy quarks. They include
the concept of confinement in an effective theory with the gluon mass parameter sent to zero
and a heuristic scenario concerning extension of the Hamiltonian approach to the dynamics of 
light quarks.
\end{abstract}
\maketitle
\section{Introduction}
\label{intro}
For precise application to the physics of hadron constituents, the Hamiltonian of QCD 
needs to be constructed as an operator acting in the space of states of virtual quarks and 
gluons in a definite form of dynamics. Among the forms identified in~\cite{Dirac:1949cp}, 
the commonly known way of describing the evolution of physical systems using time $t$ 
of some observer, is called the instant form (IF). The name originates from the space-time 
hyperplane corresponding to one value of $t$, an instant. The less familiar form, in which 
an observer describes the system evolution from one value of $x^+=t+z$ to another, is 
called the front form (FF). The name comes from the fact that a space-time hyperplane 
corresponding to one value of $x^+$ is swept by the front of a plane wave of light moving 
against the $z$-axis. The choice of the FF of dynamics for describing how quarks and gluons 
form hadrons stems from the desire to describe the hadrons in any state of motion, such as 
being at rest in a laboratory or moving with a speed nearly equal to the speed of light in 
an accelerator. In the IF, boosting hadrons is associated with complex dynamical effects 
because the quantum generators of boosts involve interactions. In the FF, there are 
3 independent Lorentz transformations that are free from interaction effects and the 
canonical dynamics is invariant with respect to them. This feature suggests a simplification 
of the task of relating the observed parton structure of the fast moving hadrons and their 
spectroscopic classification at rest in terms of states of 3 quarks or a quark-antiquark 
pair~\cite{Workman:2022ynf,ParticleDataGroup:2024cfk}. Besides offering an interesting
way to study the quantum boost problem for bound states, the FF Hamiltonian formulation 
of QCD provides the eigenvalue equations determining how the physical hadrons, 
represented by the eigenstates, are built from the quanta of quark and gluon fields.

However, the canonical FF Hamiltonian of QCD is a singular operator. One needs to identify 
a computational path that leads from the canonical theory to the finite solutions of the 
bound-state eigenvalue problems. Asymptotic freedom suggests that the binding mechanism 
for hadronic constituents involves logarithms of their relative momenta. But to reliably compute 
the logarithmic effects in bound-state dynamics, one first needs to find a way to control the 
functions more singular than the logarithms. We propose a method for doing this, including
a simpler derivation of previously obtained results~\cite{Wilson:1994fk,Perry:1994kp,
Serafin:2023pkf}. The difficulty to overcome is that the FF of the theory involves singular 
functions of two distinct kinds of variables. 

The bound-state constituent momenta measured along the $z$-axis greatly differ from the 
momenta in directions perpendicular to that axis. Consequently, the FF Hamiltonians do 
not possess any explicit three-dimensional rotation symmetry that the IF Hamiltonians have. 
Instead of the three-dimensional momenta, say $\vec p$, one works with the transverse momenta 
of quanta, denoted by $p^\perp = (p^1,p^2)$, and the longitudinal momenta, denoted by $p^+
=p^0 + p^3$. The corresponding boost-invariant variables of two kinds are the quanta relative 
transverse momenta $k^\perp = p^\perp - x P^\perp$ and the ratios $x=p^+/P^+$, where $P$ 
denotes the total momentum of the quanta involved in a Hamiltonian interaction term. Divergences 
occur due to $k^\perp$ going to $\infty$ or 0 and due to $x$ going to 0. For example, there 
are singular functions that behave as $1/x^2$ or $1/x$. The Hamiltonian requires a renormalization 
procedure for the divergences in $x$ and $k^\perp$~\cite{Wilson:1994fk}. The first step of any 
such procedure is to regulate the diverging terms.

Regularization of the Hamiltonian terms can be arranged by limiting the range of variables $k^\perp$ 
from above and $x$ from below. Suppose one does it for the transverse and longitudinal coordinates 
separately, {\it e.g.,} see~\cite{Glazek:2000dc}. In that case, the ultraviolet divergences associated with 
the longitudinal and transverse directions pile up on each other, which leads to a complex renormalization 
procedure. One can correlate the regularization cutoffs on $k^\perp$ and $x$ by limiting the total 
invariant-mass squared, denoted by $\cM^2$, of the quanta involved in an interaction term. Each of the 
quanta contributes to $\cM^2$ via $P^+$ times 
\beq
k^- \es k^0 - k^3 \rs {k^{\perp 2}  + m^2 \over x P^+} \ ,
\eeq
with $m$ being a mass assigned to a quantum in the free part of the Hamiltonian. Limiting $\cM$ by 
a cutoff from above, say $\Lambda$, one simultaneously limits $|k^\perp|$ from above and prevents 
$x$ from approaching 0 because $m^2/x$ is limited. The usage of a mass parameter in regularization 
is of interest because mass does not distinguish any direction in the space of four-momenta. The issue 
with gluons, however, is that they are massless, as befits them as gauge bosons. For them, 
\beq
\label{kperponly}
P^+ k^- \es {k^{\perp 2} \over x } \ .
\eeq
Small $x$ is not limited from below by the upper limit on the invariant mass $\cM$ when $k^\perp$ is 
correspondingly small. Therefore, the ultraviolet renormalization procedure is interfered with by the infrared 
issues of the theory. Namely, the small-$x$ region belongs to the ultraviolet regime when $k^\perp$ is 
sizable and it also contributes to the infrared regime when $k^\perp$ is small. This is where the gluon 
mass parameter, $m_g$, can be introduced as a regulator disentangling these regimes. Replacement of 
$k^{\perp 2}$ in Eq.~(\ref{kperponly}) by  $k^{\perp 2} + m_g^2$ makes small $x$ correspond to the 
ultraviolet regime even if $k^\perp$ vanishes. But when one inserts a gluon mass term in the canonical 
FF Hamiltonian of QCD for the regularization purposes, the resulting virtual quark and gluon transition 
amplitudes involve the integrands that qualitatively behave as
\beq
\label{severedivergences}
{m_g^2 \over k^{\perp 2} + m_g^2} \ {1 \over x^2} \ .
\eeq
The small-$x$ singularity is associated with a function of the transverse momentum. The singularity 
strengthens when $k^\perp$ decreases. We propose to counter such gluon mass effects by adding 
quanta of an auxiliary scalar octet field $\phi = \phi^a T^a$ to the dynamics. These quanta correspond 
to the longitudinally polarized gluons. In the limit $m_g \to 0$, the longitudinal quanta decouple from 
the theory when $m_g \to 0$. Such decoupling is known in the case of the FF of QED~\cite{Soper:1971wn,
Yan:1973qf}, but as far as the author knows has not been taken advantage of in the FF Hamiltonian of 
QCD as it is done in this paper. Since the paper is focused on the use of gluon mass for the purpose of 
regularization and renormalization of the FF Hamiltonian of QCD, it does not report on or directly refer to 
most of the rich literature on the FF formulation of quantum field theory and its applications. The literature 
can be traced following~\cite{Kogut:1972di,Yan:1973qg,Lepage:1980fj,Namyslowski:1985zq,Wilson:1994fk,
Burkardt:1995ct,Brodsky:1997de,Carbonell:1998rj,Brodsky:2014yha,Ji:2020ect}. 

Section~\ref{bareH} describes the bare FF Hamiltonian of QCD including our addition of the mass  $m_g$ 
and field $\phi$. Regularization is introduced in Sec.~\ref{regularization}. Cancellation of the small-$x$
singularities in virtual scattering amplitudes is discussed in Sec.~\ref{transition}. The renormalized 
Hamiltonian eigenvalue problem, defined using the renormalization group procedure for effective
particles (RGPEP) and computed using its weak-coupling expansion,  is described in Sec.~\ref{BSeqs}. 
The Hamiltonian dynamics of the quark-antiquark component of a heavy quarkonium is derived in
Sec.~\ref{Qdynamics}, demonstrating the cancellation of small-$x$ singularities. The limit of 
$m_g \to 0$ is obtained in Sec.~\ref{mgto0}, identifying the nature of dominant terms with the 
help of a nonrelativistic approximation for the relative motion of the constituents and commenting
on the issue of confinement. Small-$x$ singularities in components with effective gluons are 
addressed in Sec.~\ref{includinggluons}, including interactions that change the number of 
effective gluons. Discussion in Sec.~\ref{discussion} concludes the paper.
\section{Bare Hamiltonian}                       
\label{bareH}  
Canonical FF Hamiltonian of QCD is derived~\cite{Yan:1973qg,Lepage:1980fj} from the Lagrangian density
\beq
\label{LQCD}
\cL_{\rm QCD}
\es
\bar \psi \left( i / \hspace{-5pt}\partial - g / \hspace{-7pt}A - m \right) \psi 
- 
\2 {\rm ~Tr~} G_{\mu \nu} G^{\mu \nu } \ ,
\eeq
where the sum over quark flavors is assumed but not indicated. The derivation is recalled below for it is used 
later to explain our procedure for dealing with the small-$x$ divergences caused by gluons. 
\subsection{FF Hamiltonian of QCD}
\label{FFHQCD}
By minimizing action described by $\cL_{\rm QCD}$ of Eq.~(\ref{LQCD}), one obtains equations
\beq
\label{quarkEoM}
( i \f \partial - g \f \hspace{-2pt}A - m ) \psi \es 0 \ , \\
\label{gluonEoM}
\partial_\alpha G^{a\alpha \beta} 
+ 
ig [ A_\alpha , G^{\alpha \beta}]^a + J_\psi^{a \beta}
\es 0 \ ,
\eeq
where the quark current is $J_\psi^{a \beta} = -g \bar \psi 
T^a \gamma^\beta  \psi$. In the gauge $A^+=0$, 
Eq.~(\ref{quarkEoM}) relates the field $\psi_-= \Lambda_-\psi$ 
to the fields $\psi_+ = \Lambda_+ \psi$ and $A^\perp$, 
where $\Lambda_\pm = \2 \gamma^0 \gamma^\pm$. 
Namely,
\beq
\label{cpsi} 
\psi_- 
\es 
{ 1 \over i \partial^+ } [ \alpha^\perp (i \partial^\perp - g A^\perp) + \beta m ]  \psi_+ \ .
\eeq
Similarly, Eq.~(\ref{gluonEoM}) with $\beta=+$, provides the constraint 
\beq
\label{A-c}
A^- \es  {2 \partial^\perp \over \partial^+} A^\perp 
+ {2 \over \partial^{+2}} (J_A^+  + J_\psi^+) \ ,
\eeq
with $J_A^{a\mu} = - ig [  \partial^\mu A^\perp , A^\perp ]^a$. Inversion of $\partial^{+2}$ is a remote analog of 
inverting Laplasian in the Poisson equation to obtain the Coulomb potential in the IF of Hamiltonian dynamics. Generally, 
inversion of the derivative $\partial^+$ is understood here in terms of the inverse of momentum in the Fourier transform. 
Field modes constant in $x^-$ will not appear in the regulated theory, see below. When the coupling constant $g$ is set
to 0, the constraints simplify to  the constraint equations for free particles,
\beq
\label{psif-} 
\psi_{f-} 
\es 
{ 1 \over i \partial^+ } \left( \alpha^\perp i \partial^\perp + \beta m \right) \psi_+ \ , \\
\label{Af-}
A_f^- \es  {2 \partial^\perp \over \partial^+} A^\perp \ .
\eeq
One uses these equations to define the fields
\beq
\label{psi_f}
\psi_f \es \psi_+ + \psi_{f-} \ , \\
\label{A_f}
A_f \es \left(A^-_f, A^+=0, A^\perp \right) \ .
\eeq
Canonical formula for the FF Hamiltonian of QCD expressed in terms of the fields $\psi_f$ and $A_f$, is provided 
by the component $P^-$ of the total four-momentum carried by the quark and gluon fields. Thus, $P^-$  is an 
integral of Noether's energy-momentum tensor-density component $+-$ over the front,  
\beq
H_{\rm QCD} \es \int_F \  T_{\rm QCD}^{+ -} \ ,
\eeq
where $\int_F = \int d^4x \, \delta(x^+)$.
The Noether tensor is 
\beq
T_{\rm QCD}^{\mu \nu} 
\es
\sum_\chi
       { \partial \cL_{\rm QCD} \over \partial \partial_\mu \chi } \
       \partial^\nu \chi
-      
g^{\mu \nu}  \cL_{\rm QCD} \ .
\eeq
The sum extends over all fields $\chi$ in $\cL_{\rm QCD}$
with $A^+=0$. The Hamiltonian reads
\beq
\label{HrmQCD}
H_{\rm QCD} \es \int d^2x^\perp dx^- \ \cH_{\rm QCD} \ ,
\eeq
where the integrated density is, {\it cf.} \cite{Lepage:1982gd},
\beq
\label{cHcanQCD}
&&
\cH_{\rm QCD} 
\rs
\bar \psi_f { \gamma^+ (- \partial^{\perp 2} + m^2) \over 2i \partial^+ }  \psi_f
+
\2 A_{f \mu}^a  \partial^{\perp 2}  A_f^{a \mu}   
\nm
(J_{\psi f}^{a \mu}  + J_{A f}^{a \mu}) A_{f\mu}^a 
+
\2 g^2 \bar \psi_f \, \f \hspace{-2pt} A_f 
{ \gamma^+ \over i \partial^+ } 
                          \, \f \hspace{-2pt} A_f \psi_f 
\nm
\4 g^2 [A_{f\mu}, A_{f_\nu}]^a [A_f^\mu, A_f^\nu]^a  
\nm
\2 (J_{\psi f}^{a+} + J_{A f}^{a+}) {1 \over \partial^{+2}} (J_{Af}^{a+} + J_{\psi f}^{a+}) \ ,
\eeq
with currents 
\beq
\label{quarkcurrent}
J_{\psi f}^{a \mu} \es -g \bar \psi_f T^a \gamma^\mu  \psi_f \ , \\
\label{gluoncurrent}
J_{A f}^{a\mu} \es ig [  \partial^\mu A_{f \nu}, A_f^\nu ]^a \ .
\eeq
These currents depend on the fields $A^\perp$ and $\psi_+$ only, since they include 
explicit solutions for $\psi_{f-}$ and $A_f^-$. They differ from the currents
\beq
J_\psi^{a \mu} \es -g \bar \psi T^a \gamma^\mu  \psi 
\ , \\
\label{fullJA}
J_A^{a\mu}
\es
ig [ A_\alpha , G^{\alpha \mu}]^a 
\ .
\eeq
\subsection{Gluon mass and auxiliary field} 
\label{amendedH}
The gluon mass term and the associated term for the auxiliary octet scalar field $\phi = \phi^a T^a$, 
are added to the density $\cH_{\rm QCD}$ of Eq.~(\ref{cHcanQCD}) to form the amended density,
\beq
\label{cHcanQCDmg}
&&\cH^{m_g}_{\rm QCD} 
\rs
\cH_{\rm QCD} 
-
\2 m_g^2 A_{f \mu}^a   A_f^{a \mu} \\
&+&
\2 \phi^a (- \partial^{\perp 2} + m_g^2 ) \phi^a 
+
m_g \phi^a {1 \over \partial^+} (J_{Af}^{a+} + J_{\psi f}^{a+}) \ . \nnn
\eeq
The gluon mass parameter $m_g$ and auxiliary field $\phi$ introduced in Eq.~(\ref{cHcanQCDmg}) suffice 
for the purpose of regulating the gluon dynamics, see Sec.~\ref{regularization}. When $m_g$ is set to 0, 
the amended density becomes equal to the canonical $\cH_{\rm QCD}$ of Eq.~(\ref{cHcanQCD}) plus a 
density for a decoupled, free massless field $\phi$. The latter can be ignored. Thus the limit $m_g \to 0$ 
recovers the canonical FF Hamiltonian of QCD.

In Eq.~(\ref{cHcanQCDmg}), the mass term for the gluon field $A$ involves only the constrained field $A_f$ of 
Eq.~(\ref{A_f}). The kinetic term for the field $\phi$ has the standard form for scalar bosons. The coupling of $\phi$ 
to the quark current is similar in structure to the couplings introduced in massive QED~\cite{Stueckelberg:1938zz,
Soper:1971wn,Glazek:2020xpi}. The auxiliary field $\phi$ couples to the color current of gluons. However, it does 
not couple directly to itself, despite that it carries color. Our scheme differs from the generalized Stueckelberg 
formalism for massive Yang-Mills fields, in which additional constraints are imposed to eliminate terms inversely 
proportional to the gauge-boson mass parameter~\cite{Kunimasa:1967zza}. Such additional constraints are not 
needed here. 

The Fourier modes of the field $\phi$ with momentum $p$ couple to the color currents proportionally to 
$m_g/p^+$. This ratio diverges for $p^+ \to 0$. Precisely such coupling yields the cancellation of the severe 
small-$x$ divergences due to the assignment of mass $m_g$ to the transverse gluons, see Sec.~\ref{transition}.

The density $\cH^{m_g}_{\rm QCD}$ can also be derived from the Lagrangian density 
\beq
\label{LQCDmg}
\cL^{m_g}_{\rm QCD}
\es
\cL_{\rm QCD} +
\2 ( m_g A^a_\mu + \partial_\mu \phi^a )^2 \ ,
\eeq
proceeding in parallel to Sec.~\ref{FFHQCD}, keeping $A^+=0$. The term added to $\cL_{\rm QCD}$ in 
Eq.~(\ref{LQCDmg}) is not gauge invariant, and the regularization purpose it serves is not. Since the field 
$\phi$ is assumed very weakly coupled to gluons and since color is confined, any physical effects due to 
that field, if it existed, would be hard to detect. In principle, $\phi$ might contribute to the gravitational 
effects.

Quark fields $\psi$ obey the same equations in the case of $\cL^{m_g}_{\rm QCD}$ as in the case of 
$\cL_{\rm QCD}$, for the added term does not depend on the fermion fields. Therefore, the constraint 
on $\psi_-$ is the same as in QCD with $m_g=0$, Eq.~~(\ref{cpsi}). Keeping $A^+=0$, variation of $A^-$ 
yields the constraint 
\beq
\label{A-c1}
A^- \es  A_f^- 
+ {2 \over \partial^{+2}} (J_{Af}^+  + J_{\psi f}^+) + {2m_g \over \partial^+} \phi \ ,
\eeq
which differs from the canonical Eq.~(\ref{A-c}) by the last 
term only, given the identities 
$J_A^+ = J_{Af}^+$ and $J_\psi^+ = J_{\psi f}^+$ 
that follow from $\gamma^{+2}=0$. The field $\phi$ 
obeys
\beq
\Box \phi + m_g \partial_\mu  A^\mu \es 0 \ .
\eeq
The constraint Eq.~(\ref{A-c1}) implies
\beq
\label{constraintpartialA1}
\partial_\mu A^\mu \es  {1 \over \partial^+} (J_A^+ + J_\psi^+) + m_g \phi \ , \\
\label{phiJJ}
( \Box + m_g^2 ) \phi
\es
- {m_g \over \partial^+} (J_A^+ + J_\psi^+)   \ .
\eeq
The latter equation exhibits the so-called ``longitudinal'' or ``good'' currents~\cite{Fritzsch:1972jv} as the only 
sources of the field $\phi$. Evaluating $P^-$ that follows from the energy-momentum tensor density implied 
by $\cL^{m_g}_{\rm QCD}$, one obtains the amended Hamiltonian density $\cH_{\rm QCD}^{m_g}$.
\subsection{Quantum Hamiltonian}
\label{QH}
The Hamiltonian,
\beq
H_{\rm QCD}^{m_g} \es \int d^2x^\perp dx^- \ \cH_{\rm QCD}^{m_g} \ ,
\eeq
where the density $\cH_{\rm QCD}^{m_g}$ is a function of fields $\psi_f$, $A_f$, and $\phi$, given in 
Eq.~(\ref{cHcanQCDmg}), is turned into a quantum operator $\hat H_{\rm QCD}^{m_g}$ by replacing 
the fields at $x^+=0$ with operators, 
\beq
\label{barepsif}
\hat \psi_f
\es
\sum_{c ,\sigma} \hspace{-3pt} \int \hspace{-2pt} [p]
\left[  u_{p\sigma} \chi_c \hat b_{p \sigma c} e^{-ipx}
      +    v_{p\sigma} \chi_c \hat d^\dagger_{p\sigma c} e^{ipx}
\right] \hspace{-2pt} , \\
\label{bareAf}
\hat A_f^\mu
\es
\sum_{c,\sigma}  \hspace{-3pt}\int \hspace{-2pt} [p]
\left[  \varepsilon^\mu_{p\sigma} T^c \hat a_{p\sigma c} e^{-ipx}
        +  \varepsilon^{\mu *}_{p\sigma} T^c \hat a^\dagger_{p\sigma c} e^{ipx}
\right] \hspace{-2pt} , \\
\label{hatphi}
\hat \phi
\es
\sum_{c}\hspace{-3pt}
\int \hspace{-2pt} [p]
\left[  -i T^c \hat a_{p3c} e^{-ipx}
        +  i T^c \hat a^\dagger_{p3c} e^{ipx}
\right] .
\eeq
In this notation, $c$ refers to colors of $SU(3)$ and $\sigma$ denotes spin projections on $z$-axis. Momentum integration measure 
reads $[p] = dp^+ d^2p^\perp /[2p^+ (2\pi)^3]$ with $p^+ > 0$, $p^\perp = (p^1,p^2)$, $px = p^-x^+/2 + p^+x^-/2 - p^\perp 
x^\perp$. The quark spinors, $u_{p\sigma}$, $v_{p\sigma}$, and gluon polarization vectors, $\varepsilon^\mu_{p\sigma}$, are given in~\cite{Lepage:1980fj}, App. A. Flavors of quarks are kept in mind. The creation and annihilation operators, $\hat b, \hat b^\dagger, 
\hat d, \hat d^\dagger, a, a^\dagger$, will be commonly referred to as particle operators. They are assumed to satisfy the 
anti-commutation relations for fermions and commutation relations for bosons, 
\beq
\label{comrules}
\left\{ \hat b_{p \sigma c}, \hat b_{p' \sigma' c'}^\dagger \right\}
\es
\left\{ \hat d_{p \sigma c}, \hat d_{p' \sigma' c'}^\dagger \right\}
\rs
\left[ \hat a_{p \sigma c}, \hat a_{p' \sigma' c'}^\dagger \right]\\
\es
2p^+(2\pi)^3 \delta^3(p-p') \delta_{\sigma \sigma'} \delta_{cc'} \ .
\eeq
Other commutators or anti-commutators vanish. Fermion operators commute with the boson 
operators. By definition, the annihilation operators $b, d, a$ produce 0 when they act on the 
vacuum state, denoted by $|0\rangle$. Products of particle operators in all terms of the Hamiltonian 
$\hat H_{\rm QCD}^{m_g}$ are normal-ordered according to the pattern $a^\dagger b^\dagger 
d^\dagger d b a$. The normal-ordering is indicated by double dots. Thus, 
\beq
\label{EqbareH}
\hat H_{\rm QCD}^{m_g} \es \int d^2x^\perp dx^- \, 
: \cH_{\rm QCD}^{m_g}(\hat \psi_f, \hat A_f, \hat \phi): \ .
\eeq
This Hamiltonian is singular and requires regularization as a part of the renormalization procedure
described in next sections. The normal ordering will be always present but usually not indicated. 
Hats will be omitted to farther simplify the notation.
\section{Regularization}                   
\label{regularization} 
Regulating the Hamiltonian of Eq.~(\ref{EqbareH}) begins with limiting from below the range of momentum 
$p^+$ labeling creation and annihilation operators in the quantum fields in Eqs.~(\ref{barepsif}), (\ref{bareAf}) 
and (\ref{hatphi}). The lower bound on $p^+$ is denoted by $\epsilon^+ > 0$. This lower bound is meant to 
be smaller than any $p^+$ needed to be considered in any description of any physical process.
\subsection{Particle momenta and cutoff $\epsilon^+$}
\label{labels}
To introduce the cutoff $\epsilon^+$, we define the integrals over the momenta that label creation and annihilation 
operators in Eqs.~(\ref{barepsif}), (\ref{bareAf}) and (\ref{hatphi}) as discrete sums. The continuous notation 
is meant to correspond to the discrete momentum spacing that is too small to notice in physical applications. Care 
needs to be exercised because there are two limits to consider in discussing physical processes, a large-volume limit 
and a long-time limit, {\it cf.}~\cite{Gell-Mann:1953dcn}. We introduce a large cuboid extending from $-2L$ to $2L$ in 
$x^-$ and from $-L$ to $L$ in $x^1$ and $x^2$. The cuboid is assumed large enough to contain all processes of physical 
interest without difficulty. The physical processes are meant to develop over the front ``time'' $x^+$ much shorter than $4L$. 

The discrete plane-wave momenta in the quark and gluon fields are defined using the periodic boundary conditions on 
the cuboid walls at $x^+=0$,
\beq
p_n = \epsilon \, n \ , \ n \rs (n^+,n^1,n^2) \ ,
\eeq
where $\epsilon = \pi/L$, $n^+$ is a natural number while $n^1$ and $n^2$ are integers. The number $n^+$ is positive 
because a free particle of positive mass and finite energy $E$ has $p^+= E + p^z > 0$ no matter how fast and in what 
direction the particle moves. The assumption is that, in the absence of interaction, the theory describes the energy of free 
particles, which we consider to be the field quanta, {\it cf.}~\cite{Weinberg:1995mt}, p. 297. Since the box can be 
arbitrarily large, $\epsilon$ can be arbitrarily small. 

Suppose one sets an ultraviolet cutoff $\Delta$ on the particle energy~\cite{Harindranath:1987ex,Bartnik:1988fu,Glazek:1986hs,
Glazek:1997sd},
\beq
E_n \es ( p_n^+ + p_n^-)/2 < \Delta \ ,
\eeq
where $p_n^-=(p_n^{\perp 2} + \mu^2)/p_n^+$. Such cutoff implies that
\beq
\Delta - \Delta_n& < & p_n^+ < \Delta + \Delta_n \ ,
\eeq
with $\Delta_n = \sqrt{\Delta^2-p_n^{\perp 2} - \mu^2 }$. Thus, $p_n^+ > \epsilon^+$,  where $\epsilon^+ = 
(\mu^2+ p^{\perp 2}_n)/(2\Delta) > 0$ for $\Delta \gg \mu^2 \gg \epsilon^2$. Thus, the energy cutoff implies 
that the momentum $p_n^+$ for particles with positive masses is separated by the positive gap $\epsilon^+$ 
from zero. In the infrared box limit, $L \to \infty$, and the ultraviolet cutoff limit, $\Delta \to \infty$, the question 
of which quantity is greater, $\epsilon$ or $\epsilon^+$, has an answer depending on the order of limits,
\beq 
{\epsilon^+ \over \epsilon}  \es {\mu L \over 2\pi} \, {\mu \over \Delta} \ . 
\eeq
This formula relates the ratio ${\epsilon^+ /\epsilon}$ to the ratio of the number of the particle Compton wavelengths 
that fit in the cube edge and the number of free particles at rest whose total energy equals the cutoff $\Delta$. We 
assume the infrared, large $L$ limit is taken first, for a fixed mass $\mu$. In that case, $\epsilon^+ \gg \epsilon$. In 
the field expansions, the sum over momenta $p_n$ involves $p_n^+$ separated from 0 by the gap $\mu^2/(2\Delta)$. 
Therefore, the integrals in Eqs.~(\ref{barepsif}), (\ref{bareAf}), and (\ref{hatphi}), do not extend down to $p^+=0$. 
Consequently, the Hamiltonian $H_{\rm QCD}^{m_g}$ does not contain any terms that are made only of creation or 
only of  annihilation operators. Such terms would imply that a natural multiple of $\epsilon^+$ equals 0. 

The largest value of $p^+$ allowed in the operator labels in Eqs.~(\ref{barepsif}), (\ref{bareAf}) and (\ref{hatphi}), 
say $Q^+$, is assumed large enough to consider all physical processes of interest. Since every physical process is 
characterized by some conserved value of its total plus momentum, say $P^+$, and since the plus momentum 
of each participating field quantum is smaller than $P^+$, it is sufficient to assume that $Q^+$ is finite but 
greater than the total $P^+$ in any process of physical interest. 
\subsection{The weak-coupling issue of vacuum}
\label{vacuum}
The Hamiltonian of Eq.~(\ref{EqbareH}) is normal-ordered and, according to Sec.~\ref{labels}, all its terms 
contain at least one annihilation operator. Such terms annihilate the bare vacuum state, denoted by $|0\rangle$. 
Therefore, the state $|0\rangle$ is an eigenstate of $H$ with an eigenvalue 0. It is also an eigenstate with 
eigenvalue 0 of the normal-ordered operators $P^+$ and $P^\perp$, which makes it a candidate for the 
theory vacuum state.

The issue is that the interactions can cause formation of Hamiltonian eigenstates with eigenvalues smaller 
than 0, for any finite eigenvalues of $P^+$ and $P^\perp$. But when the regulated interaction terms are 
made arbitrarily small, by making the bare coupling constant $g$ arbitrarily small, all eigenstates with positive 
$P^+$ must have eigenvalues $P^-$ greater than 0. It has to be so because all the quanta of the regulated 
theory are assigned masses squared greater than 0. Therefore, all quanta necessarily contribute only positive 
amounts to the total value of free $P^-$. The interaction energy is too small to cancel that free positive 
$P^-$ when $g$ tends to 0. Hadrons are then considered to be states obtained by acting with the creation 
operators on $|0\rangle$. 

The situation changes when the coupling constant is increased and the Hamiltonian develops negative eigenvalues 
due to the interactions. Such eigenvalues would invalidate the assumption that the eigenstate $|0\rangle$ of 
eigenvalue 0 is the theory's ground state. Therefore, we assume that no such strong binding effect occurs in QCD 
with physically justified values of the coupling constant and masses of the quanta. The issue becomes whether the 
squares of masses in the renormalized effective-particle FF kinetic energies, including the adjustment of the mass-squared 
counterterms to be discussed later, are large enough to prevent interactions from generating negative eigenvalues 
$P^-$ for bound states of quarks and gluons. The author does not know any example of such negative eigenvalues 
in any realistic theory. 
\subsection{Regulating interaction terms}    
\label{regf}
Interaction terms in $H_{\rm QCD}^{m_g}$ do not converge as functions of the particle momenta. Such interactions 
cause arbitrarily large changes of momenta and produce diverging transition amplitudes. The pertinent measure of 
momentum is $p^-$, an eigenvalue of the free part of the Hamiltonian, obtained from $H_{\rm QCD}^{m_g}$ by 
setting the bare coupling constant $g$ to 0. The free part is denoted by $H_f$,
\beq
\label{hatHf}
H_f \es  
\sum_{\sigma = 1}^2 \int[p] \, {p^{\perp 2} + m^2 \over p^+} 
\left( b^\dagger_{p\sigma}b_{p\sigma} + d^\dagger_{p\sigma} d_{p\sigma} \right)
\np
\sum_{\sigma = 1}^3 \int[p] \, {p^{\perp 2} + m_g^2 \over p^+} \, a^\dagger_{p\sigma} a_{p\sigma} \ .
\eeq
It results from the first two density terms on the right-hand side of Eq.~(\ref{cHcanQCD}) and from the terms second 
and third on the right-hand side of Eq.~(\ref{cHcanQCDmg}). Quanta of the auxiliary gluon field $\phi$ are accounted 
for by summing over 3 instead of only 2 polarizations. The interaction part of $H_{\rm QCD}^{m_g}$ is denoted by $H_I$,
\beq
\label{H0}
H_{\rm QCD}^{m_g} \es H_f + H_I \ .
\eeq
With all annihilation operators commonly denoted by $a$, the structure of $H_I$ reads
\beq
\label{HIcan}
H_I \es \sum_i H_i \ , \\
\label{initialHi}
H_i \es  \hspace{-4pt}
           \left[ \prod_{m =1 }^{c_i} \hspace{-2pt} \int [p_m]  a_{p_m}^\dagger \right]  \hspace{-4pt}
           \left[ \prod_{ n = 1}^{a_i} \hspace{-2pt} \int [q_n ]  a_{q_n }              \right]  \hspace{-3pt}
           \tilde \delta_{c.a} h_i(\bar p, \bar q) ,
\eeq
where $c_i$ and $a_i$ are the numbers of creation and annihilation operators in the term $i$, respectively. The coefficient
$h_i(\bar p, \bar q)$ depends on the set of created particles' quantum numbers, $\bar p$, which includes $p_m$ with 
$m=1, ...,c_i$, and the set of annihilated particles' quantum numbers, $\bar q$, which includes $q_n$ with $n=1, ...,a_i$. 
Canonical coefficients $h_i$ are proportional to the coupling constant $g$ or its square. The symbol $\tilde \delta_{c.a}$ 
stands for the $\delta$-function that secures conservation of  the total kinematic momentum of quanta involved in a term,
\beq
\label{deltac.a}
\tilde \delta_{c.a} 
\es 2 (2\pi)^3 \delta    \left(p_c^+ - p_a^- \right) \
                    \delta^2 \left(p_c^\perp - p_a^\perp \right) \ ,
\eeq
where the subscripts $c$ and $a$ refer to quanta created and annihilated by a term, respectively,
\beq
\label{PCPA}
p_c \es \sum_{m=1}^{c_i} p_m \ , \ p_a \rs \sum_{n=1}^{a_i} q_n \ .
\eeq
Our regularization scheme sets a limit on the magnitude of change of the particle momenta that the interaction terms
can cause. We first describe the regularization of the interaction terms that are linear in $g$ and then of the terms 
that are quadratic in $g$.
\subsubsection{ Terms linear in $g$ }
\label{linearing}
Regularization of terms linear in $g$ amounts to inserting in them regulating factors $r_{\bar p, \bar q}$, 
\beq
\label{Tif}
T(\bar p , \bar q) & \to & r_{\bar p, \bar q} \ T(\bar p , \bar q) \ ,
\eeq
where, using notation of Eq.~(\ref{PCPA}), 
\beq
\label{fr}
r_{\bar p, \bar q} \es e^{ - [ r^2 \,  (\cM_c^2  - \cM_a^2) ]^2 } \ ,
\eeq
with $\cM_c^2 = p_c^2$  and $\cM_a^2 = p_a^2$. The form of $r_{\bar p, \bar q}$ in Eq.~(\ref{fr}) is an example of 
a factor that limits the invariant-mass changes and hence respects the 7 kinematic Poincar\'e symmetries of the FF of 
dynamics. Since our renormalization procedure aims to remove dependence of the effective dynamics on the 
regularization, as it is lifted by taking the limit $r \to 0$, the choice of function $r_{\bar p, \bar q}$ is {\it in principle} 
irrelevant. The choice in Eq.~(\ref{fr}) is one of many that the author tried in various computations. For example, one 
could choose $\exp[-r^2 (\cM_c^2 + \cM_a^2)]$ and obtain different expressions for the counterterms.

The insertion of regulating factors in terms linear in $g$, is illustrated below using the three-gluon interaction 
term. The term originates from the density term $-J_{A f}^{a \mu} A_{f\mu}^a$ for transverse gluons in 
Eq.~(\ref{cHcanQCD}). Before regularization, the canonical term reads
\beq
\label{HA3k12}
H_{A^3} \es g \sum_{c, \sigma} \int [123] 
\, a_1^\dagger a_2^\dagger a_3 \ Y_{123} \, \tilde \delta_{12.3} + h.c. \ , 
\eeq
where 
\beq
\label{Y123}
&&
Y_{123} \rs i f^{c_1 c_2 c_3}
\nt
\left[ \varepsilon_1^*\varepsilon_2^* \,\varepsilon_3k_{12} 
-
\varepsilon_1^*\varepsilon_3 \, \varepsilon_2^* k_{12}/x_2 
-
\varepsilon_2^*\varepsilon_3 \, \varepsilon_1^* k_{12}/x_1 \right] .
\eeq
The symbols $\varepsilon$ denote the gluon polarization four-vectors. The three-gluon vertex function 
$Y_{123}$ grows proportionally to $k^\perp_{12}$, the relative transverse momentum of the created gluons 
1 and 2, $k_{12}^\perp = x_2 p_1^\perp - x_1 p_2^\perp$, $x_1 = p_1^+/p_3^+$, $x_2 = p_2^+/p_3^+ = 
1 - x_1 $, $k_{12}^+=0$. This growth leads to the ultraviolet transverse divergences in transition amplitudes. 
The vertex also grows as $1/x_1$ or $1/x_2$ when $x_1$ or $x_2$ tend to 0, which leads to the small-$x$ 
divergences. Additional complexity results from the ratios $k_{12}^\perp/x_1$ and $k_{12}^\perp/x_2$. For 
example, a term with an infrared $k^\perp$ may become ultraviolet when $x \to 0$. We use the gluon mass 
parameter $m_g$ to untangle such mixtures of the ultraviolet and infrared divergences. The regulated 
three-gluon interaction term reads
\beq
\label{regHA3}
H_{A^3}^r \es g \sum_{c,\sigma} \hspace{-3pt} \int [123] 
\, a_1^\dagger a_2^\dagger a_3 \ Y_{123} \, \tilde \delta_{12.3} r_{12,3} + h.c.  , \\
r_{12,3} \es e^{ -[ r^2 \,  (\cM_{12}^2 - m_g^2)]^2 } \ .
\eeq
The invariant mass $\cM_{12}$ only depends on the relative motion 
variables of gluons 1 and 2. Denoting $x=x_1$ and $k^\perp = k_{12}^\perp$,
\beq
\cM_{12}^2 \es {k^{\perp 2} + m_g^2 \over x(1-x) } \ .
\eeq
When the invariant mass is finite and $r \to 0$, the regulating factor $r_{12,3}$ tends to 1. When the invariant mass 
diverges and the parameter $r$ is fixed, the factor tends to 0. The regulating suppression of the vertex occurs when 
$\cM_{12}$ exceeds the magnitude of $1/r$ or, for $|k^\perp|$ smaller than $m_g$, when $x$ decreases below 
$r^2 m_g^2$. The presence of $m_g$ guarantees that $\cM_{12} \to \infty$ when $x \to 0$. Thus the small-$x$ gluon 
divergences are turned into ultraviolet divergences. Without $m_g$, the region of small $x$ would not be regulated 
by the factor $r_{12,3}$, for one could have $k^\perp \sim x^\alpha $ with $\alpha \ge 1/2$. The non-locality of 
vertices regulated using factors $r_{12,3}$ is analyzed in~\cite{Glazek:2010zr,Glazek:2010zza}.

All terms linear in $g$ in the Hamiltonian $\hat H_{\rm QCD}^{m_g}$ are regulated using factors $r_{\bar p,\bar q}$ 
as in the above example. Such terms contain two creation operators, one for a quantum of some mass $m_1$ and 
another one for a quantum of some mass $m_2$, and an annihilation operator for some quantum of mass $m_3$, in correspondence to  $a_1^\dagger$, $a_2^\dagger$ and $a_3$ in Eq.~(\ref{regHA3}). One uses the invariant mass of 
quanta 1 and 2, squared, 
\beq
\label{cM12}
\cM_{12}^2 \es {k_{12}^{\perp 2} + m_1^2 \over x_1}
+ {k_{12}^{\perp 2} + m_2^2 \over x_2} \ .
\eeq
The regulating factor is
\beq
\label{3fields}
r_{12,3} \es e^{ - [ r^2 \, (\cM_{12}^2 - m_3^2)]^2} \ .
\eeq
\subsubsection{ Terms quadratic in $g$ }
\label{quadraticing}
Regularization of the canonical terms proportional to $g^2$ concerns products of four fields. There are two kinds
of them. One kind results from the constraint Eqs.~(\ref{cpsi}) and (\ref{A-c1}). These terms involve an inverse of 
$\partial^+$. They are traditionally called seagulls. The other kind originates from the square of $G^{a\mu \nu}$ 
in the Lagrangian density. Both kinds are treated in the same way, described below using the example of a seagull.

Constraint Eqs.~(\ref{A-c}) and (\ref{A-c1}) relate gauge field component $A^-$ to $1/\partial^{+2}$ acting on the 
color current $J$. The latter involves products of two fields. Therefore, when a color current is multiplied by the 
constrained $A^-$, the resulting density term is of the form $J (1/ \partial^{+2}) J$. Each $J$ can be though of as 
creating or annihilating a {\it gedanken} gluon that carries the momentum associated with $1/\partial^{+2}$. The 
imagined gluon together with the current $J$ form a vertex that is regulated by the factor $r_{12,3}$ in the same 
way as the three-gluon vertex is in Eq.~(\ref{regHA3}). In the fermion seagulls, the {\it gedanken} quantum is a fermion 
rather than a boson. In the Hamiltonian term due to the product of commutators, $g^2[A^i,A^j]^a [A^i,A^j]^a$, each 
commutator is treated as a current, even though no constraints are involved and the inverse of $\partial^+$ is absent.

The rgularization of terms proportional to $g^2$ is illustrated using the gluon seagull term
\beq
\label{HJJ}
H_{JJ} \es \int dx^- d^2x^\perp \ 
\2 (J_{\psi f}^{a+} + J_{A f}^{a+}) 
\nt
{-1 \over \partial^{+2}} 
(J_{Af}^{a+} + J_{\psi f}^{a+}) \ ,
\eeq
originating from the last density term in Eq.~(\ref{cHcanQCD}). It involves four terms, each of which is a product of 
four fields. All these products are regulated according to the same rule. It suffices to describe one product. We describe
in detail the regularization of the term that involves only quark and antiquark operators,
\beq
\label{HJJbeforeNO}
H_{J_\psi J_\psi} 
\es \int dx^- d^2x^\perp \ 
\2 J_{\psi f}^{a+} {1 \over (i \partial^+)^2} J_{\psi f}^{a+} \ .
\eeq
The current at $x^+=0$ is
\beq
J_{\psi f}^{a +} (x) \es -g
\sum_{c_1, c_2} \sum_{\sigma_1, \sigma_2} \ 
\chi_{c_1}^\dagger T^a \chi_{c_2}
\int[p_1p_2] 
\nt \hspace{-3pt}
\left[  
           \bar u_{p_1\sigma_1} b_{p_1\sigma_1 c_1}^\dagger  e^{ip_1x}
      +   \bar v_{p_1\sigma_1} d_{p_1\sigma_1 c_1}               e^{-ip_1x}
\right] \gamma^+ 
\nt \hspace{-3pt}
\left[  
           u_{p_2 \sigma_2} b_{p_2 \sigma_2 c_2} e^{-ip_2x}
      +   v_{p_2 \sigma_2} d^\dagger_{p_2 \sigma_2 c_2} e^{ip_2x}
\right] .
\eeq
The spinor products, in the representation of the Dirac matrices introduced in~\cite{Glazek:2013qba}, 
\beq
\label{uu1+2}
\bar u_1 \gamma^+ u_2 \es 
\bar v_1 \gamma^+ v_2 \rs 
2 \sqrt{p_1^+ p_2^+} \ \delta_{\sigma_1, \sigma_2} \ , \\
\label{uv1+2}
\bar u_1 \gamma^+ v_2 \es
\bar v_1 \gamma^+ u_2 \rs
2 \sqrt{p_1^+ p_2^+} \ \delta_{\sigma_1, - \sigma_2} \ .
\eeq
Hence the normal-ordered, regulated current is
\beq
\label{jquark1}
:J_{\psi f}^{a +} \hspace{-2pt} : \hspace{-2pt} \es \hspace{-4pt} -g \hspace{-2pt}
\sum_{c,\sigma} \hspace{-2pt}
\chi_{c_1}^\dagger T^a \chi_{c_2} \hspace{-2pt}
\int[p_1p_2] 2 \sqrt{p_1^+ p_2^+} 
\left[ : \hspace{-2pt} \{ ~ \} \hspace{-2pt} : \right]^r \hspace{-2pt} ,
\eeq
where the bracket $\left[ : \hspace{-2pt} \{ ~ \} \hspace{-2pt} : \right]^r$ is
\beq
\label{jquark2}
&&
            b_1^\dagger 
            b_2 e^{i(p_1-p_2)x}
            \left[  \theta_{1-2} r_{1,2(1-2)} + \theta_{2-1} r_{2,1(2-1)} \right]
            \delta_{\sigma_1 , \sigma_2}
\nm                    
            d^\dagger_2 
            d_1 e^{i(p_2-p_1)x}
            \left[ \theta_{2-1} r_{2,1(2-1)} + \theta_{1-2} r_{1,2(1-2)} \right]
            \delta_{\sigma_1 , \sigma_2}
\np
            b_1^\dagger  
            d^\dagger_2 e^{i(p_1+p_2)x} r_{12,(1+2)}
            \delta_{-\sigma_1 , \sigma_2}
\np
            d_1            
            b_2 e^{-i(p_1+p_2)x} r_{12,(1+2)}
            \delta_{-\sigma_1 , \sigma_2} \ .
\eeq
It carries the superscript $r$ to indicate the insertion of the regulating factors. The {\it gedanken} gluon is assigned 
the mass $m_g$ for evaluating its minus momentum. The Heaviside's $\theta$-function $\theta_{i-j} = 
\theta(p_i^+-p_j^+)$. Thus the regulated quark seagull is 
\beq
\label{HJJregulatedNO}
H_{J_\psi J_\psi}^r \es \int_F
\left[  J_{\psi f}^{a+} {1 \over (i \partial^+)^2} J_{\psi f}^{a+} \right]^r \\
\es
\int_F
:
\left[  : J_{\psi f}^{a+} : \right]^r 
{1 \over (i \partial^+)^2} 
\left[  : J_{\psi f}^{a+} : \right]^r 
: \ .
\label{HJJregulatedNO}
\eeq
The colon, a symbol of the normal-ordering, is dropped below. We display an example of the term that 
transforms a quark-antiquark pair into another such pair.
\beq
\label{seagull}
H_{J_\psi J_\psi}^{r \, q\bar q}
\es
\2 g^2
\sum_{1234 \tilde c} \  T^{\tilde c}_{12} T^{\tilde c}_{34} 
\nt
\int[1234]  \
\tilde \delta_{c.a} \, \left\{ ~~ \right\}  4 \sqrt{p_1^+ p_2^+ p_3^+ p_4^+}  \ ,
\eeq
where 
\beq
\label{currentfr}
\left\{ ~~ \right\} 
\es    
            b_1^\dagger  d^\dagger_4 d_3 b_2 \
            \theta_{1-2} r_{1,2(1-2)} \delta_{\sigma_1 , \sigma_2} \nt
           {- 2 \over (p_3^+-p_4^+)^2}
            \theta_{3-4} r_{3,4(3-4)}  \delta_{\sigma_3 , \sigma_4}
\np      
            b_1^\dagger  d^\dagger_4 d_3 b_2  \
            \theta_{2-1} r_{2,1(2-1)}  \delta_{\sigma_1 , \sigma_2} \nt
           {- 2 \over (p_3^+-p_4^+)^2}                    
            \theta_{4-3} r_{4,3(4-3)} \delta_{\sigma_3 , \sigma_4} 
\np       
            b_1^\dagger d^\dagger_2 d_3 b_4  \
            r_{12,(1+2)} \delta_{-\sigma_1 , \sigma_2} \nt
           {2 \over (p_3^++p_4^+)^2}  
            r_{34,(3+4)} \delta_{-\sigma_3 , \sigma_4} 
            \ .
\eeq
\subsubsection{Fully regulated bare Hamiltonian}                                              
\label{fullyreg}
Fully regulated QCD Hamiltonian with the gluon mass $m_g$ and field $\phi$, denoted by $H^{m_g \,  r}_{\rm QCD}$, reads
\beq
\label{EqbareHr}
H_{\rm QCD}^{m_g \, r} \es \int d^2x^\perp dx^- \, 
\cH_{\rm QCD}^{m_g \, r}(\psi_f, A_f, \phi) \ ,
\eeq
where
\beq
\label{cHcanQCDmgr}
\cH^{m_g \,  r}_{\rm QCD} 
\es
\cH_{\rm QCD}^r 
-
\2 m_g^2 A_{f \mu}^a   A_f^{a \mu}
\np  
\2 \phi^a (- \partial^{\perp 2} + m_g^2 ) \phi^a 
\np
\left[
m_g \phi^a {1 \over \partial^+} (J_{Af}^{a+} + J_{\psi f}^{a+}) \right]^r \ ,
\eeq
and
\beq
\label{cHcanQCDr}
\cH_{\rm QCD}^r 
\es
\bar \psi_f { \gamma^+ (- \partial^{\perp 2} + m^2) \over 2i \partial^+ }  \psi_f
+
\2 A_{f \mu}^a  \partial^{\perp 2}  A_f^{a \mu}   
\nm
\left[ (J_{\psi f}^{a \mu}  + J_{A f}^{a \mu}) A_{f\mu}^a \right]^r
\np
\2 g^2 \left[ \bar \psi_f \, \f \hspace{-2pt} A_f 
{ \gamma^+ \over i \partial^+ } 
                          \, \f \hspace{-2pt} A_f \psi_f \right]^r
\nm
\4 g^2 \left[ [A_{f\mu}, A_{f_\nu}]^a [A_f^\mu, A_f^\nu]^a  \right]^r
\nm
\2 \left[ (J_{\psi f}^{a+} + J_{A f}^{a+}) {1 \over \partial^{+2}} (J_{Af}^{a+} + J_{\psi f}^{a+}) 
\right]^r \ .
\eeq
The regularization brackets $[~~~]^r$ embrace all interaction terms. They will be  omitted 
to simplify notation. 
\section{Gluon mass and small $x$ in scattering amplitudes}  
\label{transition}
Insertion of a mass term for gluons in the FF Hamiltonian of QCD leads to severe small-$x$ divergences due to inverse of 
$\partial^{+2}$, {\it e.g.,} see~\cite{Wilson:1994fk}, Sec. IX A. In contrast, the Hamiltonian $H^{m_g \,  r}_{\rm QCD}$ 
of Eq.~(\ref{EqbareHr}), with density $\cH^{m_g \,  r}_{\rm QCD}$ given in Eq.~(\ref{cHcanQCDmgr}), leads to the cancellation 
of these singularities despite that $m_g > 0$. The reason is that $H^{m_g \,  r}_{\rm QCD}$ includes also a kinetic term for a 
color octet of scalar fields of the same mass and an interaction term that couples that octet field to the ``good'' color currents 
of quarks and transverse gluons, proportionally to the gluon mass. 
\subsection{Severe small-$x$ singularity}  
\label{gluonsmallx}
To identify the source of the gluon severe small-$x$ divergences, the current-gluon coupling term, {\it i.e.}, the Hamiltonian 
term resulting from the third density term on the right-hand side of Eq.~(\ref{cHcanQCDr}), is separated into its transverse 
and longitudinal parts,
\beq
- (J_{\psi f}^{a \mu}  + J_{A f}^{a \mu}) A_{f\mu}^a 
\es
(J_{\psi f}^{a j}  + J_{A f}^{a j}) A_{f}^{a j} 
\nm
\2 (J_{\psi f}^{a +}  + J_{A f}^{a +}) A_{f}^{a -} \ . 
\eeq
Only the longitudinal part involves the inverse of $\partial^+$. Using Eq.~(\ref{Af-}) for $A_f^{a-}$ and integrating by parts, the 
Hamiltonian density of Eq.~(\ref{cHcanQCDmgr}) for fields that vanish at large distances is transformed to the equivalent form
\beq
\label{HQCDmgphir}
\cH_{\rm QCD}^{m_g \, r}
\es
\bar \psi_f { \gamma^+ (- \partial^{\perp 2} + m^2) \over 2i \partial^+ }  \psi_f
\np
\2 A_f^{a j}  (- \partial^{\perp 2} + m_g^2 )  A_f^{a j}   
\np
  (J_{\psi f}^{a j}  + J_{A f}^{a j}) A_f^{a j}
+
\2 g^2 
\bar \psi_f  A_f^j \gamma^j { \gamma^+ \over i \partial^+ } \gamma^k A_f^k \psi_f 
\nm
\4 g^2 [A_f^j, A_f^k]^a [A_f^j, A_f^k]^a  
\nm
\2 (J_{\psi f}^{a+} + J_{A f}^{a+}) {1 \over \partial^{+2}} (J_{Af}^{a+} + J_{\psi f}^{a+}) 
\np
\2 \phi^a (- \partial^{\perp 2} + m_g^2 ) \phi^a 
\np
\left( \partial^\perp A_f^{a\perp} + m_g \phi^a \right) 
{1 \over \partial^+} \left(J_{Af}^{a+} + J_{\psi f}^{a+}\right) \ .
\eeq
The superscripts $j$ and $k$ take only two values, 1 and 2, for the front transverse directions. The only terms leading 
to $1/\partial^{+2}$, or $1/x^2$ for gluons, are
\beq
\label{H+1}
\cH_{+1} 
\es
\left( \partial^\perp A_f^{a\perp} + m_g \phi^a \right) {1 \over \partial^+} \eta J_f^a \ , \\
\label{H+2}
\cH_{+2}
\es
- \2  \eta J_f^a {1 \over \partial^{+2}} \eta J_f^a \ ,
\eeq
where $\eta J_f^a = J_{\psi f}^{a+} + J_{A f}^{a+}$. The four-vector $\eta$ has components $\eta^-=2$, $\eta^+=0$, and
$\eta^\perp =0$. The term $\cH_{+1}$ is proportional to $g$ and the term $\cH_{+2}$ to $g^2$. Integration of these two 
densities over the front yields the Hamiltonian terms denoted by $H_{+1}$ and $H_{+2}$, respectively. Regularization factors 
in these terms are introduced according to the rules described in the previous section. The gluon severe small-$x$ singularities 
occur due to $H_{+2}$ and square of $H_{+1}$. It is shown in the next section that these two contributions cancel out in 
the transition amplitudes for quarks and gluons despite the presence of the mass parameter $m_g$. 
\subsection{Cancellation of $1/x^2$ in scattering amplitudes}
\label{xintransition}
Scattering of quarks and gluons in the femtouniverse~\cite{Bjorken:1979hv} is described assuming that in the first 
approximation they propagate as free. Interactions are accounted for using an expansion in a series of  powers of 
the coupling constant $g$. Our discussion begins with the small-$x$ divergences in the second-order scattering 
matrix operator~\cite{Gell-Mann:1953dcn},
\beq
\label{T1}
T^{(2)} \es H_I^{(1)} {1 \over P^- - H_f + i\epsilon} H_I^{(1)} + H_I^{(2)} \ .
\eeq
Symbols $H_I^{(1)}$ and $H_I^{(2)}$ denote the Hamiltonian terms of order $g$ and $g^2$, respectively. $P^-$ denotes 
the initial-state eigenvalue of the free Hamiltonian $H_f$. 

Consider the quark-antiquark scattering in which a pair $q\bar q$ turns into $q'\bar q'$. Dropping the overall momentum 
conservation $\tilde \delta$-function, using $j_q$ and $j_{\bar q}$ to denote the quark currents, which are free from the 
gluon small-$x$ singularities, denoting the four-momentum transfer from the quark to the antiquark by $k=q-q'=\bar q' - 
\bar q$, $k^+>0$, and omitting $i \epsilon$, one obtains the transition amplitude on-shell of total $P^-$ in the form 
\beq
\langle q'\bar q'|T^{(2)} |q\bar q\rangle \es j_{q \alpha} T^{\alpha \beta} j_{\bar q \beta} \ , \\
\label{exchangecurrentsfactor}
T^{\alpha \beta} \es
{1 \over k^+(P^- - H_f)} \sum_{\sigma = 1}^3 
\epsilon_{k\sigma}^\alpha \epsilon_{k\sigma}^{*\beta} 
+ {\eta^\alpha \eta^\beta \over k^{+2}} \ .
\eeq
The first term on the right-hand side of Eq.~(\ref{exchangecurrentsfactor}) comes from the exchange of gluons between 
the quarks. The second term, proportional to $\eta^\alpha \eta^\beta$, is contributed by the seagull term, or $H_{+2}$ 
corresponding to the density $\cH_{+2}$ in Eq.~(\ref{H+2}). Both, the exchange and seagull contributions diverge as 
$1/k^{+ 2}$ when $k^+ \to 0$. The sum over gluon polarizations extends from 1 to 3 because it includes the contribution 
of the longitudinal gluons described by the field operator $\hat \phi$ in Eq.~(\ref{hatphi}). 
\beq
\label{sumeps1}
\sum_{\sigma = 1}^3 \epsilon_{k\sigma}^\alpha 
                                \epsilon_{k\sigma}^{*\beta} 
\es
- g^{\alpha \beta}
+(k_0^\alpha \eta^\beta + \eta^\alpha k_0^\beta)/k^+
\np m_g^2 \eta^\alpha \eta^\beta/k^{+2} \ .
\eeq 
The first two terms above come from the transverse gluons, with $k_0^- = k^{\perp 2}/k^+$, and the third term comes 
from the longitudinal gluons. So, the amplitude is
\beq
\label{T11}
\langle q'\bar q'|T^{(2)} |q\bar q\rangle
\es j_{q \alpha} \left[ {-g^{\alpha \beta} + \Pi^{\alpha \beta}
\over k^2 - m_g^2} \right] j_{\bar q \beta} \ , 
\eeq
where
\beq
\label{sumeps2}
\Pi^{\alpha \beta} 
\es
(k_0^\alpha \eta^\beta + \eta^\alpha k_0^\beta)/k^+
+ 
m_g^2 \eta^\alpha \eta^\beta/k^{+2}
\np
(k^2 - m_g^2) \, \eta^\alpha \eta^\beta / k^{+2} \ .
\eeq
The quark and anti-quark currents in Eq.~(\ref{T11}) are conserved, $k_\alpha \, j_q^\alpha = k_\alpha \, j_{\bar q}^\alpha = 0$.
The formula $k_0^\alpha = k^\alpha + (k_0^- - k^-)\eta^\alpha/2$ implies $\Pi^{\alpha \beta} \equiv \Pi \ \eta^\alpha \eta^\beta$,
\beq
\label{Pitransition}
\Pi
\es
(k_0^- - k^-)/k^+
+
m_g^2 /k^{+2}
+ 
(k^2 - m_g^2) /k^{+2} .
\eeq
The key result is that $\Pi$ vanishes. The terms $\sim 1/x^2$ cancel out completely. The regularization factors $r$ do not 
interfere with the cancellation because they are the same in the gluon exchange and the seagull term. The resulting scattering 
amplitude has the co-variant form of Eq.~(\ref{T11}) with $\Pi^{\alpha \beta}=0$. Moreover, the amplitude is free from any 
small-$x$ singularity, not only the severe $1/x^2$. An alternative way of stating this result is that the diagrammatic rule (R8) 
of perturbative QCD in App. A of~\cite{Lepage:1980fj}, continues to be valid even though the gluons are assigned the mass 
$m_g$. 

In order to extend the above reasoning to the transition amplitudes between intermediate states that are off-shell of total 
$P^-$,  we first note that the severe gluon singularities appear in perturbative calculations due to the operator 
\beq
\label{Ts}
T_{\rm singular} \es H_{+1} {1 \over P^- - H_f + i\epsilon} H_{+1} + H_{+2} \ ,
\eeq
which sums the exchange of a gluon between two color currents and the seagull. The divergences occur in the coefficients 
of the tensor $\eta^\alpha \eta^\beta$, contracted with the quark or gluon currents. The sum over transverse polarizations 
of the exchanged gluon contributes $k^{\perp 2}/k^{+2}$, as dictated by $H_{+1}$. The intermediate quanta of field $\phi$ contribute $m_g^2/k^{+2}$, also dictated by $H_{+1}$. Since the eigenvalues $k^-$ of $H_f$ for the quanta of fields $A$ 
and $\phi$ are the same, $k_g^- = (k^{\perp 2} + m_g^2)/ k^+$, the most singular contribution to the transition amplitude 
due to $k^+ \to 0$, takes the form
\beq
\label{cancellation11}
\langle T_{\rm singular} \rangle \es  j_1 \hspace{-2pt}
\left[  { k^{\perp 2}/k^{+2}  + m_g^2 /k^{+ 2} \over k^+(P^- - Q^- - k_g^-)} 
+ {1 \over k^{+2}} \right] \hspace{-2pt} j_2 ,
\eeq
where the second term in the bracket, $1/k^{+2}$ , comes from $H_{+2}$. The sum of eigenvalues of $H_f$ for other quanta in 
the same intermediate state, in which the exchanged gluons appear, is denoted by $Q^-$. Symbols $j_1$ and $j_2$ stand for the 
contractions of the quark or gluon currents with $\eta$. The three-momentum conservation $\tilde \delta$-function and $i\epsilon$ 
are omitted. The term $k^{\perp 2}/k^{+2}$ in the numerator in Eq.~(\ref{cancellation11}) is provided by the transverse gluons and 
the term $m_g^2 /k^{+ 2}$ by the longitudinal ones. The inverse of $k^{+2}$ would produce the severe singularity for $k^+ \to 0$ 
if the free Hamiltonian $H_f$ would include the mass $m_g$ for transverse gluons and the longitudinal gluons were absent. When their contribution is included,
\beq
\label{cancellation}
\langle T_{\rm singular} \rangle 
\es 
j_1 {  (P^- -  Q^-)/k^+ \over k^+ (P^- - Q^-) - (k^{\perp 2} + m_g^2)} \,  j_2   .
\eeq
The divergence $\sim 1/k^{+2}$ for $k^+ \to 0$ cancels out despite that $m_g > 0$. At the same time, $m_g^2 > 0$ regulates 
the infrared divergence due to $k^\perp \to 0$ when $k^+ \to 0$. 

The same mechanism works to all orders of perturbation theory. Our demonstration begins with Eq.~(\ref{H+1}). Every numerator 
term associated with propagation of a transverse gluon, $(k^\perp/k^+)^2$, is accompanied by a term due to the propagation of 
a longitudinal gluon, $(m_g/k^+)^2$. The transverse-gluon exchange and the longitudinal-gluon exchange involve the same 
denominator factor with the same value of the gluon $k_g^-$. The exchange extending over one intermediate state always appears 
in addition to a contribution of the Hamiltonian term $H_{+2}$. Therefore, the cancellation in exchanges extending over one 
intermediate state occurs as described above. The cancellation holds no matter how many additional interactions precede or follow 
the exchange. When the gluon exchange extends over more than one intermediate state, the contribution of $H_{+2}$ is absent. 
However, in that case, the gluon $k_g^-$ appears in more than one denominator. The additional denominators provide additional 
powers of $k^+$ in the numerator, which eliminate the severe singularity $1/k^{+2}$. 
\subsection{Small $x$ in self-interactions}
\label{Sigmax}
\begin{figure}[ht!]
          \caption{Self-interaction of particle of type 1 via emission and absorption of particle of type 2.}
          \label{fig:self}
\begin{center}
         \includegraphics[width=0.4\textwidth]{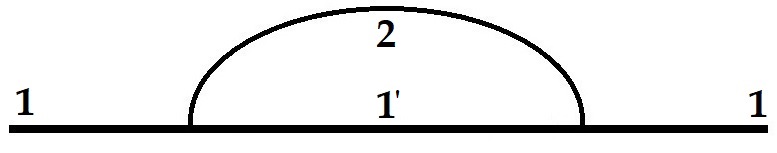}
\end{center} 
\end{figure}

In the amplitudes discussed in Sec.~\ref{xintransition}, the severe small-$x$ divergences of the gluon exchange and seagull 
terms cancel each other. But in the quark and gluon self-interactions, schematically illustrated in Fig.~\ref{fig:self}, the 
normal-ordered seagulls do not contribute. 
The second-order self-interaction sketched in Fig.~\ref{fig:self} requires a counterterm. Including the counterterm, the self-interaction 
reads
\beq
\label{selfinteraction1text}
\langle p' | p\rangle \, {\Sigma^{(2)} \over p^+} 
\es
\langle p' |  H_{+1} {1 \over P^- - Q^- - H_f } H_{+1} |  p\rangle 
\np
\langle p' | p\rangle \, {C^{(2)} \over p^+} \ ,
\eeq
where the counterterm contribution is denoted by $C^{(2)}$. The factor $\langle p' | p\rangle $ accounts for the state normalization. 
We obtain
\beq
\label{selfinteraction1quark333full}
{\Sigma_q^{(2)} \over p^+}
\es
{ g^2  C_F \over p^+}
\int {dx \, d^2k^\perp \over 2 (2\pi)^3} \, 
  \left[ {N \over D} + 4\sigma_q^{(2)} \right]  \, r^2_{qg,p} \ , \\
\label{NumSigmaQfull}
N
\es 4 { k^{\perp 2} + (1-x)m_g^2 \over x^2}  + 2 { k^{\perp 2} + x^2 m_i^2 \over  1-x } \ ,\\
\label{NumSigmaQfullD}
D
\es
x(1-x) p^+( P^- - Q^-  - p_q^- - p^-_g) \ ,
\eeq 
where $4\sigma_q^{(2)}$ stands for the integrand of the counterterm. $x$ is the fraction of the quark momentum $p^+$, carried 
by the emitted gluon, ranging between 0 and 1. $k^\perp = p_g^\perp - x p^\perp$. For $N$ colors of quarks, $C_F = (N^2-1)
(2N)$. The squared regularization factor, $r^2_{q g, q}$, comes from the double action of the Hamiltonian term $H_{+1}$. The 
severe small-$x$ singularities are contained in 
\beq
\label{selfinteraction1quark333}
{\Sigma_{q \, \rm sing}^{(2)} \over p^+} \hspace{-3pt}
\es
{ g^2  C_F \over p^+} \
\int {dx \, d^2k^\perp \over 2 (2\pi)^3} 
\nt \hspace{-6pt}
\left[ {4p^-_g/x \over x ( P^- - Q^- - p_q^- - p^-_g)} + 4\sigma_q^{(2)} \right] 
r^2_{q g, q} .
\eeq 
$p_q$ denotes the free four-momentum of a quark of mass $m$, after emitting and before absorbing a gluon.
$p_g$ denotes the free four momentum of the gluon of mass $m_g$. It is visible that to counter the singularity 
$\sim 1/x^2$, one needs the counterterm integrand 
\beq
\label{sigmaqint}
\sigma_q^{(2)} \es 1/x^2 \ .
\eeq
Inclusion of the less singular terms than $4p_g^-/x$, shows that the small-$x$ divergence $1/x^2$ and the ultraviolet quadratic 
transverse divergence in $\Sigma_q^{(2)}/ p^+$ are simultaneously countered by the counterterm whose integrand is 
\beq
\label{sigmaqintfull}
\sigma_q^{(2)} \es 1/x^2 + 1/[2(1-x)] \ .
\eeq
Similarly, the 2nd-order transverse-gluon self-interaction reads, 
\beq
\label{SigmaTgluon}
{\Sigma_{g \, \rm sing}^{(2)} \over p^+}
\es {2 g^2 C_A \over p^+} \int {dx d^2k^\perp \over  2 (2\pi)^3 } 
\left[ {N \over D}  + \sigma_g^{(2)} \right] r^ 2_{g g,g} \ , \\
N \es 
(k^{\perp 2} + m_g^2) \ \left[ 1 + {1 \over x^2} + {1 \over (1-x)^2 } \right] 
\nm
m_g^2 \left[ \2 + {1 \over x(1-x)} \right] \ , \\
D \es
x (1 - x)p^+(P^- - Q^-) - (k^{\perp 2} + m_g^2) \ .
\eeq
For $N$ colors $C_A=N$. It is visible that the counterterm integrand
\beq
\label{sigmag2}
\sigma_g^{(2)} \es 1 + {1 \over x^2} + {1 \over (1-x)^2 } \ ,
\eeq
removes the severe divergences of type $1/x^2$ for $x \to 0$. Inclusion of 1 in Eq.~(\ref{sigmag2}), secures simultaneous 
cancellation of the ultraviolet quadratic transverse divergence. The gluon self-interaction due to the intermediate quark 
pairs does not involve divergences $\sim 1/x^2$. The self-interaction of longitudinal gluons is free from such divergences
as well. The remaining singularities are only logarithmic, as promised. In orders higher than second, the one-particle irreducible 
self-interactions require more complex expressions for the counterterms. Self-interactions in the renormalized bound-state 
eigenvalue problems of the Hamiltonian $H^{m_g \,  r}_{\rm QCD}$ are discussed in the following sections.
\section{Hamiltonian eigenvalue problem }
\label{BSeqs}
Eigenstates of the Hamiltonian $H^{m_g \,  r}_{\rm QCD}$ of Eq.~(\ref{EqbareHr}) represent hadrons of QCD in the limit 
of lifting the regularization, $r \to 0$ and $m_g \to 0$. The eigenstates are the combinations 
\beq
|h\rangle \es \sum_i \psi_i |i\rangle ,
\eeq 
where $|i\rangle$ are the basis states obtained by acting on the state $|0\rangle$ with products of creation operators 
introduced in Eqs.~(\ref{barepsif}), (\ref{bareAf}) and (\ref{hatphi}). To describe a hadron, one needs to compute the relevant 
eigenvalue of the matrix $H^{m_g \, r}_{ij} = \langle i |H^{m_g \,  r}_{\rm QCD}|j\rangle$ and evaluate the corresponding set
of coefficients $\psi_i$. Analytic calculations appear not feasible because of the matrix size and complexity. To use computers, 
one needs to overcome several difficulties.

The Hamiltonian matrix has infinite size. One must replace it by some equivalent finite matrix. To begin with, one could limit 
the momenta and number of quarks and gluons in the basis states. However, setting such limits is not straightforward because 
the matrix elements $H^{m_g \, r}_{ij}$ diverge as functions of the momenta, {\it e.g.,} see Eq.~(\ref{HA3k12}) for $k^\perp \to 
\infty$ or $x \to 0$. The diverging matrix elements dominate the dynamics. They need to be dealt with first thing to establish
computational access to the finite quantities of physical interest~\cite{Wilson:1965zzb,Wilson:1970tp,Wilson:1994fk}, including 
the issue of explaining the quantum-binding mechanism for quarks and gluons. The singular interactions can change the number 
of quanta. Therefore, the states with quarks and gluons in fast relative motion are degenerate with states of less rapid but more 
numerous quanta. Traces of these complexities appear in QCD phenomenology. For example, applications of the parton model 
suggest that the gluon distribution in proton grows for small $x$ and requires some mechanism of saturation. Eventually, one 
needs a method of estimating the accuracy of a computation.

To overcome the difficulties mentioned above, the QCD bound-state eigenvalue problems are formulated not in terms 
of the canonical creation and annihilation operators introduced in Eqs.~(\ref{barepsif}), (\ref{bareAf}) and (\ref{hatphi}). 
Instead of $b$, $d$, and $a$, commonly denoted below by $q$, one uses operators $b_s$, $d_s$, and $a_s$, denoted 
by $q_s$. The latter are called the operators for effective particles. The parameter $s$ can be thought about as a size 
of the particles. The canonical operators $q$ correspond to the point-like quanta, $q = q_{s=0}$. When $s \sim 1/m$,
where $m$ is a hadron mass, the effective particles are meant to correspond to the constituent quarks and gluons. 

Evaluation of the QCD Hamiltonian in terms of operators $q_s$ is carried out using the renormalization group procedure 
for effective particles (RGPEP)~\cite{Glazek:2026xnp}. The procedure is designed to smooth out the singular canonical 
interactions. The smoothing is provided by the $s$-dependent vertex form factors obtained by solving the RGPEP 
equations. To describe how the gluon mass $m_g$ and auxiliary field $\phi$ contribute to the effective bound-state 
dynamics, we need to recall the elements of the RGPEP.
\subsection{Elements of the RGPEP}
\label{elementsRGPEP}
Quantum numbers of the phenomenological constituent quarks~\cite{ParticleDataGroup:2024cfk} are the same as 
those of the bare quarks in the canonical QCD, with the exception of the mass parameters. Therefore, the RGPEP for
QCD assumes that the effective particles are unitarily related to the bare particles, 
\beq
\label{qt}
q_s \es \cU_s^\dagger q \, \cU_s \ ,
\eeq
and the quantum numbers of $q_s$ and $q$ are the same. The Hamiltonian remains unchanged. Consequently, the 
products of operators $q_s$ in the Hamiltonian have different coefficients, say $c_s$, from the coefficients $c$ of 
the corresponding products of operators $q$, but
\beq
\label{cHctqt}
H(c_s,q_s) \es  H(c,q) \ .
\eeq
The renormalized expressions for the coefficients $c_s$, are obtained by solving a first-order differential equation in $s$, 
called here the RGPEP equation, see Eq.~(\ref{RGPEP1}) below. The initial condition at $s=0$ is provided by the canonical
coefficients, $c$, including the modifications resulting from the counterterms determined while solving the RGPEP equation,
according to the rules of the similarity renormalization group procedure (SRG)~\cite{Glazek:1993rc,Glazek:1994qc}. 
Computation of the coefficients $c_s$ is carried out using the operator
\beq
\cH \es H(c_s,q) \rs \cU_s H(c_s,q_s) \, \cU_s^\dagger \ ,
\eeq
whose definition implies that
\beq
\label{RGPEP}
\cH' \es \left[ \cG , \cH \right] \ ,
\eeq
where prime denotes the derivative $d/ds^2$; we use $s^2$ here for $s$ to have dimension of length. The anti-Hermitian 
operator 
\beq
\label{GenaH}
\cG \es \cU' \cU^\dagger 
\eeq
is called the generator of the RGPEP. Whole class of generators can be considered~\cite{Glazek:1997gt,Glazek:2006mg}. 
Hamiltonian operators we discuss have the form $\cH = \cH_f + \cH_I$, where $\cH_f$ is given in Eq.~(\ref{hatHf}). The 
commutator $[ \cH_f, \cH_I]$ is anti-Hermitian and can be used as a generator of a unitary transformation,
\beq
\label{Gcommutator1}
\cG \es [ \cH_f, \cH_I] \ .
\eeq
Evolution in $s$ with such generator tends to narrow the coefficients $c_s$ in the sense that they only allow changes of 
eigenvalues of $\cH_f$ that are limited by $\sim 1/s$.  With this choice for the generator, Eq.~(\ref{RGPEP}) takes the 
double-commutator form, 
\beq
\label{RGPEP1}
\cH' \es \left[ [ \cH_f, \cH_I] , \cH \right] \ ,
\eeq
closely resembling the Wegner flow equation for Hamiltonian matrices in condensed matter physics~\cite{Wegner:1994fdg}, 
see also~\cite{Mielke:1998ag,Wegner_2006,Kehrein:2006ti,PhysRevX.9.021037,Deift1983OrdinaryDE,doi:10.1137/1030090,
BROCKETT199179}. The operator Eq.~(\ref{RGPEP1}) serves the computation of the coefficients $c_s$ and as such does 
not exactly reduce to finite matrix equations, because the particle operators do not. However, the double commutator 
structure still implies that $\cH$ fulfills the cluster decomposition principle~\cite{Weinberg:1995mt}; there is no generation 
of the disconnected interactions. Model studies of finite matrix approximations to the RGPEP equations with various generators 
show a need for alterations of the generator $\cG$ of Eq.~(\ref{Gcommutator1}) when one seeks to achieve convergence of 
solutions for $\cH$ using the weak-coupling expansion~\cite{Glazek:2002iw,Glazek:2003xr}. Here we are concerned with the 
lowest orders of the expansion and no need arises for altering Eqs.~(\ref{Gcommutator1}) and (\ref{RGPEP1}).
\subsection{Renormalized effective eigenvalue problem}
Assuming a solution for $\cH$ as a function of $s$ is available, the QCD Hamiltonian is obtained in the form
\beq
\label{Ht}
H_s \equiv H(c_s,q_s) \es  \cU_s^\dagger [ \cH = H(c_s,q) ] \, \cU_s \ .
\eeq
The eigenvalue problem for hadrons as bound states of effective quarks and gluons reads
\beq
\label{Htep}
H_s | \psi \rangle \es P^- | \psi \rangle \ , \\
\label{Hteppsi}
|\psi \rangle \es \sum_i \psi_i(s) |i\rangle_s \ ,
\eeq 
where $P^-$ is the eigenvalue. The basis states $|i\rangle_s$ are created by applying products of the creation operators 
$q_s$ to the state $|0\rangle$. If one solved the RGPEP equation for $H_s$ exactly and found exact eigenvalues and 
eigenstates of $H_s$, the coefficients $\psi_i(s)$ and basis states $|i\rangle_s$ would depend on $s$ but the eigenvalues 
and corresponding eigenstates $| \psi \rangle $ would not~\cite{Glazek:2021vnw}. Approximate calculations yield results 
varying with $s$. The magnitude of such variation indicates how large the theoretical error of an approximate calculation
may be. Another measure is provided by the accuracy of obeying symmetries, such as the rotation symmetry.   
\section{Weak-coupling expansion for $H_s$}
\label{seriesHt}
Given the initial condition at $s=0$ in the form of the regulated Hamiltonian $H^{m_g \,  r}_{\rm QCD}$ of Eq.~(\ref{EqbareHr}), 
the differential Eq.~(\ref{RGPEP1}) can be solved order-by-order in the expansion of the coefficients $c_s$ in the series of powers 
of the coupling constant $g$. Such expansion is valid when the coupling constant is made so small that all computed interaction 
terms are small. But solutions for the coefficients $c_s$ may contain inverse powers or logarithms of the regularization parameter $r$. 
When one lifts the regularization, taking the limit $r \to 0$, such coefficients become infinite despite that $s$ is finite and $g$ small. Consequently, the corresponding matrix elements of $H_s$ between the effective basis states of finite momenta diverge and the 
eigenvalue problem is ill-defined. To formulate a soluble theory, the initial condition at $s=0$, given by $H^{m_g \,  r}_{\rm QCD}$ 
in Eq.~(\ref{EqbareHr}), needs to be modified by adding terms, called counterterms, determined by the condition that the divergences 
in $c_s$ for finite $s$ are eliminated. The finite parts of the counterterms must be fixed by comparison with data, including the 
symmetries observed at $s$. Thus, the initial condition for solving the RGPEP equation is changed from $H^{m_g \,  r}_{\rm QCD}$ 
of Eq.~(\ref{EqbareHr}) to
\beq
\label{initcond}
H^{m_g \,  r }_{\rm QCD \, \cC}
\es 
\int dx^- d^2x^\perp \, \left( \cH^{m_g \,  r}_{\rm QCD} + \cC^{m_g \,  r}_{\rm QCD} \right) \ .
\eeq
The subscript $\cC$ indicates the inclusion of the counterterms, denoted by $\cC^{m_g \,  r}_{\rm QCD}$. Perturbative solution 
of the RGPEP equation for $H_s$, including the counterterms in the initial condition at $s=0$, becomes calculable order-by-order 
for very small coupling constants, though the radius of convergence is not known. To describe hadrons, one needs to extrapolate 
the weak-coupling expansion to the realistic values of the interaction strength~\cite{Wilson:1994fk}. The realistic magnitudes for
large $s$ ought to match the running interaction strength fitted to high-energy data within the same scheme. Outlines of the 
entire procedure, using simple models, are available in~\cite{Glazek:2002iw,Glazek:2003xr,Glazek:2021vnw}.
\subsection{Solutions of lowest orders}
This section illustrates the weak-coupling approach to solving the RGPEP equation for terms up to the third order. 
The second-order formulas are used later on to discuss the bound-state eigenvalue problems in QCD of heavy quarks 
including the mass $m_g$ and field $\phi$. The third-order solution shows how the RGPEP computations proceed 
beyond the second order, {\it cf.}~\cite{Glazek:2012qj}. 

In the series expansion of the operator $\cH$ in powers of $g$,
\beq
\label{cHexpansion}
\cH \es \cH_f + \cH^{(1)} + \cH^{(2)} + \cH^{(3)} + O(g^4) \ ,
\eeq
where the terms $\cH^{(n)}$ are proportional to $g^n$. The operator $\cH^{(n)}$ obeys the equation obtained by equating 
coefficients of $g^n$ on both sides of Eq.~(\ref{RGPEP1}). For the terms of first 3 orders, besides $\cH^{(0)} = \cH_f$, 
the equations read
\beq
\label{h31'text}
\cH^{(1)'}
\hspace{-4pt} \es \hspace{-4pt}
   \left[ \left[ \cH_f , \cH^{(1)} \right] , \cH_f \right]   , \\
\label{h32'text}
\cH^{(2)'}
\hspace{-4pt} \es \hspace{-4pt}
   \left[ \left[ \cH_f ,\cH^{(2)}  \right] , \cH_f \right]  
+ \left[ \left[ \cH_f , \cH^{(1)} \right] , \cH^{(1)} \right] \hspace{-3pt} , \\
\label{h33'text}
\cH^{(3)'}
\hspace{-4pt} \es \hspace{-4pt}
   \left[ \left[ \cH_f ,\cH^{(3)}  \right] , \cH_f \right] 
+\left[ \left[ \cH_f ,\cH^{(2)}  \right] , \cH^{(1)} \right]
\hspace{-4pt} \np \hspace{-4pt}
\left[ \left[ \cH_f , \cH^{(1)} \right] , \cH^{(2)} \right]  .
\eeq
In general, the derivative of $\cH^{(n)}$ only involves operators $\cH^{(k)}$ with $k < n$, which facilitates solving 
for $\cH$ order-by-order, to all orders. 

The solution for terms of the first order reads
\beq
\label{cH1t}
\cH^{(1)}
\es
f_{LR} \, \cH^{(1)}_0 \ ,
\eeq
where the subscript 0 refers to the initial condition at $s=0$, and
\beq
\label{f1t}
f_{LR} \es e^{-(s \Delta_{LR})^2 } \ , \\
\label{f1t2}
\Delta_{LR} \es p_L^- - p_R^- \ .
\eeq
The subscript $L$ refers to the product of creation operators in a term in $\cH$, standing on the term's left-hand side, and the 
subscript $R$ refers to the product of annihilation operators in a term, standing on its right-hand side. Using the 
notation introduced in Eqs.~(\ref{initialHi}) - (\ref{fr}) for an interaction term $\cH_i$, action of $f_{LR}$ on any term in 
Eq.~(\ref{cH1t}), and on any other operator at any value of $s$, is defined by
\beq
\label{Tft}
f_{LR} \cH_i \es 
           \left[ \prod_{m =1 }^{c_i} \int [p_m]  a_{p_m}^\dagger \right] 
           \left[ \prod_{ n = 1}^{a_i} \int [q_n ]  a_{q_n }              \right] \nt
           \tilde \delta_{c.a} \ T_i(\bar p, \bar q) \, f^s_{\bar p,\bar q} \ .
\eeq
In words, the action of $f_{LR}$ on $\cH_i$ implies insertion of the factor $f^s_{\bar p,\bar q}$ in the integrand of $\cH_i$. 
The solution for terms of the second order, obtained using solutions for the first-order terms, reads
\beq
\label{cH2t}
\cH^{(2)}
\es
f_{LR} \, \cH^{(2)}_0
+ 
(f_{LR} - f_{LI} f_{IR}) \
\Delta_{LIR} \
\cH^{(1)}_0 \, \cH^{(1)}_0 \ . \nn
\eeq
The subscript $I$ refers to the labels of the annihilation operators in $f_{LI}\cH_0^{(1)}$, or to the labels of 
the creation operators in $f_{IR}\cH_0^{(1)}$.
\beq
\label{DeltaLIR}
\Delta_{LIR} \es { \Delta_{LI} - \Delta_{IR} 
\over 
\Delta_{LI}^2 + \Delta_{IR}^2 - \Delta_{LR}^2 } \ .
\eeq
When the denominator in $\Delta_{LIR}$ vanishes, the difference of the form factors, $f_{LR} - f_{LI} f_{IR}$, vanishes, too. 
The solution for the third-order terms reads
\beq
\label{RGPEP3solution3}
\cH^{(3)} \es 
( f_{LR} - f_{LI} f_{IR} ) 
\, \Delta_{LIR}  \nt
\left[ \Delta_{IJR} \, \cH^{(1)}_0 \, \cH^{(1)}_0 \, \cH^{(1)}_0 + \cH^{(1)}_0 \, \cH^{(2)}_0  \right]
\np
( f_{LR} - f_{LJ} f_{JR} )        
\, \Delta_{LJR} \nt
\left[ \Delta_{LIJ} \, \cH^{(1)}_0 \, \cH^{(1)}_0 \, \cH^{(1)}_0 + \cH^{(2)}_0 \, \cH^{(1)}_0 \right]  
\nm
( f_{LR} - f_{LI} f_{IJ} f_{JR} )  \, \Delta_{LIJR} 
\,  \cH^{(1)}_0 \, \cH^{(1)}_0 \cH^{(1)}_0 \ ,  \nn
\eeq
where
\beq
\Delta_{LIJR} 
\es
{  (\Delta_{LJ} - \Delta_{JR}) \Delta_{LIJ}
+ (\Delta_{LI} - \Delta_{IR}) \Delta_{IJR} 
\over 
    \Delta_{LI}^2 + \Delta_{IJ}^2 + \Delta_{JR}^2 - \Delta_{LR}^2 } . \nn
\eeq
Subscripts $I$ and $J$ refer to the intermediate configurations of the particle operators involved in the interaction terms. 
Solutions for the Hamiltonian terms of orders higher than 3rd are obtained following the pattern shown by these examples.
\subsection{Counterterms}
\label{cC}
Computation of the counterterms in the initial Hamiltonian, $H^{m_g \,  r }_{\rm QCD \, \cC}$, is outlined above 
Eq.~(\ref{initcond}). It is illustrated below in the case of the quark self-interaction, which contributes to the quark 
bound-state dynamics. The self-interaction appears first in the second-order effective Hamiltonian in Eq.~(\ref{cH2t}). 
The right-hand side contains a product of two first-order terms, $\cH^{(1)}_0 \, \cH^{(1)}_0$. Consider the part of that 
product corresponding to Fig.~\ref{fig:self}, where a quark 1 emits and absorbes gluons 2.  The part involves the 
product $b_1^\dagger a_2 b_{1'} \ b_{\tilde 1}^\dagger a_{\tilde 2}^\dagger b_{\tilde 1'}$. Commuting $b_{1'}$ 
to the right of $b_{\tilde 1}^\dagger$ and $a_2$ to the right of $a_{\tilde 2}^\dagger$ yields $\delta_{1' \tilde 1} 
\delta_{2 \tilde 2}$, according to the commutation rules specified in Eq.~(\ref{comrules}). Integration $\int [\tilde 1 
\tilde 2 \tilde 1']$ results in the operator $\hat C=\int[1] \, c_1 \, \hat b_1^\dagger \hat b_1$, with the coefficient
\beq
\label{c1quark}
c_1 \es 
{ g^2  C_F \over p_1^+}
\int {dx \, d^2k^\perp \over 2 (2\pi)^3} \, \left( f_{LR} - f_{LI} f_{IR}  \right) {N \over D}  \, r^2_{qg,p} \ .
\eeq 
The numerator $N$ and denominator $D$ are given in Eqs.~(\ref{NumSigmaQfull}) and (\ref{NumSigmaQfullD}). 
Integration variables $x$ and $k^\perp$ are defined below Eq.~(\ref{NumSigmaQfullD}). The color factor $C_F$ 
equals 4/3 for $SU(3)$. The operators $\hat b_1^\dagger$ and $\hat b_1$ carry the same quantum numbers 1. 
Therefore, Eq.~(\ref{f1t}) impllies $f_{LR}=1$. In contrast, the momenta $p_2$ and $p_{1'}$ of the contracted quark 
and gluon operators yield $f_{LI} = f_{IR} = \exp \left\{ - (s/p_1^+)^2 [m_1^2 - (p_2 + p_{1'})^2]^2\right\}$, which 
results in
\beq
\label{fLIfIRDs}
f_{LI} f_{IR} \es \exp \left\{ - 2 (s D/p_1^+)^2 /[x(1-x)]^2 \right\} \ .
\eeq
This factor vanishes for small $x$ and large $k^\perp$. Therefore, the integrand involving $f_{LI} f_{IR}$ contributes a finite 
quantity to the coefficient $c_1(s>0)$ in the limit $r \to 0$. 

The only source of severe small-$x$ singularity in the coefficient $c_1$ is the term $4(k^{\perp 2} + m_g^2)/x^2$ in the 
numerator $N$, in the integrand with $f_{LR}=1$. This term is also singular as a function of $k^\perp \to \infty$. Both 
singularities are regulated by the factor $r^2_{qg,p}$. When the regularization is lifted by going to the limit $r \to 0$, 
the quark self-interaction diverges to negative infinity proportionally to $1/r$. But the solution for $\cH^{(2)}$ in Eq.~(\ref{cH2t}) 
contains also the term $f_{LR} \, \cH^{(2)}_0$, where $\cH^{(2)}_0$ is the initial condition for solving the RGPEP equation.
The initial condition can be set to contain a part canceling the divergent dependence of the self-interaction on $r$ in
the effective Hamiltonian. The diverging integral is already calculated in Sec.~\ref{Sigmax}. We can remove that divergence 
from the effective Hamiltonian by including an opposite term, called the self-interaction counterterm, in $\cH^{(2)}_0$.
The inclusion amounts to the replacement of the integrand $f_{LR} N/D$ in the expression for the effective Hamiltonian 
by $f_{LR}[ N/D+ 4 \sigma_q^{(2)} ]$, where $\sigma_q^{(2)}$ is given in Eq.~(\ref{sigmaqintfull}). This is how the RGPEP 
explains the guess made in the scattering theory in Sec.~\ref{Sigmax}. The resulting integrand,
\beq
\label{SigmaRGPEP}
f_{LR} [N/D+ 4 \sigma_q^{(2)}] \es - (4 m_1^2 + 2 m_g^2)/D \ ,
\eeq
integrates in the limit $r \to 0$ to a logarithmically divergent term,
\beq
\label{leftlog}
-\delta m^2_{1\ln} \es - {g^2  C_F \over  (2\pi)^2 } m_1^2 
\left[ \ln (2 r^2 m_1^2) + \gamma_E/2 \right] \ .
\eeq
Consequently, the Hamiltonian $\cH_0^{(2)}$ is further supplied with a logarithmic mass-squared counterterm 
$\delta m^2_{1 \ln}$. This means that not only the self-interaction divergence is removed from the effective 
Hamiltonian with finite $s$ but also that a choice is made for the finite part of the counterterm.  The choice 
made here is adopted to obtain the result that the quark masses appearing in the eigenvalue problems for 
the finite second-order effective Hamiltonian, such as in Eq.~(\ref{HLR1pom1text}) in Sec.~\ref{Hin2bodySpace}, 
are the ones introduced in $H_f$ in Eq.~(\ref{hatHf}). Similar analysis can be carried out for the gluon self-interaction 
counterterms, but we do not discuss the eigenvalue equations for gluonia in this work.
\section{Gluon mass and small $x$ in heavy quarkonia}
\label{Qdynamics}
To discuss the cancellation of gluon singularities $\sim 1/x^2$ in the QCD eigenvalue problems with $m_g > 0$, we consider
the theory with only heavy quarks, {\it i.e.}, the theory in which $\Lambda_{\rm QCD}$ of the RGPEP scheme is set to values
much smaller than the quark masses, {\it cf.}~\cite{Serafin:2023pkf}. In such a theory, the running coupling constant for the 
parameter $s \lesssim 1/m$, where $m$ is the quark mass, can be made sufficiently small for using the weak-coupling expansion 
of Sec.~\ref{seriesHt} to approximately compute the Hamiltonian $H_s$ in Eq.~(\ref{Ht}). 
\subsection{Dynamics in dominant component}
According to Eqs.~(\ref{Htep}) and \ref{Hteppsi}), the quarkonium eigenstate of the Hamiltonian $H_s$ has infinitely many components 
with various numbers of the effective quarks, antiquarks, and gluons, {\it cf.}~\cite{Glazek:2021vnw}. However, the interaction terms that 
change the number of effective particles include a small coupling constant and the vertex factors $f^s$, which implies that a quarkonium 
eigenstate of the lowest mass is dominated by its quark-antiquark component,  
\beq
\label{Pnt1}
|P \rangle \es \int[12] \, \psi^s_P(1,2) \, |12\rangle_s \ .
\eeq
The effective basis states, $|12\rangle_s = b_{s1}^\dagger d_{s2}^\dagger |0\rangle$, and the wave function $\psi^s_P(1,2)$ correspond to 
the RGPEP evolution parameter $s$. The Hamiltonian eigenstate is labeled by $P$ to indicate that $|P\rangle$ is simultaneously an eigenstate 
of the operators $\hat P^+$ and $\hat P^\perp$ with the eigenvalues $P^+$ and $P^\perp$, respectively. In other words, the matrix elements 
of the Hamiltonian $H_s$ between the effective quark-antiquark basis states are assumed to form the central part of the window matrix whose eigenstates approximately represent quarkonia in the theory; the concept of window Hamiltonians in the RGPEP is described in detail in Sec. 4.3 of~\cite{Glazek:2026xnp}. Its essence is that one identifies a small subspace of the space of states, called the window, in which the eigenstate in question can be approximated well by neglecting its components outside that subspace. The assumption of dominance of the two-body component is successfully used in interpreting quarkonium data~\cite{ParticleDataGroup:2024cfk,Appelquist:1974zd,Appelquist:1974yr,
Eichten:1978tg,Eichten:1979ms,Wilson:1974sk,Bodwin:1994jh,Brambilla:2010cs}. In our approach, the two-body component can, as a first approximation, be described by solutions to the eigenvalue problem for the two-body part of the operator $H_s$ whose form is
\beq
\label{HQbarQ}
H_{Q\bar Q} \es \int[122'1'] \, {_s\langle 12|} H_s | 1'2' \rangle_s  \  b_{s1}^\dagger d_{s2}^\dagger d_{s2'} b_{s1'} \ . 
\eeq
The integration includes summing over the discrete quantum numbers. Without losing generality, the wave function is given the form
\beq
\label{psix1k12}
\psi^s_P(1,2) \es P^+ \tilde \delta_{12.P} \ \psi^s_{1 2}(x_1,k_{12}^\perp) \ ,
\eeq 
where $\tilde \delta_{12.P}$ constrains the quarks to carry together $P^+$ and $P^\perp$, {\it cf.} Eq.~(\ref{deltac.a}). The relative momentum variables $x_1$ and $k_{12}^\perp$ are defined as in the case of an interaction term in Eq.~(\ref{HA3k12}). The variables are defined in the
same way no matter what two-body system one considers. The eigenvalue equation for the wave function reads
\beq
\label{EV2}
\int[1'2'] \, _s\langle 12| H_s | 1'2' \rangle_s \, \psi^s_P(1',2') \es E \, \psi^s_P(1,2) ,
\eeq
where $E(P^+,P^\perp) = (P^{\perp 2} + M^2)/P^+$ is the eigenvalue and $M$ denotes the quarkonium mass, $E(M,0^\perp)=M$. 
In the leading order of the weak-coupling expansion, $H_s = H_f + H_s^{(2)}$, where $H_s^{(2)}$ is obtained from $\cH^{(2)}$ of 
Eq.~(\ref{cH2t}) using Eq.~(\ref{Ht}). Hence, 
\beq
\label{2bodyMatrixHs}
_s\langle 12| H_s^{(2)} | 1'2' \rangle_s 
\es  
\langle 12|\cH^{(2)} |1'2'\rangle .
\eeq
This direct relationship between the effective-particle matrix elements of $H_s$ and matrix elements of $\cH$ is a general 
feature of solutions to the RGPEP equation. 

All matrix elements of all Hamiltonian terms acting in the effective two-body space include the factor $\tilde \delta_{12.1'2'}$, 
which universally secures conservation of $P^+$ and $P^\perp$ by the dynamics. Unless it needs to be mentioned, this 
$\delta$-function will be omitted. We also point out that there are no first-order terms in Eq.~(\ref{2bodyMatrixHs}), corresponding 
to $\cH^{(1)}$ in Eq.~(\ref{cHexpansion}) because they change the number of quanta. Such changes in the effective dynamics 
are addressed in Sec.~\ref{includinggluons}, after the pertinent analysis of the dynamics conserving the number of effective 
particles.
\subsection{$Q\bar Q$ Hamiltonian matrix }
\label{Hin2bodySpace}
According to Eqs.~(\ref{cH2t}) and (\ref{2bodyMatrixHs}), the $Q\bar Q$ Hamiltonian matrix in Eq.~(\ref{EV2}) has matrix elements
\beq
\label{HLR1pom1text}
&&
_s\langle 12| H_s | 1'2' \rangle_s
\rs \langle 12|  \left[ {\cH_f} + f_{LR} \, \cH_r^{(2)}  
\right. 
\nm 
\left.
f_{LI} f_{IR} \, \Delta_{LIR} \, \cH_0^{(1)}  \, \cH_0^{(1)} \right]  | 1'2' \rangle \ , 
\eeq
where 
\beq
\label{cHr}
\cH_r^{(2)} \es \cH^{(2)}_0 + \Delta_{LIR} \  \cH^{(1)}_0 \, \cH^{(1)}_0 \ .
\eeq
In Eq.~(\ref{HLR1pom1text}), the factor $f_{LI}f_{IR}$  suppresses all singularities generated by the interactions. The factor 
$f_{LR}$ does not produce such suppression for the singularities in $\cH_r^{(2)}$. Instead, these singularities cancel out. 
The divergent quark self-interactions, resulting from $\Delta_{LIR} \  \cH^{(1)}_0 \, \cH^{(1)}_0$ in $\cH_r^{(2)}$, are canceled 
by the self-interaction counterterms introduced in $\cH^{(2)}_0$, as described in Sec.~\ref{cC}. The singular gluon-exchange 
term in $\cH_r^{(2)}$, generated by $\Delta_{LIR} \ \cH^{(1)}_0 \, \cH^{(1)}_0$ in Eq.~(\ref{cHr}), diverges like $1/x^2$, where 
$x$ is the fraction carried by the exchanged gluon of the bound-state momentum $P^+$. In that case, the opposite divergence 
is contributed by the gluon seagull term in $\cH^{(2)}_0$. Details of the latter cancellation are described in the next section.
\subsection{Cancellation of $1/x^2$ in dominant component}           
\label{xinboundstates}
We now describe how the singularities $\sim 1/x^2$ in the gluon-exchange parts of $\Delta_{LIR} \  \cH^{(1)}_0 \, \cH^{(1)}_0$ 
and $\cH^{(2)}_0$, both contained in $\cH_r^{(2)}$, cancel each other in the leading-order bound-state eigenvalue problem 
for the quarkonium dominant component. The cancellation occurs despite that $m_g$ is not zero. 
\begin{figure}[ht!]
          \caption{Notation for the Hamiltonian matrix elements $\langle 12| f_{LR}\cH_r^{(2)} | 1'2' \rangle$ involving two 
                        complementary orderings of the gluon exchange and the seagull term. Small circles with letter r symbolize 
                        the regularization factors $r_{\overline 1,\underline 1 \rm g}$ and $r_{\overline 2, \underline 2 \rm g}$ in the 
                        tensor $h^{\mu \nu}$ in Eqs.~(\ref{hmatrix}) and (\ref{exchangefr}).}
          \label{fig:ABC}
\begin{center}
         \includegraphics[width=0.35\textwidth]{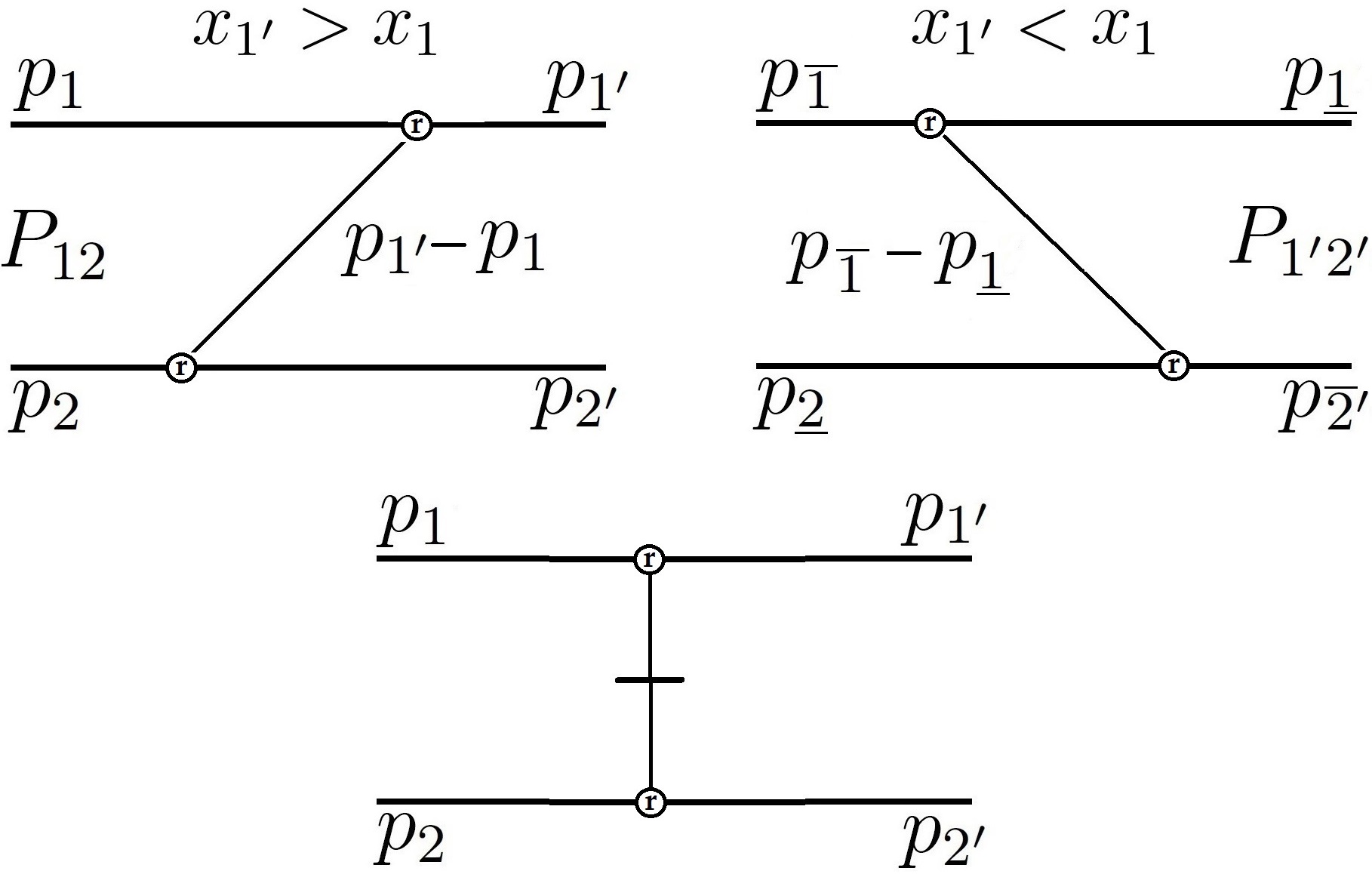}
\end{center} 
\end{figure}

Omitting the momentum-conserving $\tilde \delta$-function, the exchange matrix elements $\langle 12| f_{LR}\cH_r^{(2)} | 1'2' \rangle$
form the matrix
\beq
\label{hmatrix}
h^f \es j_{1\mu} h^{\mu \nu} j_{2 \nu} \ , 
\eeq
where the superscript $f$ indicates inclusion of the factor $f_{LR}$ in the tensor $h^{\mu \nu}$. The quark and antiquark currents, obtained 
from Eq.~(\ref{quarkcurrent}), are denoted by $j_1$ and $j_2$, and
\beq
\label{exchangefr}
h^{\mu \nu} \es  
f_{LR} \left( d^{\mu \nu} \Delta_{LIR}  + \eta^\mu \eta^\nu / q^{+2} \right) 
r_{\overline 1,\underline 1 \rm g} r_{\overline 2, \underline 2 \rm g} \ .
\eeq
This formula is illustrated in Fig.~\ref{fig:ABC} using diagrams. Two upper diagrams correspond to the gluon exchange interaction 
described by the term containing $d^{\mu \nu}$ in the bracket of Eq.~(\ref{exchangefr}). The lower diagram corresponds to the term 
containing $\eta^\mu \eta^\nu$ in the bracket, which is due to the seagull. Besides the labels $1$, $1'$, $2$, and $2'$, the notation 
we use includes labels $\overline 1$, $\underline 1$, $\overline 2$, and $\underline 2$, defined by the equations $\overline 1 = 1' 
\, \theta(p_{1'}^+ - p_{1}^+)  + 1 \, \theta(p_{1}^+ - p_{1'}^+)$, $\underline 1 = 1' \, \theta(p_{1}^+ - p_{1'}^+) + 1 \, \theta(p_{1'}^+ 
- p_{1}^+)$ for the quarks and by similar equations for the antiquarks. There are two momentum transfers, one for the quark and another 
one for the antiquark, 
\beq
q_1 \es p_{\overline 1} - p_{\underline 1} \ , \  q_2 \rs p_{\overline 2} - p_{\underline 2} \ ,
\eeq
where $p$ refers to the four-momentum of a quark of mass $m_1$ or an antiquark of mass $m_2$. We use
\beq
\label{rhosdefined}
\rho_1 \es m_g^2 - q_1^2 \ , \  \rho_2 \rs m_g^2 - q_2^2 \ .
\eeq
For the components $+$ and $\perp$, $q_1 = q_2 \equiv q$. The minus components of $q_1$ and $q_2$ differ when $p_1^- + 
p_2^- \neq p_{1'}^- + p_{2'}^-$. According to Eq.~(\ref{DeltaLIR}),
\beq
\label{DeltaLIR   pom}
\Delta_{LIR} \es - \2 \left({1 \over \rho_1} + {1 \over \rho_2}\right) \ .
\eeq
The tensor $d^{\mu \nu}$ multiplying $\Delta_{LIR}$ in Eq.~(\ref{exchangefr}),
\beq
\label{dmunuq}
d^{\mu \nu} \es  - g^{\mu \nu} + \eta^\mu \eta^\nu (\rho_1 + \rho_2)/ (2q^{+2}) \ ,
\eeq
results from summing over 3 polarizations of the exchanged gluon, 2 transverse and 1 longitudinal. The latter is contributed
by the quanta of the auxiliary field $\phi$. 

The term $-g^{\mu\nu}$ on the right-hand side of Eq.~(\ref{dmunuq}), leads to the quark-antiquark Coulomb potential with 
the Breit-Fermi spin dependence, described by the matrix 
\beq
\label{hC}
h^C \es  j_{1\mu} \, f_{LR} (- g^{\mu \nu}) \Delta_{LIR} \, j_{2 \nu} \, 
r_{\overline 1,\underline 1 \rm g} r_{\overline 2, \underline 2 \rm g} \ .
\eeq
The potential is slightly altered to the Yukawa form with the gluon mass parameter $m_g$. The long-distance Coulomb 
form is recovered when $m_g \to 0$. The Coulomb term is also altered at large momentum transfers, by the factor $f_{LR}$ 
of Eq.~(\ref{f1t})~\cite{Glazek:2020xpi,Glazek:2014ria,Glazek:2010zr}, 
\beq
\label{fLRC}
f_{LR} \es e^{- [s(\rho_1 - \rho_2)/q^+]^2} \ .
\eeq
The alteration occurs because the effective particles of size $s$ do not interact at short distances as the point-like particles do. 
Since the Coulomb term is finite, one can lift the regularization in it, sending $r \to 0$. Thus, the product of the regularization 
factors $r_{\overline 1,\underline 1 \rm g} r_{\overline 2, \underline 2 \rm g}$ is replaced by 1. The resulting matrix, denoted by 
$h_C$, is called the effective Coulomb interaction.
 
The second term on the right-hand side of Eq.~(\ref{dmunuq}) is obtained using the quark current conservation, as described 
in the case of Eq.~(\ref{sumeps2}) in Sec.~\ref{xintransition}. That term diverges like $1/x^2$ when $ x = q^+/P^+$ tends to 
0. Similar divergence appears in the seagull term $\sim \eta^\mu \eta^\nu$, the second term in the bracket in Eq.~(\ref{exchangefr}). 
The terms $\sim 1/x^2$ in the tensor $h^{\mu \nu}$ sum up to
\beq
\label{rho1-1}
h^{\mu \nu}_\eta  \es
f_{LR} \left[ {(\rho_1 + \rho_2)^2 \over 4 \rho_1 \rho_2}  -  1 \right]
\, { \eta^\mu \eta^\nu \over q^{+2}} \, r_{\overline 1,\underline 1 \rm g} r_{\overline 2, \underline 2 \rm g} .
\eeq
Using the identity $\rho_1-\rho_2  = (x_{1'}-x_1) (\cM_{12}^2 - \cM_{1'2'}^2 )$, one arrives at the expression 
\beq
\label{newterm}
h^{\mu \nu}_\eta  \es
f_{LR} \ { (\cM_{12}^2 - \cM_{1'2'}^2 )^2  \over \rho_1 \rho_2} \ { \eta^\mu \eta^\nu \over 4 P^{+2}} 
r_{\overline 1,\underline 1 \rm g} r_{\overline 2, \underline 2 \rm g} \ ,
\eeq
which demonstrates that the interaction is free from the small-$x$ divergences, as promised. We denote  the matrix involving 
$j_{1\mu} h_\eta^{\mu \nu} j_{2 \nu}$ by $h^\eta$. The regularization can be lifted in $h^\eta$ as in the effective Coulomb 
interaction. 

The product $\rho_1 \rho_2$ in the denominator of $h^\eta$ in Eq.~(\ref{newterm}) suggests that, in the limit $m_g \to 0$, a 
singularity $\sim 1/q^4$ is obtained. However, the difference $\cM_{12}^2 - \cM_{1'2'}^2$ tends to 0 when $q^+$ and $q^\perp$ 
tend to zero. Therefore, the net dependence of $h^\eta$ on $q \to 0$ in the limit $m_g \to 0$ is of the type $1/q^2$ instead of 
$1/q^4$. The former is integrable in the eigenvalue equation with the measure $dq^+ d^2q^\perp$.

The result in Eq.~(\ref{newterm}) implies that the need for the counterterm identified in Eq.~(36) of~\cite{Serafin:2023pkf} is 
eliminated, {\it cf.}~\cite{Serafin:2026ree}, simplifying the formalism. Instead of the counterterm, the divergence is countered 
by the contribution of quanta of the auxiliary field $\phi$ for any value of $m_g$. Moreover, the cancellation mechanism is 
rooted in using $m_g$ together with $\phi$, which is set up already at the stage of defining the bare theory Hamiltonian in Eq.~(\ref{cHcanQCDmg}). It thus appears simpler to implement than the program of computing counterterms in a sequence of successive approximations~\cite{Glazek:1994qc}. 

The factor $\eta^\mu \eta^\nu$ implies that $h^\eta$ is diagonal in the quark spins. As such, it could have a classical limit. 
However, it does not have any counterpart among classical interactions because it vanishes for $|\vec k_{12}\,| = |\vec k_{1'2'}\,|$, 
which is the on-shell condition for conservative classical limits of quantum interactions. The magnitude of matrix elements of 
$h^\eta$ for $|\vec k_{12}\, | \neq |\vec k_{1'2'}\,|$ is much smaller than the magnitude of the Coulomb interaction matrix elements 
for the same momentum transfers, roughly by the factor $\sim (k/m)^2$, where $k$ refers to the typical relative momentum of the 
quarks in a quarkonium and $m$ to the scale of their masses. As a result,
\beq
\label{hCheta}
h^f \es h^C + h^\eta \ ,
\eeq
where the matrix $h^\eta$ can be neglected in comparison with $h^C$ in the case of non-relativistic relative motion of quarks.
\subsection{Finite $Q \bar Q$ eigenvalue problem with $m_g > 0$}
\label{FeffInt}
The bound-state eigenvalue problem in Eq.~(\ref{EV2}), for the effective quark-antiquark Hamiltonian matrix of 
Eq.~(\ref{HLR1pom1text}), written in terms of the wave function $\psi^s_{1 2}(x_1,k_{12}^\perp)$, defined in~Eq.~(\ref{psix1k12}) 
and abbreviated to $\varphi_{k}$ for simplifying the notation, omitting the discrete quark labels, has the form 
\beq
\label{QbarQevp}
\sum_{i=1}^2  {m_i^2 + k^{\perp 2}_{12} + \sigma_i \over p_i^+   } \, \varphi_{k} 
+ \hspace{-4pt} \int_{k'} \hspace{-.05in} {H_{kk'} \over P^+}\, \varphi_{k'} \hspace{-4pt} 
\es \hspace{-4pt} {M^2 \over P^+} \, \varphi_{k} .
\eeq
Subscript $i=1$ corresponds to the quark and $i=2$ to the antiquark. The self-interaction terms $\sigma_i$ result solely from 
$-\langle 12| f_{LI} f_{IR} \, \Delta_{LIR} \, \cH_0^{(1)}  \, \cH_0^{(1)} | 1'2' \rangle$, since $\Sigma_i + \delta m^2_{i \ln} = 0$ as 
a consequence of the adjustment of the mass-squared counterterms derived in Sec.~\ref{cC}, see Eqs.~(\ref{SigmaRGPEP}) 
and (\ref{leftlog}). The gluon momentum variables in the self-interactions are $x= p_g^+/p_i^+$, ranging from 0 to 1, and 
$k^\perp = p_g^\perp - x p_i^\perp$, which is unlimited. We obtain 
\beq
\label{deltai}
\sigma_i 
\es
{g^2 C_F \over  16\pi^3} \int {dx \, d^2k^\perp \over x(1-x)} \, 
\ { j_{i\mu} d^{\mu \nu} j_{i\nu}^* \over p_i^+ \Delta p_i^-}  \ e^{-2 (s \Delta p_i^-)^2}\ ,
\eeq
where $j_{1\mu} = \bar u_{p_1} \gamma_\mu u_{p_q}$ or $j_{2\mu} = \bar v_{p_q} \gamma_\mu v_{p_2}$, and
\beq
\Delta p_i^- \es p_g^- +  p_q^- - p_i^- \rs \rho_i/p_g^+ \rs \Delta \cM_i^2/p_i^+ , \\
d^{\mu \nu} \es - g^{\mu \nu} + \eta^\mu \eta^\nu \rho_i/p_g^{+2} , 
\eeq
with $\rho_i = m_g^2 - (p_i-p_q)^2$. The kernel $H_{kk'}$ is 
\beq
\label{finitekernel}
H_{kk'} \es  h^f_{kk'} + h^{f \hspace{-2pt}f}_{kk'} \ ,
\eeq
where $h^f_{kk'}$ denotes the matrix elements of $h^f$ in Eq.~(\ref{hCheta}). The kernel $h^{f \hspace{-2pt}f}_{kk'}$ comes from the 
gluon-exchange in $-\langle 12| f_{LI} f_{IR} \, \Delta_{LIR} \, \cH_0^{(1)}  \, \cH_0^{(1)} | 1'2' \rangle$ of Eq.~(\ref{HLR1pom1text}). 
The superscript ${f \hspace{-2pt}f}$ indicates the inclusion of the factors $f_{LI} f_{IR}$. We have
\beq
\label{exchhkk'}
h^{f \hspace{-2pt}f}_{kk'} \es - g^2 C_F \
e^{-s^2(\rho_1^2 + \rho_2^2)/p_g^{+2}} \
{\rho_1 + \rho_2  \over 2\rho_1 \rho_2} \nt
j_{1\mu}  \left[ - g^{\mu \nu} + \eta^\mu \eta^\nu {\rho_1 + \rho_2 \over 2p_g^{+2}} \right] j_{2\nu} \ ,
\eeq
with $\rho$s defined in Eq.~(\ref{rhosdefined}). The color factor $C_F=4/3$, the same as in the self-interaction 
terms $\sigma_i$, is obtained for $Q\bar Q$ singlets of the color $SU(3)$.  The integral over  $k'$ denotes 
\beq
\label{evpint}
\int_{k'} \es \int {dx_{1'} \, d^2k^\perp_{1'2'} \over 16\pi^3 x_{1'}x_{2'} }
\rs
\int  {d^3 k' \over 16\pi^3 } \, { \cM_{1'2'} \over E_{1'} E_{2'}}  ,
\eeq
where $E_{i'} = \sqrt{ m_{i'}^2 + k'^{\perp 2} + k'^{z 2}}$  and  the components of the three-vector $\vec k'$ are $k'^\perp = 
k_{1'2'}^\perp$,  $k'^z=x_{1'} p^-_{2'} - x_{2'} p_{1'}^-$.  In the rest frame of  constituents $1'$ and $2'$, the quark carries a 
three-momentum $\vec k'$ and the antiquark carries $-\vec k'$. Analogous change of variables applies for the quarks carrying 
momenta $p_1$ and $p_2$ in the same kernel, defining the relative momentum $\vec k$. Such relative momenta can  be 
defined in any two-body system.
\section{$Q\bar Q$ dynamics for $m_g \to 0$}
\label{mgto0}
The eigenvalue problem in Eq.~(\ref{QbarQevp}) involves effective interaction terms depending on the mass parameter $m_g$, 
which is meant to be sent to 0 at the end of a calculation. The functions $f_{LI}$ and $f_{IR}$, obtained by solving the RGPEP 
Eq.~(\ref{RGPEP1}), behave as
\beq
\exp[ - ( s \Delta p^-)^2]  \es \exp[ - (s \rho/p_g^+)^2 ] \ .
\eeq
Their dependence on $\rho/x$, where $x = p_g^+/P^+$, makes them decrease as $m_g^2/x$ increases. They vanish faster 
than any power of $x$ when the gluon $x$ tends to 0. The inverse powers of $x$ are thus harmless in their presence. However, 
when $m_g \to 0$, the small-$x$ divergences reappear as logarithms of $m_g$, for the divergences $\sim 1/x^2$ cancel out. 
We show, keeping track of the details, that the logarithms of $m_g$ cancel out as well, but only in the eigenvalue equation for 
the colorless quark-antiquark states. Further useful features of the limit $m_g \to 0$ are discussed in subsequent sections. 
\subsection{Terms with $-g^{\mu\nu}$}
\label{mgdivergencesA}
The integrand of the self-interaction $\sigma_i $ in Eq.~(\ref{deltai}) and the gluon exchange kernel $h^{f \hspace{-2pt}f}_{kk'}$ in 
Eq.~(\ref{exchhkk'}), are sums of two terms, one with $-g^{\mu \nu}$ and another one with $\eta^\mu \eta^\nu$. One can write 
\beq
\label{sigmaiandhff}
\sigma_i  \es \sigma_i ^g + \sigma_i ^\eta \ , \ h^{f \hspace{-2pt}f }_{kk'} \rs h^{f \hspace{-2pt}f g}_{kk'} + h^{f \hspace{-2pt}f \eta}_{kk'} \ .
\eeq 
The division facilitates the relativistic analysis that follows. All the terms with $-g^{\mu \nu}$ are finite when $m_g \to 0$. In the case of 
self-interaction, 
\beq
\label{deltaig}
{\sigma_i ^{g} \over p_i^+}
\es
{g^2 C_F \over  (4\pi)^2} \, { m_i^2 \over  p_i^+} \nt
\int_0^\infty \hspace{-8pt} dz \hspace{-2pt}
\left[  1 - {6 \over z+1} + {1 \over (z+1)^2} \right] e^{- (a z)^2 } ,
\eeq
where $a = \sqrt{2} \, s \, m_i^2/p_i^+$, the variable $z = (E_i^2 - m_i^2)/m_i^2$, and $E_i = E_{qi} +E_g$, according to the 
choice of integration variables introduced in Eq.~(\ref{evpint}). The integral converges for $s > 0$ and, for small $s$, yields 
the leading term $(p_i^+/ m_i^2)\sqrt{2\pi} /(4s)$, which contributes to Eq.~(\ref{QbarQevp}) through
\beq
\label{deltatot}
{\sigma_1^g \over p_1^+} + {\sigma_2^g \over p_2^+}
\es  
{g^2 C_F \over  (4\pi)^2 } {\sqrt{2\pi} \over 2 s } \ .
\eeq
There are also contributions depending on $s$ logarithmically. The magnitude of the self-interaction terms can be estimated
in the quarkonium rest frame assuming a non-relativistic relative motion of the quarks with equal masses, $m_1 = m_2 = m$, 
for $s \sim 1/m$. In these circumstances, the complete result on the right-hand side in Eq.~(\ref{deltatot}), including the integral 
over $z$ of all terms in Eq.~(\ref{deltaig}), is on the order of $ - \alpha m/3$, not exceeding about 10\% reduction of the bound-state 
mass eigenvalue for $\alpha \sim 0.3$ favored by the phenomenology of quarkonia~\cite{Workman:2022ynf,
ParticleDataGroup:2024cfk}. In the part of $h^{f \hspace{-2pt}f}_{kk'}$ containing $- g^{\mu \nu}$, in Eq.~(\ref{exchhkk'}),  
\beq
\label{eqdiv-g}
h^{f \hspace{-2pt}f g}_{kk'}
\es 
g^2 C_F \ e^{-s^2(\rho_1^2 + \rho_2^2)/q^{+2}} \ { \rho_1 + \rho_2 \over 2 \rho_1 \rho_2 } \ j_1^\mu  j_{2 \mu} \ ,
\eeq
the product of quark currents is finite for the quark states with finite invariant masses. The $\rho$s depend only on the quark 
momenta. Large values of $\rho$s are suppressed by the factors $ff$, not indicated. In comparison with the effective Coulomb 
interaction, Eqs.~(\ref{hC}) and (\ref{fLRC}), the term $h^{f \hspace{-2pt}f g}_{kk'}$ is smaller by the factor of $\exp[ -2s^2 \rho_1\rho_2/
q^{+2} ] $. It is included in the effective dynamics in combination with the finite terms containing $\eta^\mu \eta^\nu$ in 
Sec.~\ref{Veffandconfinement}. 
\subsection{Terms with $\eta^\mu \eta^\nu$}
\label{mgdivergencesB}
The terms proportional to the tensor $\eta^\mu \eta^\nu$ in $\sigma_i $ of Eq.~(\ref{deltai}) and in the kernel 
$h^{f \hspace{-2pt}f}_{kk'}$ of Eq.~(\ref{exchhkk'}), are 
\beq
\label{deltaieta}
\sigma_i ^\eta
\es
{g^2 C_F \over 16\pi^3} \int {dx \, d^2k^\perp \over x(1-x)} \, 
\ { j_i^+ j_i^{+*} \over x p_i^{+2} } \ e^{-2 (s \rho_i/p_g^+)^2}  , \\
\label{hffkk'eta}
h^{f \hspace{-2pt}f \eta}_{kk'} \es \hspace{-3pt} - g^2 C_F 
{ (\rho_1 + \rho_2)^2  \over 4  \rho_1 \rho_2} 
{j_1^+j_2^+ \over p_g^{+2}} e^{-s^2(\rho_1^2 + \rho_2^2)/p_g^{+2}} .
\eeq
The self-interaction does not depend on the quark spin. It is a number with the engineering  dimension of momentum squared
and it multiplies the wave function $\varphi_k$. The kernel $h^{f \hspace{-2pt}f \eta}_{kk'}$ is dimensionless but it is integrated 
together with the wave function $\varphi_{k'}$ using the measure of Eq.~(\ref{evpint}), whose engineering dimension is momentum 
squared. The result of integration is independent of the quark spin. We obtain 
\beq
\label{deltaieta1}
\sigma_i ^\eta \varphi_k \hspace{-3pt}
\es \hspace{-3pt}
{g^2 C_F \over 16\pi^3} \varphi_k  \int dx \, d^2k^\perp { 4 \over x^2 } e^{-2 (s \rho_i/p_g^+)^2} , \\
\label{hffkk'eta1} \hspace{-8pt}
\int_{k'} h^{f \hspace{-2pt}f \eta}_{kk'} \, \varphi_{k'} 
\hspace{-3pt} \es \hspace{-3pt}
- g^2 C_F \int {dx_{1'} \, d^2k_{1'2'}^\perp \over 16\pi^3 x_{1'}x_{2'}} \, 
{ (\rho_1 + \rho_2)^2  \over 4  \rho_1 \rho_2} \nt
{ 4 \sqrt{x_1 x_{1'} x_2 x_{2'}} \over (x_{1'}-x_1)^2 } \ e^{-s^2(\rho_1^2 + \rho_2^2)/p_g^{+2}} \varphi_{k'}  .
\eeq
These interactions contribute to the eigenvalue problem the sum
\beq
\label{eqdiv2}
S \es
\left( {\sigma_1 ^\eta \over p_1^+} + {\sigma_2^\eta \over p_2^+} \right) \varphi_k
+
{1 \over P^+}  \int_{k'} h^{f \hspace{-2pt}f \eta}_{kk'} \, \varphi_{k'} \ .
\eeq
To show that the sum is finite in the limit $m_g \to 0$, we introduce $\sigma_{12}^\eta = \sigma_1^\eta /x_1 
+ \sigma_2^\eta /x_2$ and write
\beq
\label{cancellation}
P^+S  \es \sigma^\eta_{12} \varphi_k + \int_{k'} h^{f \hspace{-2pt}f \eta}_{kk'} \varphi_{k'} 
\rs
\left( \sigma^\eta_{12}  + \int_{k'} h^{f \hspace{-2pt}f \eta}_{kk'} \right)  \varphi_k 
\np
\int_{k'} h^{f \hspace{-2pt}f \eta}_{kk'} \left( \varphi_{k'} - \varphi_k \right) \  ,
\eeq
where the wave function $\varphi_{k}$ is a vector with 4 components corresponding to the 4 possible combinations of the quark spins. 
The kernel $h^{f \hspace{-2pt}f \eta}_{kk'}$ is a matrix proportional to $I_{4\times 4}$. The wave function with argument $\vec k$ is 
subtracted from the wave function with argument $\vec k\,'$ under the integral over $\vec k \,'$. Simultaneously, the integral of the 
kernel over $k'$ is multiplied by the wave function $\varphi_{k}$ and added to the self-interaction terms.  
\begin{figure}[ht!]
          \caption{Diagrams illustrating the self-interaction terms $ \sigma_i^\eta$ of Eq.~(\ref{deltaieta1}) together 
                         with the associated parts of the interaction kernel $h^{f \hspace{-2pt}f \eta}_{kk'}$ of Eq.~(\ref{hffkk'eta1}), 
                         arranged according to Eq.~\ref{cancellation}). The black blobs represent the RGPEP vertex factors $f_{LI}$ and $f_{IR}$.}
          \label{fig:selfexch}
\begin{center}
         \includegraphics[width=0.3\textwidth]{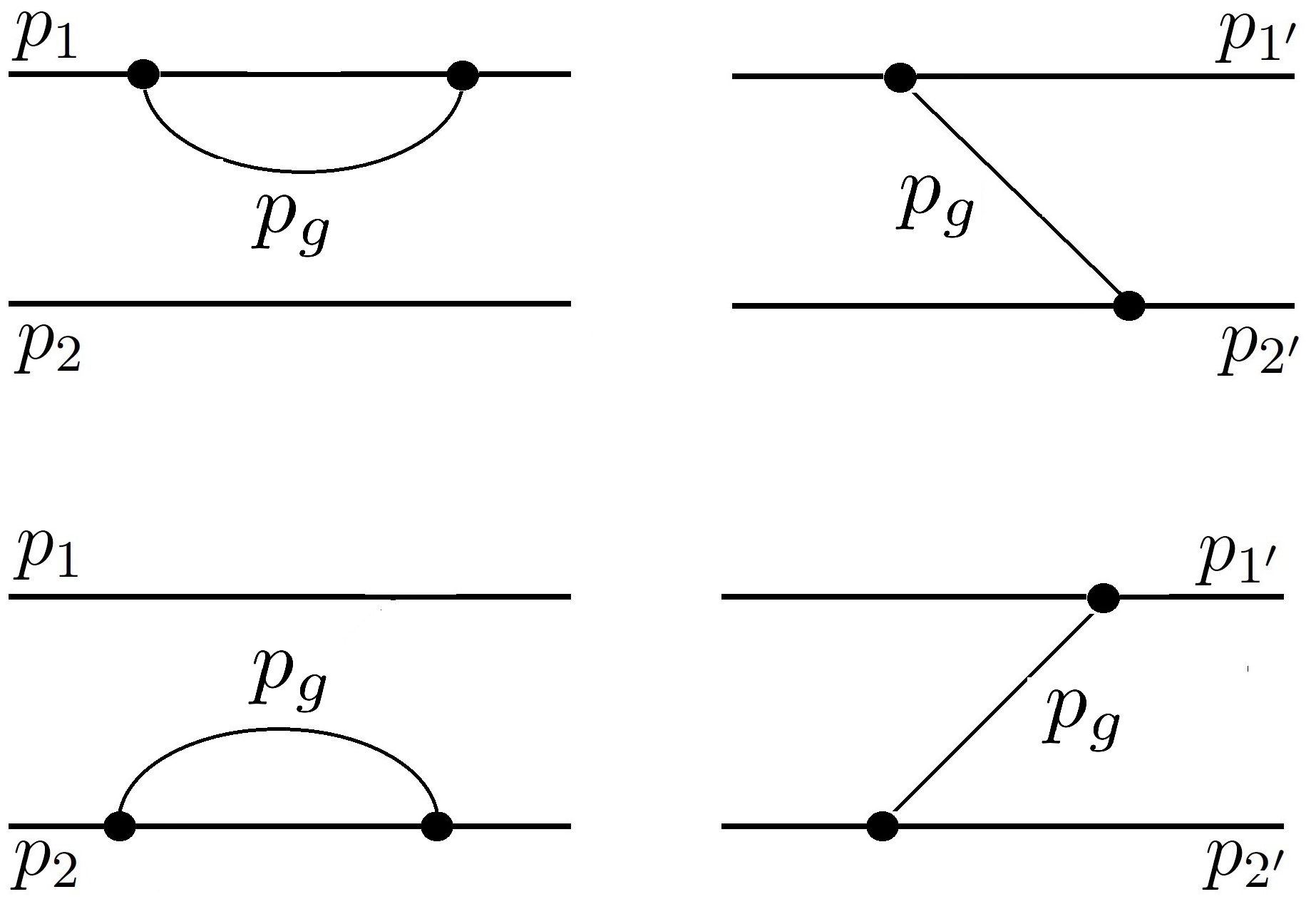}
\end{center} 
\end{figure}
\subsubsection{Self-interaction}
\label{mglimitVIIIB1}
We now discuss the resulting self-interaction terms including the kernel integrals, which multiply the wave function 
$\varphi_k$ on the right-hand side of Eq.~\ref{cancellation}), 
\beq
\tilde \sigma \es \tilde \sigma_1 + \tilde \sigma_2 \rs  \sigma^\eta_{12} + \int_{k'} h^{f \hspace{-2pt}f \eta}_{kk'} \ .
\eeq
Integration of the kernel over $\vec k'$ is split into two parts corresponding to the two self-interaction terms, $\tilde \sigma_1$ 
and $\tilde \sigma_2$. The quark self-interaction $\tilde \sigma_1$, illustrated with the upper left diagram of Fig.~\ref{fig:selfexch}, 
is paired with the integral of the kernel $h^{f \hspace{-2pt}f \eta}_{kk'}$ over $p_{1'}^+$ from 0 to $p_1^+$, illustrated by the 
upper right diagram in the same figure. The integral in the antiquark self-interaction, shown graphically in the lower left diagram, 
is paired with the integral of the kernel $h^{f \hspace{-2pt}f \eta}_{kk'}$ over $p_{2'}^+$ from 0 to $p_2^+$, represented by the 
lower right diagram. The integrals of the quark self-interaction and the paired kernel part behave the same way for $m_g \to 0$ 
as the integrals of the antiquark self-interaction and its paired kernel part. Only the quark terms are described in more detail. Using 
Eqs.~(\ref{deltaieta1}) and  (\ref{hffkk'eta1}), the sum of the pertinent integrals is
\beq
\tilde \sigma_1 
\es {\sigma_1^\eta \over x_1}  +  \int_{k'} h^{f \hspace{-2pt}f \eta}_{kk'} \ \theta(x_1-x_{1'})  \\
\es 
{G \over x_1} \int_0^1\hspace{-4pt} dx \, d^2k^\perp { 4 \over x^2 } \ e^{-2 (s \rho_1/p_g^+)^2} 
\nm
G
\int_0^{x_1} \hspace{-4pt} { dx_{1'}\, d^2k_{1'2'}^\perp \over  x_{1'}x_{2'}} \, { (\rho_1 + \rho_2)^2  \over 4  \rho_1 \rho_2}
\nt
{ 4 \sqrt{x_1 x_{1'} x_2 x_{2'}} \over (x_{1'}-x_1)^2 } \, e^{-s^2(\rho_1^2 + \rho_2^2)/p_g^{+2}} \ ,
\eeq
with $G = g^2 C_F/(16\pi^3)$. The integrals over transverse momenta extend over the entire transverse plane. The limit of $\tilde \sigma_1$ when $m_g \to 0$ 
is found by changing variables in the exchange integral,
\beq
\label{change1}
x_g \es x_1-x_{1'} \rs x x_1 , \ p_g^\perp \rs x k_{12}^\perp + k^\perp  , \\
\hspace{-10pt} x_{q'} \es x_{1'} \rs \hspace{-3pt} (1-x) x_1 , \ p_{q'}^\perp \rs  \hspace{-3pt} (1-x) k_{12}^\perp - k^\perp  \hspace{-3pt} , \\
x_{\bar q'} \es x_{2'} = x_2 + x x_1 \ , \ p_{\bar q'}^\perp \rs  
-k^\perp_{12} + p_g^\perp  \ . 
\eeq
The Jacobian is $x_1$ and
\beq
\label{change2}
\int_0^{x_1} \hspace{-4pt} { dx_{1'} \, d^2k_{1'2'}^\perp \over x_{1'} x_{2'} } 
\es \hspace{-3pt} \int_0^1{ dx \, d^2k^\perp \over (1-x)  (x_{2} +x x_1) }\ .
\eeq
In the exchange integral, $\rho_1$ and $\rho_2$ are the ones defined in Eq.~\ref{rhosdefined}). Proceeding in analogy to 
Eqs.~(\ref{rho1-1}) and (\ref{newterm}), using the identity $\rho_1-\rho_2  = - x x_1 (\cM_{12}^2 - \cM_{1'2'}^2 )$, we obtain 
\beq
x_1 \tilde \sigma_1/G
\es 
\int_0^1\hspace{-4pt} dx \, d^2k^\perp { 4 \over x^2 } \ e^{-2 (s \rho_1/p_g^+)^2} 
\nm
\int_0^1 \hspace{-6pt}  \, { dx \, d^2k^\perp \sqrt{x_2} \over  \sqrt{  (x_{2} +x x_1) (1-x)} }\ e^{-s^2(\rho_1^2 + \rho_2^2)/p_g^{+2}}
\nt
 \left[ { 4 \over x^2 } +  { x_1^2  (\cM_{12}^2 - \cM_{1'2'}^2 )^2  \over \rho_1 \rho_2} \right]  .
\eeq
The discussion following Eq.~(\ref{newterm}) applies here as well and the only terms potentially generating a divergence 
when $x \to 0$ for $m_g = 0$ are the ones with $4/x^2$, denoted by $\tilde \sigma_{1x}$. We substitute $q^\perp = 
k^\perp/\sqrt{x}$ and obtain
\beq
\label{x1tildesigma_1x}
&& x_1 \tilde \sigma_{1x} /G
=
\int_0^1\hspace{-4pt} dx \, d^2q^\perp { 4 \over x } \ e^{-s_1^2(\rho_1^2 + \rho_2^2)/x^2} 
\nt
\left[  e^{- s_1^2 (\rho_1^2 - \rho_2^2)/x^2}  - { 1 \over  \sqrt{  (1 +x x_1/x_2) (1-x)} }  \right]\hspace{-3pt}  ,
\eeq
where $s_1 = s/p_1^+$ and
\beq
\label{rho1indelta1}
\rho_1/x \es \hspace{-5pt} { q^{\perp 2} +  m_1^2 x \over 1-x} + {m_g^2 \over x} \ , \\
\hspace{-5pt} \rho_2/x 
\es  \hspace{-5pt} {  (q^\perp + \sqrt{x} k_{12}^\perp/x_2 )^2 + m_2^2  x x_1^2/x_2^2 \over 1 + x x_1/x2 } + {m_g^2 \over x}  .
\eeq
The factor $4/x$ in Eq.~(\ref{x1tildesigma_1x}) appears to produce a logarithmic divergence in the limit $m_g \to 0$, since in that limit 
the terms $m_g^2/x$ are eliminated from $\rho_1/x$ and $\rho_2/x$. But when $m_g=0$, for fixed $x_1$ and $k^\perp_{12}$, in the 
limit $x \to 0$,  $\rho_1/x \to q^{\perp 2} + O(xm^2)$, $\rho_2/x \to q^{\perp 2} + O(\sqrt{x}\,m^2)$, and the exponential function 
inside the square bracket approaches 1 when $x \to 0$. The approach to 1 is secured because the exponential function outside the square 
bracket limits the range of integration over $q^\perp$. The second term in the square bracket in Eq.~(\ref{x1tildesigma_1x}) approaches 1, 
too. Consequently, the square bracket vanishes when $x \to 0$ and the logarithmic divergence is absent. The self-interactions $\tilde 
\sigma_1$ and $\tilde \sigma_2$ are finite in the limit $m_g \to 0$. 

Assuming equal quark masses, $m$, $p_i^+ \sim x_i m$, $s \sim 1/m$ and neglecting $k^\perp_{12}$, a calculation shows 
that $\tilde \sigma_1$ and $\tilde \sigma_2$ amount in the eigenvalue problem to modifying the mass $m$ by 5\% or less 
when $x_1$ ranges from 0.2 to 0.8. Outside that range, the wave function $\varphi_k$ is expected to be negligible.
\subsubsection{Exchange}
\label{mglimitVIIIB2}
Now we describe the second term on the right-hand side of Eq.~(\ref{cancellation}),
\beq
\label{intKphi'-phi}
h^{f \hspace{-2pt}f \eta} \Delta \varphi \es \int_{k'} h^{f \hspace{-2pt}f \eta}_{kk'} \left( \varphi_{k'} - \varphi_k \right) \  ,
\eeq
in the limit $m_g \to 0$. Quark spins are conserved. The integration over $x_{1'}$ and $k_{1'2'}^\perp$ is carried out using 
the variables $z$ and $q^\perp$,
\beq
\label{zqperp}
x_{1'} \es x_1+ z \ , \ k_{1'2'}^\perp = k_{12}^\perp+ q^\perp \ ,
\eeq
for fixed $x_{1}$ and $k_{12}^\perp$. We introduce $s_P = s / P^+$ and write
\beq
\label{ExchPsi2POM}
h^{f \hspace{-2pt}f \eta} \Delta \varphi \es
{- g^2 C_F  \over 2 (2\pi)^3}  \int_{-x_1}^{x_2} \hspace{-10pt} dz \hspace{-3pt} \int \hspace{-3pt} d^2q^\perp \,
e^{-s_P^2(\rho_1^2 + \rho_2^2)/z^2} 
\nt \hspace{-4pt}
{ ( \rho_1 + \rho_2)^2  \over 4 \rho_1 \rho_2 } 
{ 4 \sqrt{ x_{1} x_{1'} x_{2} x_{2'}} \over x_{1'} x_{2'} \, z^2  } (\varphi_{k'} - \varphi_k)  .
\eeq
Using $p_1^\perp = x_1 P^\perp + k_{12}^\perp$, $p_{1'}^\perp = x_{1'} P^\perp + k_{1'2'}^\perp$ {\it etc.},  and introducing 
\beq
\label{qzktext}
q^\perp \es \sqrt{|z|} \, k^\perp \ ,
\eeq
we obtain
\beq
\label{rho1 etaeta}
{\rho_1 \over |z|} \hspace{-4pt}\es \hspace{-4pt} {m_g^2 \over |z|} + {m_1^2 |z|/x_1^2 
+ (k^\perp  - k_{12}^{\perp} s_z \sqrt{|z|} /x_1 )^2 \over 1 + z / x_1 }  , \\
\label{rho2 etaeta}
{\rho_2 \over |z|}\hspace{-4pt} \es \hspace{-4pt} {m_g^2 \over |z|} + {m_2^2 |z|/x_2^2 
+ (k^\perp + k_{12}^{\perp} s_z \sqrt{|z|} /x_2 )^2 \over 1  - z / x_2 }  ,
\eeq
where $s_z$ denotes the sign of $z$. The integral in Eq.~(\ref{ExchPsi2POM}) 
is finite for $m_g >0$ because the exponential factor suppresses the divergence 
due to the factor $1/z^2$. We inspect how the integral behaves for $m_g$ set to 0. The measure $d^2q^\perp$ provides a factor 
$|z|$, $d^2q^\perp = |z|d^2k^\perp$, and turns  the singularity $\sim 1/z^2$ to $\sim 1/|z|$. The wave function difference 
$\varphi_{k'}-\varphi_k$ vanishes in the limit of $z$ and $q^\perp$ tending to 0. But Eq.~(\ref{qzktext}) implies that $q^\perp$ 
vanishes like $\sqrt{|z|}k^\perp$ when $|z|$ tends to 0, provided $k^\perp$ is limited. The latter is the case thanks to the exponential 
factor. Hence, the difference $\varphi_{k'} - \varphi_k$ eliminates the singularity $\sim 1/|z|$ and the exchange term is finite in the 
limit $m_g \to 0$.
\subsection{Summary of the limit $m_g \to 0$}
\label{EVPsummarymgto0}
The renormalized effective two-body dynamics, encoded in the matrix elements $_s\langle 12| H_s^{(2)} | 1'2' \rangle_s$ of Eq.~(\ref{2bodyMatrixHs}), has the form of Eq.~(\ref{QbarQevp}), with the specified self-interactions $\sigma_i$ of 
Eq.~(\ref{sigmaiandhff}) and kernel $H_{kk'}$ of Eq.~(\ref{finitekernel}),
\beq
\label{effEVP}
&&
\sum_{i=1}^2  {m_i^2 + k^{\perp 2} + \sigma_i^g + \sigma_i^\eta \over x_i  } \, \varphi^s_{k} 
\np
\int_{k'} \hspace{-.05in}
\left( h^{f g}_{kk'} + h^{f \eta}_{kk'} + h^{f \hspace{-2pt}f g}_{kk'} + h^{f \hspace{-2pt}f \eta}_{kk'} \right)\, \varphi^s_{k'} \nn
\es M^2 \, \varphi_{k}^s \ .
\eeq
The self-interactions $\sigma_i^g$ are finite and in the limit $m_g \to 0$ contribute a small amount to the eigenvalue in comparison 
to the quark masses, see Sec.~\ref{mgdivergencesA}. The self-interaction term $\sigma_i^\eta$ and the exchange term with the kernel
$h^{f \hspace{-2pt}f \eta}_{kk'}$, {\it i.e.,} the last term under the integral, are analyzed together in Sec.~\ref{mgdivergencesB}. The spin-conserving 
terms are the only ones involving logarithms of the gluon mass, $m_g$. It is shown in Secs.~\ref{mglimitVIIIB1} and \ref{mglimitVIIIB2}
that the logarithms of $m_g$ in these terms cancel out. The term $h^{f g}_{kk'}$ provides the Coulomb interaction including the associated 
spin dependent factors. The small-$x$ singularities due to the gluon-exchange and seagull terms in $h^{f\eta}_{kk'}$ are shown in Sec.~\ref{xinboundstates} to cancel out in the presence of $m_g > 0$, leaving behind a small correction when $m_g = 0$, Eq.~(\ref{newterm}) 
and $h^\eta$ in Eq.~(\ref{hCheta}). The kernel $h^{f \hspace{-2pt}f g}_{kk'}$ is suppressed in comparison to the Coulomb term, as explained 
in Sec.~\ref{mgdivergencesA}. The parameter $s$ is reinstated in the wave function labeling as a reminder that Eq.~(\ref{effEVP}) is obtained 
by solving the RGPEP Eq.~(\ref{Htep}), using the weak-coupling expansion of second-order.
\subsection{Nonrelativistic approximation to $Q\bar Q$ interactions}
\label{Veffandconfinement}
Previous sections show that the relativistic eigenvalue problem of Eq.~(\ref{effEVP}) has a finite limit when $m_g \to 0$. We now identify the 
main features of the dynamics for $m_g=0$. To describe the motion of quarks in a quarkonium, we use their relative momenta, $\vec k$ and 
$\vec k'$, defined below Eq.~(\ref{evpint}). The average quark mass, $(m_1+m_2)/2$, is denoted by $m$. We discuss 
the effective dynamics at $s \sim 1/m$. The vertex factors limit the difference between $|\vec k\,|$ and $|\vec k'\,|$ that the interactions can 
cause. The difference does not exceed the scale set by $1/s \sim m$. The smallest value of the invariant mass of quarks 1 and 2, $2m$, 
corresponds to $\vec k = 0$, or  $k^\perp = 0$ and $x_1 = x_{10} = m_1/(m_1+m_2)$, $x_{20} = m_2/m_1+m_2)$. Since the invariant mass 
of the quarks rises with the magnitude of their relative momentum, and the off-diagonal matrix elements of the interaction involve a small coupling constant, the quark relative momentum in an eigenstate with a low mass eigenvalue does not exceed the scale of $m$~\cite{Glazek:2026xnp}. The small term $h^\eta$ in $h^f$ in Eq.~(\ref{hCheta}) is neglected. The quarkonium mass eigenvalue is written 
in the form $M = 2m + B$,  in which $B$ denotes the binding energy considered small in comparison to $2m$. Consequently, $M^2$ is
approximated by $4m(m + B)$, neglecting $B^2$. The quarks' invariant mass reads
\beq
\cM_{12}^2 \es {m_1^2 + k^{\perp 2}  \over x_1  } + {m_2^2 + k^{\perp 2}  \over x_2  } \\
                   \es \left[  m_1 + m_2  + {\vec k\,^2  \over 2 \mu } + O(k^4/m^3) \right]^2 \ ,
\eeq
where $\mu$ denotes the reduced mass for the quarks. Neglecting $B^2/(4m)$ in Eq.~(\ref{effEVP}), one  obtains
\beq
\label{effEVPnr}
   {\vec k\,^2  \over 2\mu }  \varphi^s_{k} + \int_{k'} \hspace{-.05in} V^C_{kk'} \varphi^s_{k'} 
+ [V \varphi^s ]_k \es B \, \varphi_{k}^s \ ,
\eeq
with 
\beq 
\label{[Vvarphi]}
4m [V \varphi^s ]_k \es \left( {\sigma_1  \over  x_1 } + {\sigma_2  \over  x_2 } \right) \, \varphi^s_{k} 
+ \int_{k'} \hspace{-.05in}  h^{f \hspace{-2pt}f }_{kk'} \varphi^s_{k'}  \ .
\eeq
The term with matrix $V^C = h^C/(4m)$ is the Coulomb interaction stemming from Eq.~(\ref{hC}). 
The term $[V \varphi^s ]_k$ is described below.
\subsubsection{Interaction $V \varphi^s$}
\label{Vphis}
The interaction $V \varphi^s$ of Eq.~(\ref{[Vvarphi]}) in the limit $m_g \to 0$, is discussed following the steps made in 
Sec.~\ref{mgdivergencesB}, where it is shown that the terms $\sim \ln m_g$ cancel out in the full relativistic formula. 
Knowing that the singular terms cancel out, we can identify elements of the remaining finite terms. We take advantage 
of the simplifications occurring when one uses the non-relativistic approximation to describe the relative motion of 
quarks. Following Sec.~\ref{mgdivergencesB}, we write $[V \varphi^s ]_k$ as a sum of two parts corresponding to the 
upper and lower diagrams in Fig.~\ref{fig:selfexch},
\beq
[V \varphi^s ]_k \es [V \varphi^s ]_k^1 + [V \varphi^s ]_k^2  \ ,
\eeq
where, using $z = x_1'-x_1$, 
\beq 
\label{[Vvarphi 1]}
4m [V \varphi^s ]_k^1 \es {\sigma_1  \over  x_1 } \, \varphi^s_{k} 
+ \int_{k'} \hspace{-.05in}  h^{f \hspace{-2pt}f }_{kk'} \varphi^s_{k'} \, \theta(-z)  \ , \\
\label{[Vvarphi 2]}
4 m [V \varphi^s ]_k^2 \es  {\sigma_2  \over  x_2 } \, \varphi^s_{k} 
+ \int_{k'} \hspace{-.05in}  h^{f \hspace{-2pt}f }_{kk'} \varphi^s_{k'} \, \theta(z)  \ .
\eeq
We focus on $[V \varphi^s ]_k^1$. Details of $[V \varphi^s ]_k^2$ are similar. The self-interactions $\sigma_i$ are given 
in Eq.~(\ref{deltai}). They preserve the quark spins. We have
\beq
\label{deltaiC}
\sigma_i 
\es
G \int {dx \, d^2k^\perp \over 1-x} \, e^{-2 (s \rho_i/p_g^+)^2}
\nt 
{ j_{i\mu} \left[ - g^{\mu \nu} + \eta^\mu \eta^\nu \rho_i/p_g^{+2} \right] j_{i\nu}^* \over \rho_i}  \ ,
\eeq
where $G = g^2 C_F/[ 2(2\pi)^3]$, $x=p_g^+/p_i^+$ ranges  from 0 to 1, $k^\perp = p_g^\perp - x p_i^\perp$ is unlimited, 
and $\rho_i = - (p_i-p_q)^2$. The dominant parts of the quark currents are $j_i^+ = 2 \delta_{\rm spin} \sqrt{ p_i^+ p_q^+}$ and 
$j_i^-= 2\delta_{\rm spin} m_i^2/\sqrt{ p_i^+ p_q^+}$, being on the order of the quark mass. The other parts of the currents are 
on the order the quark relative momenta~\cite{Lepage:1980fj}. Neglecting the momenta in comparison to the masses, the 
self-interactions amount to the multiplication of the wave function $\varphi^s_{k}$ by
\beq
\label{deltaiG1}
\sigma_i 
\es
4G \int {dx \, d^2k^\perp \over 1-x} \, 
{ - m_i^2 + p_i^+ p_q^+ \rho_i/p_g^{+2} \over \rho_i} e^{-2 (s \rho_i/p_g^+)^2} . \nn 
\eeq
In the case of $[V \varphi^s ]_k^1$, Eq.~(\ref{exchhkk'}) for $h^{f \hspace{-2pt}f }_{kk'} $ implies 
\beq
\int_{k'} \hspace{-.05in}  h^{f \hspace{-2pt}f }_{kk'} \varphi^s_{k'} \, \theta(-z) 
\es
 - G \int_0^{x_1} {dx_{1'} \, d^2k^\perp_{1'2'} \over x_{1'}x_{2'} }
{\rho_1 + \rho_2  \over 2\rho_1 \rho_2} \nt
j_{1\mu}  d^{\mu \nu} j_{2\nu} \,
e^{- (s \rho_1/p_g^+)^2 - (s \rho_2/p_g^+)^2}  \varphi^s_{k'} , \nn
\eeq
where, neglecting the transverse relative momenta in comparison to the masses, the dominant term in the product of currents 
conserves the quark spins and reads 
\beq
j_{1\mu}  d^{\mu \nu} j_{2\nu} 
\es
4 m_1 m_2 \left[ -1 +  { \sqrt{x_1x_{1'}x_2 x_{2'} } \over 2 m_1 m_2} \,  (\rho_1 + \rho_2)/z^{2} \right] \ .    ~~~~~~~~~~~  {\red OK}  
\eeq
From Eqs.~(\ref{rho1 etaeta}) and (\ref{rho2 etaeta}) for $m_g=0$, with accuracy to terms bilinear in the quark relative momenta, 
\beq
\rho_1 \es \rho_2 \rs \vec q\,^2 \ , \ \vec q \rs \left[ k_{1'2'}^\perp - k_{12}^\perp , z( m_1+m_2) \right]  \ . 
\eeq
Consequently,
\beq
\label{hffphi}
\int_{k'} \hspace{-.05in}  h^{f \hspace{-2pt}f }_{kk'} \varphi^s_{k'} \, \theta(-z) 
\es
 - G \int_0^{x_1} {dx_{1'} \, d^2k^\perp_{1'2'} \over x_{1'}x_{2'} }
{1 \over \rho} \nt
j_{1\mu}  d^{\mu \nu} j_{2\nu} \,
e^{- 2(s \vec q\,^2/p_g^+)^2 }  \varphi^s_{k'}  ,
\eeq
where 
\beq
j_{1\mu}  d^{\mu \nu} j_{2\nu} 
\es
4 m_1 m_2 \left( -1 +  {\vec q\,^2 \over q^{z 2}} \right) \ .    
\eeq
The last equation follows from the approximations $x_1 = x_{1'} = x_{10}$ and $x_2 = x_{2'} = x_{20}$.
Now we use Eqs.~(\ref{deltaiG1}), (\ref{hffphi}),  and the relations $|q^z| = x m_1 = |z|(m_1+m_2)$,
which hold in the leading approximation neglecting the quark relative momenta in comparison to 
their masses. Eq.~(\ref{[Vvarphi 1]}) is obtained in the form
\beq
\label{[Vvarphi 1] pom 1}
4m [V \varphi^s ]_k^1 
\es
{4G \over x_1} \int_0^1 dx \, d^2k^\perp  m_1^2  \, 
\ \left[ {1 \over  q^{z2}  } - { 1 \over (1-x) \vec q\,^2}  \right] \nt e^{-2 (s \vec q\,^2/p_g^+)^2} \ \varphi^s_{k} 
\nm
4G \int_0^{x_1} {dx_{1'} \, d^2k^\perp_{1'2'} \over x_{1'}x_{2'} }
m_1 m_2 \left( {1\over q^{z 2}}  - {1 \over \vec q\,^2}  \right) \nt
e^{- 2(s \vec q\,^2 /p_g^+)^2 }  \varphi^s_{k'}  \ .
\eeq
We change the integration variable $k^\perp$ to $q^\perp$, $p_q^\perp = p_1^\perp - q^\perp$,
$k_{1'2'}^\perp = k_{12}^\perp - q^\perp$, and $x_{1'} = x_1 + z = x_1 - |z|$. We also substitute $p_g^+ 
= |z| P^+$ and introduce $s_p = s/P^+$. Using $q^z = x \, m_1 = |z| \, (m_1+m_2)$, $dx \, m_1 \to dq^z$ and $dx_{1'} \, (m_1 + m_2) \to dq^z$. For $s = 1/m$ and $P^+= 2m$, 
\beq
e^{-2 (2 m s_p \vec q\,^2/q^z)^2} = e^{-2 (\vec q\,^2/ m q^z)^2} = e^{-(2/\cos^2 \hspace{-2pt} \theta)  (q/m)^2 }
\hspace{-3pt} ,
\eeq
which implies that $q/m$ can be limited to values smaller than 1, validating the non-relativistic approximation. Extending 
the integrals freely beyond the mass scale because of the exponential functions quick falloff, and approximating $x_i \sim 
x_{i'} \sim x_{i0}$, we arrive at
\beq
[V \varphi^s ]_k^1 
\es
- 2G \int  d^3q \, \theta( q^z) \, \left( {1\over q^{z 2}}  - {1 \over \vec q\,^2}  \right) 
\nt e^{- 2(2m s_p \vec q\,^2/q^z)^2}  
\left( \varphi^s_{k'} -  \varphi^s_{k}  \right) .
\eeq
The interaction $[V \varphi^s ]_k^2$ provides the complementary part corresponding to $\theta(-q^z)$ and the combined result is
\beq
[V \varphi^s ]_k \es 
- g^2 C_F \int  {d^3q \over (2\pi)^3} \,  \left( {1\over q^{z 2}}  - {1 \over \vec q\,^2}  \right) 
\nt e^{- 2(2m s_p \vec q\,^2/q^z)^2}  \left( \varphi^s_{k'} -  \varphi^s_{k}  \right) \ .
\eeq
This result matches Eq.~(67) in~\cite{Serafin:2023pkf}, obtained there using an additional counterterm. In our approach, such 
counterterm is not needed, which we explain in Sec.~\ref{xinboundstates} below Eq.~(\ref{newterm}). Expanding the wave-function 
in a Taylor series in $\vec q$, 
\beq
\label{expansion}
\varphi^s_{\vec k+ \vec q} -  \varphi^s_{\vec k} \es 
\2 q^i q^j {\partial \over \partial k^i} {\partial \over \partial k^j}  \varphi^s_{\vec k} + O(q^4) \ ,
\eeq
one obtains the quadratic term, 
\beq
[V \varphi^s ]_k \es -  t^{ij} {\partial \over \partial k^i} {\partial \over \partial k^j} \varphi_{k}^s  \ , \\
t^{ij} \es G \, { \pi^{3 \over 2} \over 12 (\sqrt{8} \, m s_p)^3}  \, \delta^{ij} \ .
\eeq
The eigenvalue Eq.~(\ref{effEVPnr}) thus takes the form
\beq
\label{effEVPnr osc Deltak}
{\vec k\,^2  \over 2\mu }  \varphi^s_{k} + \int_{k'} \hspace{-.05in} V^C_{kk'} \varphi^s_{k'} 
 - \2 \mu \omega^2  \Delta_k \varphi_{k}^s  \es B \, \varphi_{k}^s \ ,
\eeq
including the rotation invariant harmonic oscillator potential with
\beq
\omega^2 \es {\alpha C_F \over 48 \sqrt{2\pi}} { (P^+/s)^3 \over m_1 m_2 (m_1 + m_2)^2 } .
\eeq
Thus our approach to QCD of heavy quarks reproduces the frequency obtained in Eq.~(73) of~\cite{Serafin:2023pkf}. 

The effective Hamiltonian eigenvalue Eq.~(\ref{effEVPnr osc Deltak}), containing a superposition of the Coulomb and 
harmonic oscillator potentials, and similar equations derived for three-quark states, are known to lead to reasonable 
values of masses of ground and lowest excited mesons and baryons  made of heavy quarks~\cite{Serafin:2018aih}.
For example, after fitting the masses of quarks $c$ and $b$ to the heavy meson mass spectrum, the predicted masses 
of baryons $\Omega_{\rm ccc}(3/2+)$ and $\Omega_{\rm ccc}(3/2-)$ are about 4797 and 5103 MeV/$c^2$, 
respectively~\cite{Serafin:2018aih}. Recent lattice computations~\cite{Dhindsa:2024erk} produce 4793 and 5094 
MeV/$c^2$, correspondingly. On one hand this agreement is surprising because of the crude approximations made 
in the second-order RGPEP approach. On the other hand it is remarkable because the simple RGPEP calculation not 
only predicts the baryon masses but also provides expressions for their wave functions. There exist independent
arguments for expecting that an effective FF theory may involve a quadratic potential~\cite{Trawinski:2014msa}.
\subsubsection{$Q \bar Q$ component and quark confinement}
\label{QQconf}
The harmonic oscillator potential in Eq.~(\ref{effEVPnr osc Deltak}) is obtained at short quark-antiquark distances after the 
cancellation of small-$x$ logarithmic divergences is accounted for in the eigenvalue problem for the Hamiltonian of 
Eq.~\ref{HQbarQ}). The diverging terms involve the logarithms $\ln (s m_g^2/P^+)^{-1}$ and color factors. The cancellation 
occurs in the globally colorless states of the effective quark and antiquark, because the color factors in the quark self-interactions 
and in the gluon exchange terms are opposite to each other, $4/3$ and $-4/3$, respectively.  But in the color-octet effective 
quark-antiquark states, the gluon exchange color factor is $1/6$ instead of $-4/3$ while the quark self-interaction terms stay 
the same. Therefore, the logarithms of the gluon mass do not cancel out and the color-octet quarkonium mass eigenvalues 
tend to infinity when $m_g \to 0$. Moreover, if instead of the quark-antiquark states one considers states with the quantum 
numbers of a single quark, the diverging quark self-interaction term has no second-order Hamiltonian term to cancel with. 
These results are in agreement with the absence in the Hamiltonian spectrum of single-quark eigenstates with finite mass 
eigenvalues, while the colorless quarkonia have finite masses. One can thus hope that the RGPEP scheme for computing 
renormalized, scale-dependent Hamiltonians in QCD of heavy quarks, understood as a limit of a theory with $m_g > 0$ when 
$m_g \to 0$, is a candidate for deriving realistic quantum mechanical eigenvalue problems for hadrons. We wish to stress
at this point that independently of the arguments given in this paper, defining confinement as the absence of colored states 
with finite masses requires some parameter to carry out a mathematically well defined, finite calculation before one considers 
the infinite masses of colored states as the result of taking that parameter to its suitable limit. The remaining discussion 
concerns some aspects of our approach, such as the cancellation of small-$x$ singularities in the quarkonium components 
including effective gluons in addition to the heavy quark pair, and comparison with QED~\cite{Bloch:1937pw}, where the 
cancellation of the photon small-$x$ singularities does not result in any residual harmonic force between electrons. Some of 
these aspects have been addressed in the past in various other ways~\cite{Wilson:1994fk,Perry:1994kp,Perry:1994mv, Brisudova:1996vw}. 
The discussion that follows is focused on approaching QCD through the limit $m_g \to 0$. 

The interaction terms discussed above appear in the renormalized Hamiltonian for the heavy effective quarks corresponding 
to the scale parameter $s \sim 1/m$, instead of the quarks of bare QCD, corresponding to $s=0$. Interactions of the effective 
quarks at a scale $s$ cause only limited changes of the FF free energy, the limit being $\sim 1/s$. In contrast, the interaction terms 
in the bare QCD Hamiltonian can directly produce arbitrarily large changes of the FF energy and thus they can reach quark-antiquark 
states that are degenerate in energy with the multi-particle states. Consequently, the bare interactions are able to generate significant 
eigenstate components with many bare quanta, producing that way a complex image of the hadronic structure. In the alternative 
representation in terms of the effective particles, the probability of the multiparticle components is diminished. This feature is of 
interest because the phenomenology of heavy quarkonia assures us that the quark-antiquark component provides a physically 
reasonable approximation~\cite{Workman:2022ynf, ParticleDataGroup:2024cfk}. Nevertheless, the effective Hamiltonian computed 
using the RGPEP still includes the interactions changing the number of the effective quanta, albeit they are suppressed by the 
vertex factors with a small FF-energy width when $s$ is large. 

The effective interactions discussed above are merely of the second order in the expansion in a series of powers of the coupling 
constant $g$. The coupling constant is plainly assumed to be small. However, in the FF phenomenology of heavy hadrons, one 
uses $\alpha_s = g_s^2/(4\pi)$ that depends on $s$, {\it e.g.,} see~\cite{Serafin:2018aih}. The realistic coupling constant $\alpha_s$ 
for $s \sim 1/m$ is on the order of 1/3. Such value corresponds to the running coupling constant measured at the electroweak 
mass scale. Rigorous extrapolation of the second-order picture to the realistic values of the coupling constant requires a demonstration 
that the terms of orders higher than second in the expansion in powers of $g_s$ involve relatively small operator coefficients. The 
expectation that for $s \sim 1/m$ the quarkonium components other than the quark and antiquark are relatively small needs to be 
verified using Hamiltonians computed in the RGPEP of orders higher than second~\cite{Glazek:2021vnw}.

The rotation invariant effective quadratic potential is the first term in the Taylor expansion in Eq.~(\ref{expansion}). The sum of 
the Taylor series yields an effective quark-antiquark potential varying logarithmically with $\vec r$ at large distances~\cite{Serafin:2023pkf}. 
But the rotation symmetry is increasingly broken as the distance increases toward and above the quarkonium radius. Namely, at 
large distances, the coefficient of $\ln |r^\perp|$ for $r^z=0$ is twice larger than the coefficient of $\ln |r^z|$ for $r^\perp = 0$. 
This result matches the results obtained in~\cite{Perry:1994mv,Brisudova:1996vw}. The reason for the rotation symmetry breaking 
to increase with the inter-quark distance is likely to be the fact that the quark-antiquark component of large potential energy is 
degenerate with the components containing additional gluons. Therefore, one may expect the $Q \bar Q$ component with large 
potential energy to be strongly mixed by the interactions with the components containing additional gluons~\cite{Brisudova:1997rv}. 
The mechanism of breaking and the gluon cure for recovering the rotation symmetry needs to be assessed using the RGPEP, by 
computing the renormalized Hamiltonian terms including effective gluons at scales $s \sim 1/m$ in the limit $m_g \to 0$. The 
effective gluons may develop masses that grow as $s$ increases.

The second-order seed of confinement cannot appear in QED. But the second-order QCD interactions differ from the analogous 
second-order QED interactions between the charged fermions by replacement of $e^2$ with $(4/3)g^2$ only. The issue is that 
the second-order RGPEP additions to the Coulomb interaction between a fermion and an antifermion are canceled in QED when 
one accounts for the interactions that couple the fermion-antifermion and fermion-antifermion-boson components of a bound 
state. Thus the second-order quark-confinement seed can be valid in QCD and absent in QED if the coupling between the components 
$Q \bar Q$ and $Q \bar QG$ is suppressed in QCD ($G$ denotes an effective gluon). The suppression may result from the FF energy 
gap generated by the non-Abelian interactions in the three-body component. There are no such interactions between photons and 
charged fermions. Hence there is no such gap in QED. In the next section, we provide an initial discussion of the interaction terms in 
the quark-antiquark-gluon components of quarkonia using our RGPEP approach with $m_g$ and $\phi$ in the limit of $m_g \to 0$.
\section{Small-$x$ singularities in components with effective gluons} 
\label{includinggluons}
The quarkonium two-body component is coupled to the three-body component containing an effective gluon in addition to the 
quark pair, 
\beq
\label{QbarQGstates}
 |Q_{s1} \bar Q_{s2} G_{s3}\rangle \es b_{s1}^\dagger d_{s2}^\dagger a_{s3}^\dagger |0\rangle \ ,
\eeq
and to the states with more effective gluons. States including additional quark-antiquark pairs are assumed so heavy that they 
do not significantly contribute to the ground and low excited quarkonium eigenstates. They can be included in precise RGPEP 
calculations. All these components  correspond to the basis states $| i \rangle_s$ in Eq.~(\ref{Hteppsi}), while their coefficients 
$\psi_i(s)$ are provided by the eigenstate wave functions. 

The RGPEP effective eigenvalue problem differs from the equations derived using the Tamm-Dancoff approximation~\cite{Tamm:1945qv,Dancoff:1950ud}. That approximation requires counterterms to the regularization effects, 
including the counterterms depending on the Fock sector~\cite{Perry:1990mz,Perry:1991ny}. In contrast, the renormalized 
Hamiltonian $H_s$ is finite and independent of the regularization parameter $r \to 0$. The effective interaction terms in $H_s$ 
include the vertex form factors $f^s_{\bar p,\bar q}$ computed according to Eqs.~(\ref{cH1t}) to (\ref{Tft}). The eigenvalue 
problem for $H_s$ with finite $s$ does not require any counterterms. Nevertheless, its eigenstates have components with 
different numbers of effective constituents,
\beq
|\psi \rangle \es |2\rangle_s + |3 \rangle_s + . . . \ ,
\eeq
where dots indicate the components $|n\rangle_s$ with $n> 3$. Below, we demonstrate the cancellation of the gluon small-$x$ 
singularities among the Hamiltonian terms acting in the effective-particle three-body component.
\subsection{Effective interactions in the three-body component}
\label{H_3body seagull etc}
According to Eq.~(\ref{cH2t}), using abbreviated notation for the states of effective particles, $|Q_{s1} \bar Q_{s2} G_{s3}\rangle \equiv
|123\rangle_s$, and taking  advantage of an identity analogous to Eq.~(\ref{2bodyMatrixHs}), 
\beq
\label{3bodyMatrixHs}
_s\langle 123| H_s^{(2)} | 1'2'3' \rangle_s 
\es  
\langle 123|\cH^{(2)} |1'2'3'\rangle \ ,
\eeq
the Hamiltonian matrix elements in the $Q\bar Q G$ component of a quarkonium are
\beq
\label{HLR1pom1text  pom}
_s\langle 123| H_s | 1'2'3 \rangle_s
\es \langle 123|  \left[  {\cH_f} + f_{LR} \, \cH_r^{(2)}  
\right.
\nm
\left. 
 f_{LI} f_{IR} \, \Delta_{LIR} \, \cH_0^{(1)}  \, \cH_0^{(1)} \right]  | 1'2'3' \rangle , \nn
\eeq
where $\cH_r^{(2)} = \cH^{(2)}_0 + \Delta_{LIR} \  \cH^{(1)}_0 \, \cH^{(1)}_0$. The matrix elements of the operator $f_{LR} \, \cH_r^{(2)}$ 
include the self-interactions and the gluon exchanges, both due to $\Delta_{LIR} \ \cH^{(1)}_0 \, \cH^{(1)}_0$ and the self-interaction 
counterterms and seagulls in $\cH^{(2)}_0$. These terms involve singularities $\sim 1/x^2$. Matrix elements of $f_{LI} f_{IR} \, \Delta_{LIR} 
\, \cH_0^{(1)}  \, \cH_0^{(1)}$ do not diverge for states of finite total $P^+$ when $s > 0$, as long as $m_g > 0$. Our description of the 
interactions in the effective three-body component parallels the case of two-body component discussed in Sec.~\ref{Qdynamics}. 
\subsection{Cancellation of $1/x^2$ in terms $f_{LR}\cH_r^{(2)}$}
\label{3body1x2}
\begin{figure}[ht!] 
          \caption{Illustration of the Hamiltonian $Q\bar QG$ matrix elements $\langle 123| f_{LR}\cH_r^{(2)} | 1'2'3' \rangle$ 
                         in Eq.~(\ref{HLR1pom1text  pom}), involving two complementary orderings of an exchange of a gluon 4 
                         between a gluon 3 and a quark 1, panels $\rm a_1$,  $\rm a_2$, and an associated seagull interaction 
                         between the gluon 3 and the quark 1, panel $\rm a_3$. Panels $\rm b_1$, $\rm b_2$, $\rm b_3$ illustrate
                         the matrix elements of the analogous interactions of the gluon 3 with antiquark 2. Exchange of the gluon
                         4 between the quark 1 and antiquark 2 and the associated seagull are illustrated in panels $\rm c_1$, 
                         $\rm c_2$, $\rm c_3$. Cancellation of the small-$x$ singularities among the interactions illustrated by 
                         the triple panels $\rm a$,  $\rm b$, and $\rm c$, is described in the text.}
                         \label{Fig3-bodyfHcancel}
\begin{center}
         \includegraphics[width=0.3\textwidth]{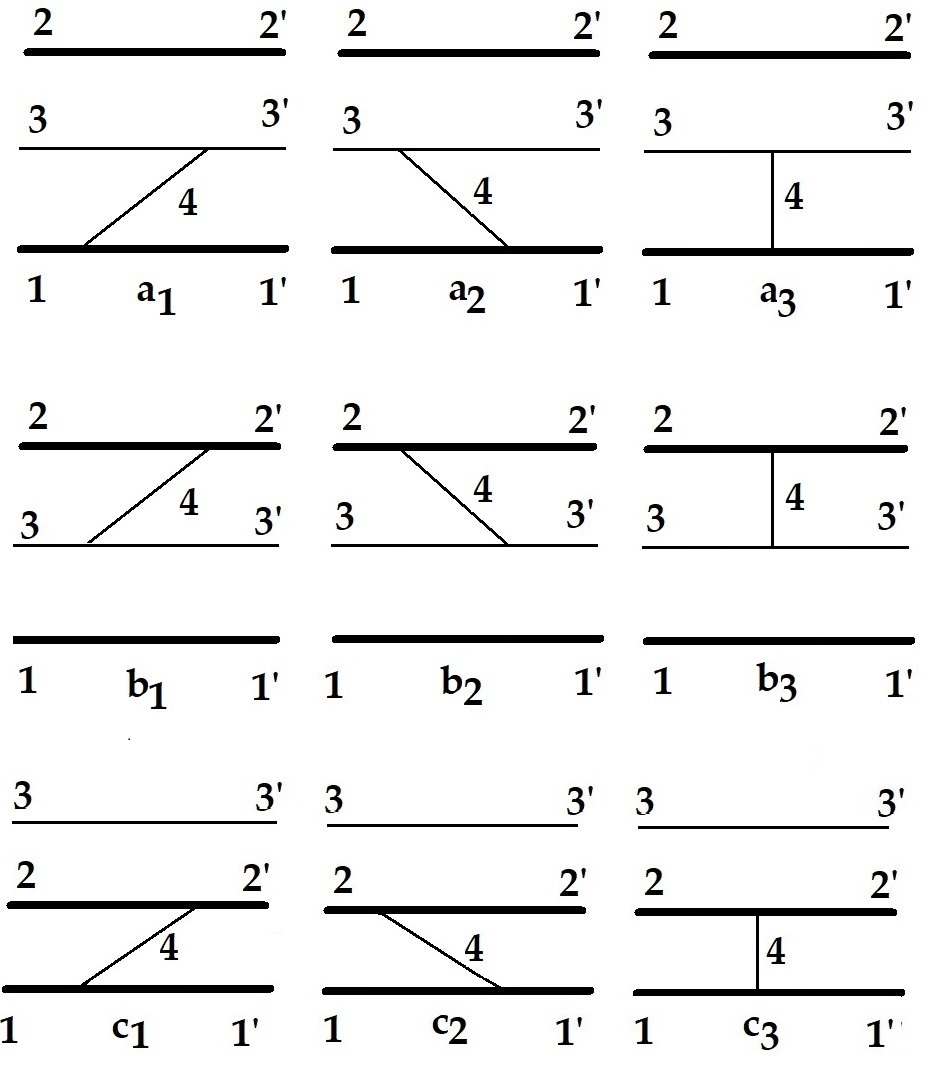}
\end{center} 
\end{figure}

One source of the divergences $\sim 1/x^2$  in $f_{LR}\cH_r^{(2)}$ are the self-interactions. These are canceled by the self-interaction counterterms. These counterterms act in all sectors of the virtual Fock space equally, as a consequence of the RGPEP operator calculus, 
see Sec.~\ref{cC}. Therefore, the self-interaction divergences in the three-body component are countered in the same way as in the 
two-body component. Divergences $\sim 1/x^2$ appear in $f_{LR} \, \cH_r^{(2)}$ also due to the gluon exchange and gluon seagull 
terms. These are illustrated graphically in Fig.~\ref{Fig3-bodyfHcancel}. The gluon exchange terms proportional to $g^{\mu \nu}$ are 
finite. The singularities $\sim 1/x^2$ appear in the terms proportional to $\eta^\mu \eta^\nu$ only.  They cancel out  
within the 3 subsets of terms illustrated by the 3 rows of panels in Fig.~\ref{Fig3-bodyfHcancel}. The cancellation occurs
despite the assignment of mass $m_g$ to gluons thanks to the contribution of quanta of the auxiliary field $\phi$. 

The mechanism of cancellation of singularities $\sim 1/x^2$ in the case of the quark-antiquark gluon exchange and 
the corresponding seagull term, panels $\rm c$ in Fig.~\ref{Fig3-bodyfHcancel}, is the same as the one described in 
Sec.~\ref{xinboundstates},  Eqs.~(\ref{rho1-1}) and (\ref{newterm}). The quark-antiquark interactions in the three-body 
and two-body components differ only by the color factor, $-1/6$ instead of $4/3$, and thus do not require further 
discussion.

The interaction of a gluon with a quark or an antiquark in the 3-body space, illustrated by panels $\rm a_1$,
$\rm a_2$, $\rm b_1$, and $\rm b_2$ in Fig.~\ref{Fig3-bodyfHcancel}, result from the gluon exchange, caused 
by the double action of the terms stemming from $- (J_{\psi f}^{a \mu}  + J_{A f}^{a \mu}) A_{f\mu}^a$ in 
$\cH_{\rm QCD}$ in Eq.~(\ref{cHcanQCD}) and from $m_g \phi^a {1 \over \partial^+} (J_{Af}^{a+} + J_{\psi f}^{a+})$ 
in Eq.~(\ref{cHcanQCDmg}). The currents $J_{\psi f}^{a \mu}$ and $J_{A f}^{a\mu}$ are defined in 
Eqs.~(\ref{quarkcurrent}) and (\ref{gluoncurrent}). The divergences $\sim 1/x^2$ in these exchanges are canceled 
by the seagull terms originating from the last Hamiltonian density term in Eq.~(\ref{cHcanQCD}). In terms of the
operators, the singularity $1/x^2$ in $f_{LR} \ \Delta_{LIR} \ f_r \cH^{(1)}_0 \, f_r \cH^{(1)}_0$ is canceled by the 
seagull $f_{LR} f_r f_r H_{\rm sgq}$, where 
\beq
\cH^{(1)}_0 
\eqv
- \int  \left( J_{\psi f}^{a \mu} + J_{A f}^{a \mu} \right) \left( A_{f\mu}^a + {m_g \eta_\mu  \over \partial^+} \phi^a \right) \ , \\
\label{HJJ POM 21}
H_{\rm sgq} \es - \int J_{A f}^{a+} {1 \over \partial^{+2}} J_{\psi f}^{a+}  \ ,
\eeq
and $\int = \int d^2x^\perp  dx^-$ for $x^+=0$. Ignoring pair creation and annihilation, the operator $\cH^{(1)}_0$ 
is a sum of 4 terms, 
\beq
\cH^{(1)}_0 \eqv \cH_{\psi A} + \cH_{\psi \phi} + \cH_{A A} + \cH_{A \phi} \ . 
\eeq
Including sums over discrete quantum numbers in the integration sign, one has
\beq
\label{cH10line1}
\cH_{\psi A} \es - \int \left[: J_{\psi f}^{a \mu}   A_{f\mu}^a :\right]^r \\
\es g
\int[11'4] \, \tilde \delta_{c.a} \, r_{\overline 1, \underline 1 4} 
\ \chi_1^\dagger  T^4 \chi_{1'}  \ \bar u_1 \gamma_\nu u_{1'}       
\nt
\left(  \varepsilon^\nu_4 \ b_1^\dagger  b_{1'}  a_4 + \varepsilon^{\nu *}_4  a^\dagger_4 b_1^\dagger  b_{1'}  \right)
\nm
g
\int[22'4] \, \tilde \delta_{c.a} \, r_{\overline 2, \underline 2 4} 
\ \chi_{2'}^\dagger T^4 \chi_2 \ \bar v_{2'} \gamma_\nu v_2 \nt  
\left(  \varepsilon^\nu_4 \ d^\dagger_2 d_{2'} a_4 + \varepsilon^{\nu *}_4  \ a^\dagger_4 d^\dagger_2 d_{2'} \right) \ , 
\eeq
\beq
\label{cH10line2}
\cH_{\psi \phi} \es - \int \left[ :J_{\psi f}^{a \mu}  {m_g \eta_\mu  \over \partial^+} \phi^a :\right]^r \\
\es g
\int[11'\tilde 4] \, \tilde \delta_{c.a} \, r_{\overline 1, \underline 1 \tilde 4} 
\ \chi_1^\dagger T^{\tilde 4} \chi_{1'} 
\ \bar u_1 \gamma^+ u_{1'} 
\nt
\ {m_g  \over p_{\tilde 4}^+} 
\left( b_1^\dagger b_{1'}  a_{\tilde 4} + a^\dagger_{\tilde 4} b_1^\dagger b_{1'}  \right) 
\nm
g
\int[22'\tilde 4] \, \tilde \delta_{c.a} \, r_{\overline 2, \underline 2 \tilde 4}  
\ \chi_{2'}^\dagger T^{\tilde 4} \chi_2
\ \bar v_{2'}    \gamma^+ v_2
\nt {m_g  \over p_{\tilde 4}^+}
\left( d^\dagger_2d_{2'}  a_{\tilde 4} + a^\dagger_{\tilde 4}d^\dagger_2d_{2'}  \right) \ , 
\eeq
\beq
\label{cH10line3}
\cH_{A A} \es - \int \left[: J_{A f}^{a \mu}  A_{f\mu}^a  :\right]^r  \\
\es
\label{HA343'check3summaryfinal} 
g \int [343']  \left[  r_{34,3'} a_3^\dagger a_4^\dagger a_{3'} \ Y_{343'} \, \tilde \delta_{34.3'} 
\right.
\np
\left.
r_{3,43'} \,a_{3}^\dagger  a_4 a_{3'} \ Y_{3'43}^* \, \tilde \delta_{3.43'} \right] \ , 
\eeq
\beq
\label{HA343'check4}
& &  Y_{343'} \equiv i f^{c_3 c_4 c_{3'}}
\left[ \varepsilon_3^* \varepsilon_4^* \,\varepsilon_{3'} p_3 
-
\varepsilon_3^*\varepsilon_{3'} \, \varepsilon_4^* ( p_{3'} + p_{3} )  \right]  \ , \nn
\eeq
\beq
\label{cH10line4}
\cH_{A \phi}  \es - \int \left[ :J_{A f}^{a \mu}  {m_g \eta_\mu  \over \partial^+} \phi^a : \right]^r \\
\es g m_g
\int [433'] \, \tilde \delta_{c.a} \ r_{43,3'} \, i f^{c_4 c_3 c_{3'}} 
\nt
{p_3^+ + p_{3'}^+ \over p_4^+}  \ 
\varepsilon^*_3 \varepsilon_{3'}     \ \tilde a^\dagger_{4} a^\dagger_{3}    a_{3'}    \  + \ h.c. \ . 
\eeq
The relevant seagull terms are contained in the operator of Eq.~(\ref{HJJ}). From Eqs.~(\ref{jquark1}) 
and (\ref{jquark2}), we obtain the gluon-quark and gluon-antiquark seagull terms
\beq
\label{Hsqg2}
H_{sgq} \es -g^2 \int [33'11'] \ \tilde \delta_{c.a} \ \delta_{\rm spins} 
r_{\overline 1,\underline 1 (\overline 1 - \underline 1)} 
r_{\overline 3,\underline 3 (\overline 3 - \underline 3)} 
\nt
{ ( p_3^+ + p_{3'}^+ )  \ 2 \sqrt{p_1^+ p_{1'}^+}  \over (p_{3'}^+ - p_3^+)^2} 
\nt 
\chi_{1}^\dagger T^a \chi_{1'} \, i f^{abc} \  a^\dagger_{3b} a_{3'c}  b_1^\dagger  b_{1'} \ , \\
\label{Hsantiqg2}
H_{sg\bar q} \es+ g^2 \int [33'22'] \ \tilde \delta_{c.a} \ \delta_{\rm spins}
r_{\overline 2,\underline 2 (\overline 2 - \underline 2)} 
r_{\overline 3,\underline 3 (\overline 3 - \underline 3)} 
\nt
{ ( p_3^+ + p_{3'}^+ ) \ 2 \sqrt{p_{2}^+ p_{2'}^+} \over (p_{3'}^+ - p_3^+)^2} 
\nt
 \chi_{2'}^\dagger T^a \chi_{2} \, i f^{abc} \ a^\dagger_{3b} a_{3'c}  d^\dagger_{2} d_{2'}  \ , 
\eeq
corresponding to the panels $\rm a_3$ and $\rm b_3$ in Fig.~\ref{Fig3-bodyfHcancel}. We focus on the gluon 3 interaction with 
the quark 1 and omit discussion of the gluon 3 interaction with the antiquark 2. Cancellation in the latter works in the same way. 
Since the sign and color factors for exchanges and seagulls in the panels $\rm a$ and $\rm b$ are the same, $-3/2$, the cancellation 
rests on the momentum and spin factors in Eqs.~(\ref{cH10line1}) to (\ref{Hsantiqg2}). The divergence $\sim 1/x^2$, with $p_4^+
=|x|P^+$, comes from the squares of $\varepsilon_4^-$ and $m_g/p_4^+$ in the exchange and from $1/p_4^{+2}$ in the seagull. The 
gluon exchanges illustrated in the panel $\rm a$ of Fig.~\ref{Fig3-bodyfHcancel} include two complementary orderings, one for 
positive and another one for negative $x =x_{1'} - x_1$, 
both producing the divergence $\sim 1/x^2$. It suffices to consider the cancellation for one sign of $x$. We describe details of
the case $x_1>x_{1'}$. It resembles the case described in Sec.~\ref{xinboundstates} for the two-body component, see the panels 
upper-left and lower in Fig.~\ref{fig:ABC}.

Using the abbreviation $\theta_{1-1'} = \theta(p_1^+-p_{1'}^+)$, the operator generating the exchange of the transverse gluon 
4, a quantum of the field $A$, between the gluon 3 and quark 1, reads
\beq
\label{L1final}
H_A \es g^2 \sum_4 \int[11'33'] \,  \theta_{1-1'} \, { \tilde \delta_{13.1'3'} \over p_4} \, \Delta_{LIR}
\, \chi_1^\dagger  T^4 \chi_{1'} \, 
\nt  
\bar u_1 \gamma_\nu u_{1'} \ i f^{c_{3} c_4 c_{3'}} 
\ Y_{131'3'} \ b_1^\dagger  b_{1'} \, a_{3}^\dagger a_{3'}  \ , 
\eeq
where $Y_{131'3'} = 2 \varepsilon^\nu_4 \left( \varepsilon_3^* \varepsilon_{4}^* \,\varepsilon_{3'} p_3
+
\varepsilon_{4}^*\varepsilon_{3'} \, \varepsilon_3^* p_{4} - \varepsilon_3^*\varepsilon_{3'} \, \varepsilon_{4}^* p_{3'} \right)$.
The singularity $1/x^2$ comes from the last term in $Y_{131'3'}$. The regularization factors $r$ are not displayed here 
and below to simplify the formulas. They are the same in all terms of panel ${\rm a}$ and can be reinstated according to 
Eqs.~(\ref{cH10line1}) to (\ref{Hsantiqg2}) or (\ref{currentfr}). The operator contributing the exchange of the field-$\phi$ 
quanta  is
\beq
\label{L3final}
H_\phi
 \es g^2
\int[11'33'] \, \theta_{1-1'} \, { \tilde \delta_{13.1'3'} \over p_4^+} \, \Delta_{LIR}
\ \chi_1^\dagger T^{4} \chi_{1'} 
 \bar u_1 \gamma_\nu u_{1'} 
\nt
 {m_g^2 \eta^\nu \over p_{4}^{+2}}  
\ \ 
i f^{c_{4} c_3 c_{3'}} (p_3^+ + p_{3'}^+)  \ \varepsilon^*_3 \varepsilon_{3'}   \ b_1^\dagger b_{1'}   \, a^\dagger_{3}    a_{3'}  \ .
\eeq
The factor $\Delta_{LIR}$ enters the operators according to Eq.~(\ref{HLR1pom1text  pom}). In precise analogy with 
Sec.~\ref{xinboundstates}, where we discuss the quark-antiquark interaction, we now have for the gluon-quark interaction 
\beq
\label{Deltaqgsummary}
{1 \over p_4^+}\Delta_{LIR} \es  - \, \2 \left( {1 \over \rho_1} + {1 \over \rho_3 } \right) \ ,
\eeq
where $\rho_3 = m_g^2 - (p_3-p_{3'})^2$. When $p_4^+$ tends to 0,
\beq
{1 \over p_4^+}\Delta_{LIR} \tl {-1 \over p_4^{\perp 2} + m_g^2} \ .
\eeq
The gluon-quark seagull term with $\theta_{1-1'}$ reads
\beq
\label{Hsqg2Hs}
H_s \es \hspace{-4pt} -g^2 \int [33'11'] \, \tilde \delta_{c.a} \, \delta_{\rm spins} \, \theta_{1-1'}  \,
{ ( p_3^+ + p_{3'}^+ )  \, 2 \sqrt{p_1^+ p_{1'}^+}  \over (p_{3'}^+ - p_3^+)^2} 
\nt
 \chi_{1}^\dagger T^a \chi_{1'} \, i f^{abc} \ b_1^\dagger  b_{1'} \, a^\dagger_{3b} a_{3'c}   \ .
\eeq
The common factor in Eqs.~(\ref{L1final}), (\ref{L3final}), and (\ref{Hsqg2Hs}), is
\beq
CF \es
g^2 \sum_4 \int[11'33'] \,  \theta_{1-1'} \, { \tilde \delta_{13.1'3'} \over p_4} 
\, \chi_1^\dagger  T^4 \chi_{1'} 
\nt  
\bar u_1 \gamma_\nu u_{1'} \ i f^{c_{3} c_4 c_{3'}} \ a^\dagger_{3} b_1^\dagger  b_{1'}  a_{3'} \ .
\eeq
Using $CF$, one can write the sum of terms involving $1/x^2$ in $f_{LR} \cH^{(2)}_r$ in Eq.~(\ref{HLR1pom1text  pom}) as
\beq
&& f_{LR} \left( H_A + H_\phi  + H_s \right) \\
\es f_{LR}\, CF
 \left[ {1\over p_4^+}\Delta_{LIR} ( L_A + L_\phi ) + L_s  \right] \ ,
\eeq
where 
\beq
L_A \es  2 \sum_4 \varepsilon^\nu_4 
\left[ - \varepsilon_3^*\varepsilon_{3'} \, \varepsilon_{4}^* p_{3'} \right]  \ , \\
L_\phi \es 
- {m_g^2 \eta^\nu \over p_{4}^{+2}}  \ (p_3^+ + p_{3'}^+)  \ \varepsilon^*_3 \varepsilon_{3'}  \ , \\
L_s \es
- \varepsilon^*_3 \varepsilon_{3'} \
{ p_3^+ + p_{3'}^+  \over p_4^{+ 2}} \ \eta^\nu \ .
\eeq
Since $\varepsilon_4 = \eta \varepsilon_4^-/2 + \varepsilon_4^\perp$, the small-$p_4^+$ singularity of $L_A$ comes solely from 
$\varepsilon_4^- = 2 \varepsilon_4^\perp p_4^\perp/p_4^+$, and $\varepsilon_{4}^* p_{3'} =  \varepsilon_{4}^*  p_3$, the  singular
part of transverse gluon exchange is only 
\beq
L_A & \to &
- \varepsilon_3^*\varepsilon_{3'} \ { p_4^{\perp 2} \over p_4^{+2} } \, \eta^\nu \,  (p_{3'}^+ + p_3^+) \ .
\eeq
All terms diverging as $1/p_4^{+2}$ for $p_4 \to 0$ add up to 
\beq
\label{3bod1x2cancellation}
&& f_{LR} \left( H_A + H_\phi  + H_s \right)^{\rm div} \\
\es  f_{LR} \, CF \ (- \varepsilon_3^*\varepsilon_{3'}) 
\, \eta^\nu \, {p_{3'}^+ + p_3^+ \over p_4^{+2} } 
\nt
\left\{  {-1 \over p_4^{\perp 2} + m_g^2}  \left[ \left( p_4^{\perp 2} \right)_A + \left( m_g^2 \right)_\phi \right] 
+ \left( 1 \right)_s  \right\}  \ ,
\eeq
where the subscripts indicate the terms origin. The curly bracket is 0. This result completes our demonstration that the severe 
small-$x$ singularities cancel out in the effective $Q_1\bar Q_2$ and $Q_1\bar Q_2 G_3$ components of the quarkonium 
eigenvalue problem for the renormalized Hamiltonian of QCD for heavy quarks, when the gluons are assigned the mass 
$m_g$ and the Hamiltonian includes a coupling to the auxiliary field $\phi$. The  terms remaining in $f_{LR}\cH_r^{(2)}$ after
the cancellation can be computed using the analogy between the interactions in the three-body component arranged 
pairwise as shown in Fig.~\ref{Fig3-bodyfHcancel} and the two-body interactions derived in Sec.~\ref{xinboundstates}, 
Eqs.~(\ref{hC}), (\ref{rho1-1}), and (\ref{newterm}), replacing $\rho_2$ by $\rho_3$. The resulting terms have finite limits 
as functions of the finite constituent momenta when $m_g$ tends to 0.
\begin{figure}[ht!]   
          \caption{Illustration of the Hamiltonian $Q\bar QG$ matrix elements 
                        $\langle 123| f_{LI} f_{IR} \, \Delta_{LIR} \, \cH_0^{(1)}  \, \cH_0^{(1)} | 1'2'3' \rangle$ 
                        in Eq.~(\ref{HLR1pom1text  pom}). Panels $\rm a_1$ and $\rm a_3$ correspond to an exchange of a gluon 4
                        between the quark 1 and gluon 3 or quark 1 and quark 2, which together cancel the logarithmic dependence 
                        of the self-interaction of the quark 1 shown in panel $\rm a_2$ on the gluon mass $m_g$. Panels $\rm b_1$ 
                        and $\rm b_3$ correspond to an exchange of a gluon 4 between the antiquark 2 and gluon 3 or antiquark 2 
                        and quark 1, which together cancel the logarithmic dependence on $m_g$ in the self-interaction of the 
                        antiquark 2 shown in panel $\rm b_2$. Panels $\rm c_1$ and $\rm c_3$ illustrate the exchanges of gluon 4 
                        between the gluon 3 and quark 1 and antiquark 2, which together cancel the logarithm of $m_g$ in the 
                        self-interaction of gluon 3 shown in panel $\rm c_2$. Details are described in the text.}
                         \label{Fig3-bodyffHcancel}
\begin{center}
         \includegraphics[width=0.3\textwidth]{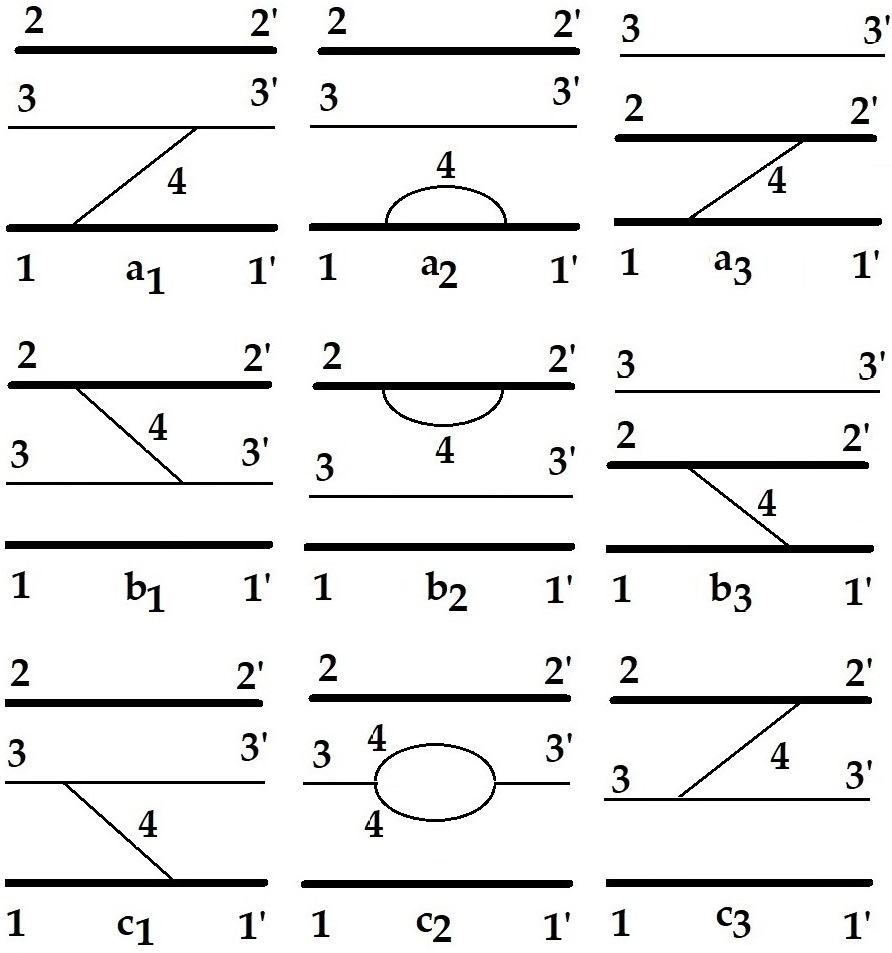}
\end{center} 
\end{figure}
\subsection{$Q \bar Q G$ interactions with factors $f_{LI}f_{IR}$}
The pertinent Hamiltonian terms in Eq.~(\ref{HLR1pom1text  pom}) are $- f_{LI} f_{IR} \,  \Delta_{LIR} \, \cH_0^{(1)} \cH_0^{(1)}$.
They are illustrated in Fig.~\ref{Fig3-bodyffHcancel}. As in the case of the quark-antiquark component discussed in Sec.~\ref{mgto0}, 
these terms are finite for $m_g > 0$ despite the presence of factors $1/x_4^2$ or $1/x_4$, because for $x_4 \to 0$ the factors $f_{LI}$ 
and $f_{IR}$ depend on the square of $p_4^-$ exponentially,
\beq
\label{fsp41}
f_{LI} \sim f_{IR} & \sim & e^{ - (s p_4^-) ^2 }  \rs e^{ - [ (s/P^+)  (p_4^{\perp 2} + m_g^2)/x] ^2 } \ .
\eeq
This behavior of $f_{LI}$ and $f_{IR}$ extends to gluon exchanges in all components of the quarkonium eigenstate. Further, as in 
Sec.~\ref{mgto0}, the effective interactions in Fig.~\ref{Fig3-bodyffHcancel} are finite when $m_g \to 0$ for fixed $|p_4^\perp|$.
But for $|p_4^\perp| \lesssim m_g$, the argument of the exponential becomes $ - [ (s m_g^2/P^+) /x_4] ^2$ and the region of 
small $x_4 \equiv x$ is suppressed only for $m_g > 0$. The small-$x$ singularities translate to the singular dependencies of the 
interaction terms on $m_g$ as in the two-body component. Below, we outline the argument that the singular dependencies on 
$m_g$ cancel out in the globally color-singlet states $|Q_{s1}\bar Q_{s2} G_{s3}\rangle$. The outline closely follows the analysis 
carried out in Sec.~\ref{mgdivergencesB}. We focus attention on the terms with $\eta^\mu \eta^\nu$ in the sum over gluon 
polarizations, since terms with $g^{\mu \nu}$ are finite and small.

Interactions illustrated in Fig.~\ref{Fig3-bodyffHcancel} involve only subsystems of two particles, or a single particle in the case 
of self-interactions. The quark and antiquark self-interactions are the same as in the component $Q_{s1} \bar Q_{s2}$. The row 
$\rm c$ of panels features the gluon self-interaction, which does not appear in the component $Q_{s1} \bar Q_{s2}$. Singularity 
of a self-interaction in a panel row cancels out with the singularities of two exchanges in the same row. The color factors of the 
three terms in each of the rows, sum to 0. To demonstrate the cancellation, it suffices to show that the momentum singularities 
of the interactions in one row are the same. 

We describe details in the case of row $\rm a$. Following the case of the two-body component, the integral over transverse momentum 
$q^\perp$ carried by the gluon in panel $\rm a_1$, is rewritten in terms of $k^\perp = q^\perp/\sqrt{x}$, where $x = p_4^+/(p_1^+ 
+ p_3^+)$, {\it cf.} Eq.~(\ref{qzktext}). This way the power of $x$ in the denominator is reduced by 1, leaving only the logarithmic 
singularity to deal with. The gluon exchange of panel $\rm a_1$ is integrated with the wave function $\varphi_{1'23'}$. Following the 
example of two-body Eq.~\ref{cancellation}) , we integrate the gluon exchange kernel, denoted here by $h^{f \hspace{-2pt}f }$, with 
$\varphi_{1'23'} - \varphi_{123}$, instead of just with $\varphi_{1'23'}$. The difference $\varphi_{1'23'} - \varphi_{123}$ vanishes for 
$p_4^+ \to 0$, the reason being explained in the last paragraph of Sec.~\ref{mglimitVIIIB2}. Hence, the integral of the difference 
with the kernel $h^{f \hspace{-2pt}f }$ has a finite limit when $m_g \to 0$. Simultaneously, we add the integral of $\varphi_{123}$ 
with $h^{f \hspace{-2pt}f }$ to the self-interaction in panel $\rm a_2$. It remains to show that the singularities due to small $p_4^+$ 
in the integrands of the self-interaction and the kernel $h^{f \hspace{-2pt}f }$ are equal to each other. Our argument proceeds 
{\it a la} Eq.~(\ref{x1tildesigma_1x}). But the equality in the three-body case cannot occur as it does in 
the two-body case because the $Q\bar Q$-component gluon-exchange kernel contains a product of two quark currents. In contrast, 
the $Q\bar Q G$ component gluon exchange in panel $\rm a_1$ involves one quark current and one gluon current. The key to the 
cancellation of $\ln m_g$ is that in the terms with tensor $\eta^\mu \eta^\nu$ the fermion and boson currents provide the factors 
$2 (p_1^+ p_{1'}^+)^{1/2}$ and $p_3^+ + p_{3'}^+$, respectively. When the gluon $p_4^+$ vanishes, {\it i.e.,} the gluon $x \to 0$, 
the product of these factors equals $2p_1^+ 2p_3^+$,  which is the same as the product of two fermion currents, or two boson 
currents. Since the kernel $h^{f \hspace{-2pt}f }$ has the color factor $-3/2$ and the self-interaction has $4/3$, one needs to
take into account the color factor $1/6$ for the gluon exchange in the octet pair to obtain the cancellation. The subsystem of quark 
1 and antiquark 2 in the three-body component is analyzed as in Sec.~\ref{mgdivergencesB}, except for the color factors that differ.
Analysis of the interactions illustrated in the panel row $\rm b$ in Fig.~\ref{Fig3-bodyffHcancel} proceeds as in the case of the panel 
row $\rm a$. In the panel row $\rm c$, the cancellation between two kernel integrals, each with a quark and a gluon current, and 
the product of two gluon currents is expected because of the equality of products of these currents when $p_4^+ \to 0$. The color 
factors mentioned above, leading to the cancellation of $\ln m_g^{-1}$, are obtained for coloreless $Q\bar Q G$ states. If a three-body 
state is not a color singlet, the diverging self-interactions are not canceled. They grow logaritmically as functions of $1/m_g$ to infinity 
when $m_g \to 0$, signifying that only the color singlets may lead to finite bound-state mass eigenvalues. 

In the integrands including the exchange kernel and wave-function difference, such as $h^{f \hspace{-2pt}f }(\varphi_{1'23'} - 
\varphi_{123})$ above, one can apply the Taylor expansion to the difference, as suggested by the case of the two-body component 
discussed in Sec.~\ref{Vphis}. For gluons, the non-relativistic approximation does not apply. Nevertheless, in the transverse directions, 
one can proceed using the expansion analogous to Eq.~(\ref{expansion}), substituting $k'^\perp = k^\perp + q^\perp$. Thus the 
kernels involving $\eta^\mu \eta^\nu$, which are even in $q^\perp$, can generate transverse harmonic oscillator potentials with explicit 
formulas for their frequencies. However, the three-body eigenvalue problem is more complex than the two-body one and the author 
has not analyzed its longitudinal dynamics enough to discuss it in this paper. The transverse oscillators are expected to contribute 
mass squared gaps. The corresponding eigenfunctions should fall off when the constituents relative momenta become extreme,
since the Hamiltonian part $H_f$ grows extremely for such momenta. 
\subsection{Interactions changing the number of gluons}
\label{changingnumber}
The bound-state effective Fock space components are coupled by the interactions that change the number of constituents. 
These interactions can be considered perturbations since they include the small coupling constant and RGPEP factors $f^s$, 
the latter significantly reducing the strength of the effective emission and absorption processes in comparison to their canonical 
counterparts. On one hand, the coupling constant extrapolation to realistic values at $s \sim 1/m$ increases the importance of 
the mixing of the Fock sectors in the eigenvalue problems for hadrons. On the other hand, the effective potentials are expected 
to further weaken the mixing. For example, one can write the  three-body wave function of the component $Q_s \bar Q_s G_s$ 
in a quarkonium as a combination of the oscillator wave functions and seek coefficients of such combination numerically from 
the eigenvalue condition for a superposition of the two-body and three-body components. The harmonic frequencies would 
contribute to the masses squared of the two-body subsystems in the three-body system. The three-body component invariant 
mass would thus increase in comparison to the free invariant mass. The probability amplitude for the component would decrease. 

If the interaction between the effective gluon and quarks in the $Q\bar Q G$ component were absent, the transitions from 
the two-body to three-body component and back would cancel the interaction terms proportional to $f_{LI}f_{IR}$ in 
Eq.~(\ref{HLR1pom1text}). The cancellation mechanism would be analogous to the cancellation mechanism between the 
self-interaction of an electron and emission of low energy photons~\cite{Bloch:1937pw}. In QCD, the situation is different 
precisely because gluons carry the color charge and constantly interact with quarks. Photons carry no electric charge and 
behave as free in the presence of electrons, neglecting effects such as the photon-photon interaction. There is no 
second-order effective oscillator potential at short distances between the electron and positron in positronium. 

In the Fock sector with two gluons, there would be an additional interaction between the gluons themselves, besides the 
gluon-quark and gluon-antiquark interactions. The suppression effects associated with the effective harmonic interactions 
would appear for all components. Dominance of the quark-antiquark component can be modeled by giving the gluon in 
the $Q\bar Q G$ component a sizable mass $m_G$ and forgetting all components with more than 1 gluon. The quarkonium 
eigenstate is thus replaced by two components, the dominant quark-antiquark component and the quark-antiquark-gluon 
component in which the gluon is assigned the ansatz mass $m_G$. The three-body component can then be eliminated, 
assuming it is small. Such modeling yields reasonable results for the masses of quarkonia and baryons in QCD with only 
heavy quarks~\cite{Serafin:2018aih,Serafin:2023pkf}. In the approach with the gluon mass $m_g$ and auxiliary field $\phi$, 
the components coupled to the dominant one in a hadron are open to study, including the issue of evaluating the effective 
mass for gluons as constituents of a heavy hadron. Excited states are expected to have increased probability of components 
with gluons. Extension of the study to light quarks requires inclusion of the running coupling constant in the Hamiltonian, 
which means extension of the RGPEP weak-coupling expansion for the Hamiltonian to 4th order. The reason is that the 
running appears first in the 3rd order. The running 3rd-order vertex needs to be combined with another interaction, bringing 
the order to 4th. The 6th order would be required for precisely including effects of the running coupling constant squared, 
which corresponds to two loops in summing over contributing quanta. In any case, the self-interactions proportional to just 
$g^2$ in the second order calculation are expected to grow as $g_s^2$ when $s$ increases, likely to produce running quark 
and gluon masses. The oscillator frequencies are also going to increase as $g_s$ increases. The coherent  evolution of various 
terms could thus define an effective low-energy Hamiltonian of QCD with the characteristic elements: the Coulomb and 
harmonic oscillator effective potentials, constituent masses, large coupling combined with narrow vertex factors, 
and finite eigenvalues only for colorless states in the limit $m_g \to 0$. 
\section{Discussion}
\label{discussion}
The RGPEP with $m_g$ and $\phi$ facilitates a formulation of the Hamiltonian eigenvalue problem for hadrons in QCD
by simplifying the treatment of divergences stronger than logarithmic and providing useful logarithms of $m_g$.
The logarithms lead to a computational definition of confinement as a feature of the Hamiltonian spectrum. Namely, 
the eigenvalues corresponding to colorless eigenstates have finite limits when $m_g \to 0$, while the eigenvalues for 
colored eigenstates tend to infinity. This definition is not new but it is here arrived at by solving the RGPEP equation 
for the derivative of the Hamiltonian operator with respect to the scale parameter $s$. The initial condition at $s=0$ 
is provided by the canonical QCD. The auxiliary field $\phi$ secures cancellation of the singularities $\sim 1/x^2$ that 
the gluon mass $m_g$ leads to. Despite the limitations of the presented calculations, they do suggest a form of the 
first approximation to an effective theory for QCD hadrons. At the scales $s$ corresponding to the binding mechanism, 
the renormalized effective Hamiltonian structure includes the free part, the Coulomb potentials, and the harmonic 
oscillator potentials. The potentials are coefficients of products of creation and annihilation operators of the 
effective quarks and gluons corresponding to the scale $s$. 

The heuristic picture is derived in QCD of heavy quarks only and in the weak-coupling expansion for the running
Hamiltonian including terms merely up to the second order. In that order, the coupling constant does not run. 
However, the approximate calculation suggests that the running of the coupling constant may extend the approximate 
picture to light quarks. As the coupling constant grows and the growth accelerates as $s$ increases, the quark and 
gluon self-interactions proportional to $g_s^2$ grow as well, hopefully leading to the masses on the order of 1/3 of
the nucleon mass. The effective potentials would increase in strength, too, but their momentum range would be limited 
by the strengthened form factors $f^s$. The potentials would hence limit the relative motion of the constituents. Such 
coherence of the interaction terms, if found, would simplify calculations of the renormalized effective Hamiltonian in 
4th and higher orders of the weak coupling expansion, adjusting the RGPEP generator introduced in Sec.~\ref{elementsRGPEP} 
to most efficiently take the running of the coupling constant into account.

The Hamiltonian evolution with the scale parameter $s$ relates description of a hadron in its center of mass system 
(CMS) in terms of constituents and in the infinite momentum frame (IMF) in terms of the QCD partons. The RGPEP 
form factor
\beq
f^s \es e^{- (  s \Delta p^-)^2} \rs e^{- \left[ (s/P^+) \Delta \cM^2/x \right]^2} ,
\eeq
has the invariant mass width depending on the ratio $s/P^+$. Suppose one sets $s$ to a value on the order of
the inverse of $\Lambda_{\rm QCD}$ of the proper magnitude for the RGPEP scheme. In the CMS of a hadron, $P^+$ 
would equal the hadron mass $m$. The effective invariant mass squared width of the RGPEP vertex factor would 
be $m \Lambda_{\rm QCD}$. The renormalized effective Hamiltonian would be expressed in terms of the creation 
and annihilation operators corresponding to constituent quarks. Their wave functions would describe the relative 
motion with limited momenta. But in the IMF, where $P^+ \to \infty$, the width of $f^s$ would become infinite, 
equivalent to $s=0$, and $f^s$ would become 1. The effective Hamiltonian interaction terms would appear equal 
to the canonical QCD Hamiltonian interaction terms. The wave functions would correspond to the parton model 
picture of the hadron. Since the invariant mass squared width is proportional to $x$, the small-$x$ partons could 
only interact with small invariant mass changes, perhaps freezing in a saturated configuration. In other words, 
interactions of the quarks and gluons with large $p^+$ are little evolved in the RGPEP. Identified as partons, they 
interact according to the canonical QCD. By comparison, interactions of the quarks and gluons with small $p^+$, 
on the order of a hadron mass or its sizable fraction, are evolved a lot in the RGPEP towards the picture of 
constituent models.

The examples of effective Hamiltonian terms derived using our method suggest a way of thinking about the issue 
of vacuum effects in the FF of Hamiltonian dynamics. Such effects are extensively discussed in the literature, {\it e.g.,} see~\cite{Casher:1973vh,Kogut:1972di,Shifman:1978bx,Shifman:1978by,Gao:2021dbh,Ji:2024oka,Shuryak:2026pqt}. 
In the FF of dynamics, one is inclined to expect that the vacuum modes are associated with momentum $p^+=0$, 
for the vacuum cannot carry momentum. Since the singularities for $p^+ \to 0$ require regularization and the associated 
divergences need counterterms, one may expect that the finite parts of these counterterms may be associated with 
the vacuum effects considered in the IF of dynamics~\cite{Wilson:1994fk}. However, the IF vacuum effects are primarily 
considered for the purpose of deriving the associated quark and gluon interactions. We note that the RGPEP does 
yield the effective interactions for $s > 0$ that grow with the distance, such as between a quark and an antiquark. 
Such interactions are obtained even if no vacuum effects are included at $s=0$, which is the case in our approach 
with $m_g$ and $\phi$. We can hence hypothesize that the effective interactions that are associated with the vacuum 
in the IF of dynamics may appear in the FF of dynamics as a result of the RGPEP evolution of the effective Hamiltonian 
at $s>0$, while no vacuum-induced interactions are included in the initial condition at $s=0$. If so, instead of studying 
the vacuum effects through the regularization and computation of the counterterms, one may study the effective 
interactions using the RGPEP at $s>0$ in the theory with $m_g \to 0$. 
\bibliography{ReferencesMassiveGluons2026}
\end{document}